\documentclass[reqno,10pt]{amsart} 
\usepackage[english]{babel}
\usepackage{a4wide}
\usepackage{amsmath}
\usepackage{amsfonts}
\usepackage{amssymb}
\usepackage{amsthm}
\usepackage{bbm}
\usepackage[utf8]{inputenc}
\usepackage[hidelinks]{hyperref}
\usepackage[capitalise]{cleveref}
\usepackage{appendix}
\usepackage{booktabs}
\usepackage[margin=2.3cm]{geometry}
\usepackage{comment}

\usepackage{enumitem}

\setlist{nolistsep}
\usepackage{mathtools}
\usepackage{mathrsfs}
\usepackage{lettrine}
\usepackage{mfirstuc}
\usepackage{multicol}
\usepackage{multirow}
\usepackage{makecell}
\usepackage[table]{xcolor}
\usepackage{float}
\usepackage[maxbibnames=99]{biblatex}
\usepackage{makecell}
\usepackage{array}
\renewcommand*{\P}{\mathbb{P}}
\newcommand*{\E}{\mathbb{E}}
\newcommand*{\R}{\mathbb{R}}

\newcommand*{\Z}{\mathbb{Z}}

\newcommand*{\N}{\mathbb{N}}

\renewcommand*{\Xi}{\varXi}
\renewcommand*{\epsilon}{\varepsilon}
\renewcommand*{\theta}{\vartheta}
\renewcommand*{\Theta}{\varTheta}

\renewcommand*{\Delta}{\varDelta}

\newtheorem{theorem}{Theorem}[section]

\newtheorem{lemma}[theorem]{Lemma}
\newtheorem{assumptionA}{Assumption}

\theoremstyle{remark}
\newtheorem{remark}[theorem]{Remark}

 \title[Metastates for $d$-dimensional spherical spins]{
 $d$-dimensional spherical ferromagnets in random fields: Metastates, continuous symmetry breaking, and spin-glass features}
 \date{\today}
 \author{Kalle Koskinen$^1$}
 \address{$^1$Gran Sasso Science Institute, {\tt kalle.koskinen@gssi.it}}
 \author{Christof K\"ulske$^2$}
 \address{$^2$Ruhr University Bochum, {\tt christof.kuelske@ruhr-uni-bochum.de}}

\begin{document}
\begin{abstract} We study the large-volume behavior of the spherical model for $d$-dimensional local 
spins, in the presence of $d$-dimensional random fields, for $d\geq 2$. 
We compare two models, one with volume-scaled random fields, 
and another one with non-scaled random fields, 
on the level of Aizenman-Wehr metastates, Newman-Stein metastates, as well as  overlap distributions. We show that in $d\geq 2$ the metastates are fully supported on a continuity of random product states, with weights which we describe, 
for both models. For the non-scaled random fields, the set of a.s. cluster points of Gibbs measures contains these product states, but behaves differently 
in the 'recurrent' spin dimension $d=2$  where it also contains non-trivial mixtures of tilted measures.  
For the scaled model, moreover the overlap distribution displays spin-glass characteristics, as it is non-self averaging, and shows replica symmetry breaking, although it is ultrametric if and only if $d=1$. For $d\geq 2$ it oscillates chaotically on a set of
continuous distributions for large volumes, while 
the limiting set contains only discrete distributions in $d=1$. Our results are based on concentration estimates, analysis of Gibbs measures in finite but large volumes, and the asymptotics of $d$-dimensional random walks and their spherical projections. 

\medbreak\noindent
\emph{\keywordsname:} Gibbs measures; random symmetry breaking; metastates; overlaps; non self-averaging; chaotic size dependence \\
\emph{MSC2020:} Primary 60K35; Secondary 60G55, 60G60, 82B20, 82B21.
\end{abstract}
\maketitle


\section{Introduction}\label{sec_introduction}


Metastates \cite{Newman1997, Aizenman1990, Bovier2006} are concepts to describe the asymptotic volume dependence of 
spin models with quenched disorder, which are meaningful and non-trivial when there is 
the phenomenon of \textit{random symmetry breaking}  appearing. 
By random symmetry breaking for a disordered system we understand firstly  that 
there is more than one infinite-volume Gibbs measure, for almost every realization w.r.t to the disorder measure $\P$. 
Secondly, that there is no natural obvious preselection in terms of boundary conditions which would determine 
in which of the possible infinite-volume Gibbs measures the system will be, and there is a 
\textit{competition between states} for dominance at a given particular large volume. 
Examples for such systems are spin-glasses with deterministic or free boundary conditions, and 
random field systems with free boundary conditions \cite{Edwards1975, Sherrington1975,  Newman2022, Aizenman1990}.
See the recent \cite{ chatterjee2024featuresspinglassrandom}
for a discussion of the random-field Ising model from 
a spin-glass perspective.  
We mention that there is a closely related study 
of systems with deterministic couplings in the bulk, but random boundary conditions \cite{Endo2024}.

Let us fix a realization of the disorder, here generically denoted by $h$. (In our present paper $h=(h(i))_{i\in \N}$ will 
be a collection of random fields, but for the sake of this general exposition we can think of it as denoting
something more general, for instance, the collection of all coupling constants in the infinite volume for a spin glass model.)  
Letting the finite volume of the system, indexed by a natural number $n$ for some fixed type of boundary condition (e.g. open
boundary), tend to infinity,  one obtains a sequence of finite-volume Gibbs measures $\mu_n^h$. 
It is a signature of \textit{random symmetry breaking} that these measures may oscillate between the various different available 
Gibbs measures for large enough $n$, along the volume sequence.  
Depending on system parameters, dimensionality of the system, and type of randomness, adding randomness to a deterministic system may wash out a phase 
transition, perturb the phase transition 
or even create new types of phases. 
The phase transition is washed out e.g. 
in the lattice random field Ising model in
two spatial dimensions, and for continuous 
spins in four spatial dimensions, see \cite{Aizenman1990}. In a recent resurgence of interest in random field models, quantitative refinements 
in the random field model and other systems 
of this phenomenon have since been obtained 
in \cite{barnir2022upperlowerboundscorrelation, bowditch2021twodimensionalcontinuumrandomfield, Dario2024}. In our present context we assume that the system does have distinct ordered phases, which 
will be proved for our model system below.

How to describe the asymptotics of such a sequence for large volumes? 
In a direct approach one may look at cluster points of the random sequence, in the space of infinite-volume Gibbs states, 
for the particular realization of the disorder variables.  This yields certainly valuable information. However, in a more sophisticated approach one looks at measures (of the infinite-volume states) which 
describe the asymptotic behavior \textit{quantitatively} by assigning weights to Gibbs measures which occur in such sequences. 
These measures will be called \textit{metastates.}

\textit{Newman-Stein metastate. } As one possible construction of a  metastate one may look a large $N$-limit of the empirical measures 
$$\overline{\kappa}^h_N=\frac{1}{N}\sum_{n=1}^N \delta_{\mu_n^h}$$ 
along volume sequences, which we will call the Newman-Stein metastates (NS-metastate).  
This notion goes back to 
Chuck Newman and Dan Stein, who were motivated by the study empirical measures in dynamical systems, 
having an analogy between the index of the volume sequence in statistical mechanics, and a "time" in 
dynamical systems in mind. 

\textit{Aizenman-Wehr metastate.} 
One may also look at the so-called Aizenman-Wehr metastate $\kappa^h(d\mu)$ which  again
is a probability distribution on the random Gibbs measures of the system for fixed disorder $h$. 
It can most directly be obtained by a conditioning procedure, to be described below. 
(For reasonable systems it coincides with the NS-metastate, (only) if the latter is constructed  along  
volume sequences which need  to be sparse enough. This is necessary 
to have enough decorrelation between different volumes in order to have some quasi-independence on the level of the weights. The study of the asymptotics of the NS-metastate for non-sparse volume sequences yields 
different interesting information, which will be discussed  in the study of our present model system below. 
) For more generalities on metastates see \cite{Stein2013, Koskinen2023, Cotar2018, Arguin2014, Iacobelli2010, Arguin2010, Kuelske1997}. For related spin glass research see also \cite{Talagrand2011, Panchenko2013, Talagrand2006, Auffinger2014, Chen2025}.

It is the purpose of this paper to investigate and compare the description of the large-volume asymptotics in two relevant 
but non-equivalent views: full metastates and overlaps, in their dependence on spin-dimension.

\subsection{The ferromagnetic spherical model for $d$-dimensional spins in i.i.d. random fields}
In this article we consider the vector-valued ferromagnetic spherical model,  in the presence of quenched $\R^d$-valued random fields $h=(h(i))_{i\in \N}$, 
which are i.i.d w.r.t a disorder distribution $\P$, of which  we assume 
expectation zero, and existence of the absolute third moments. Throughout the paper we assume for the spin-dimension that $d\geq 2$, but we will also make comparisons to the case $d=1$ which was treated in \cite{Koskinen2023}. 
The conditions for the random field are explicitly detailed in \cref{assump:A}.

The $n$-spin Hamiltonian $H_n^h$ at fixed realization of the random fields $h$ is of the form 
$$ 
H_n^h (\phi) := -\frac{1}{2 n}\sum_{1\leq i,j \leq n}\langle \phi(i),\phi(j)\rangle -\sum_{1\leq i\leq n}\langle h(i),\phi(i)\rangle
$$
We assume that the standard deviation of the $j$-component, i.e. 
$s_j=\sqrt{\E (h_j(i)^2)}$ exists for all components $j=1,\dots, d$, and that it is strictly positive for all 
components. We allow that it may possibly be different for different components, but impose that the 
distribution is non-degenerate in the sense that  the covariance matrix $\Sigma := \big(\E (h_j(i) h_k(i))\big)_{1\leq j,k \leq d}$ has 
full rank $d$.

We then consider the quenched finite-volume Gibbs distributions $\mu_n^h(d\phi(1),\dots,d\phi(n))$ 
 for  
vector-valued local spin variables $\phi(i)$ taking values in $\R^d$, and sitting at sites $i\in \{1,\dots,n\}$.  
The measure $\mu_n^h$
is defined by putting the overall spherical constraint $$\sum_{i=1}^n \langle \phi(i),\phi(i)\rangle=n$$
on the sums of two-norms in the fibers, over all sites, 
and by putting the exponential Boltzmann-Gibbs factor with inverse temperature $\beta>0$ in front of the negative 
Hamiltonian, relative to the surface measure on the resulting sphere. Formally, the finite-volume Gibbs states are given by their actions on observables
\begin{align*}
f \mapsto \mu_n^{h} [f] := \frac{1}{Z_n^h} \int_{\left( \mathbb{R}^d \right)^n} d \phi \ e^{- \beta H_n^h (\phi)} \delta \left( \sum_{i=1}^n \left< \phi(i), \phi(i) \right> - n \right) f (\phi) . 
\end{align*}
See \cref{sec:constructionOfStates} for a more detailed explicit construction of the finite-volume Gibbs states, and see \cref{def:mixture_representation} for the primary representation of the finite-volume Gibbs states as integral mixtures of shifted microcanonical probability measures, see \cref{def:FVMCM} and \cref{def:FVMCMprobabilistic}, with exponential weights given by the finite-volume exponential tilting functions, see \cref{def:FVETF}.

Note that the spherical constraint has an even bigger symmetry than the interaction, 
as the site label $i$ and component label $j$ play the same role for the constraint, while 
they play different roles for the interaction. A joint rotation of all random fields and all spins 
by an element in $O(d)$, the orthogonal group of $d \times d$ real matrices, leaves the Hamiltonian invariant, but as we said, we allow the random field distribution 
to be also possibly anisotropic. Clearly, the model with 
$d$-dimensional spins is not equal to the model for $d=1$ with $d n$ spins, 
and also the results are very different, as there is a \textit{continuous 
symmetry breaking for $d\geq 2$} with uncountably many 
pure states. 
Our study of $d$-dimensional spins in random fields nevertheless builds on the one-dimensional case \cite{Koskinen2023}, 
but we will find new difficulties, new phenomena and a richer structure than in the one-dimensional case, 
which we will display for the sake of comparison. Our present treatment of the actual proofs 
contains a number of simplifications and improvements which can also be applied for $d=1$. The primary improvement being the relaxation of the condition of existence of moments of  the order $4+\varepsilon$ to only requiring the existence of second moments in dimension $d\geq3$ and the existence of the absolute third moments in dimension $d=2$. Most results hold with only a finite second moment assumption, but refined arguments with the assumption of a finite absolutely third moment were necessary in $d =2$ to prove e.g. the results on the a.s. set of cluster points for Newman-Stein metastates.  
\medskip
\subsection{Main results}
The first important fact is that there is ferromagnetic order in a regime of large enough inverse temperature, 
and small enough disorder. Symmetry breaking for the total magnetization occurs,
which may point in all possible directions of $\R^d$, with a fixed size of the magnetization vector.  
Thus, for $d\geq 2$ a continuity of infinite-volume Gibbs states 
occurs, which may be indexed by an angle $\Omega \in \mathbb{S}^{d-1}$.

As there is no natural preselection 
by boundary conditions which would enter the Hamiltonian, but we are dealing with free boundary conditions,  there will be a random symmetry breaking w.r.t the random field distribution, as all states for different angles 
may occur (and would be equivalent w.r.t the random field distribution if the latter is isotropic). 

For the presentation of our main limit results on random symmetry breaking and large-volume asymptotics, let us  
restrict to this regime of ferromagnetic order, which turns out to be given by the condition 
$$\sqrt{1-\frac{d}{\beta}}>\Vert s \Vert, $$ 
with $\Vert s \Vert^2 := \sum_{j=1}^d \E h_j(i)^2 $ measuring disorder strength. Note that one part of \cref{assump:A} is that properties of the random field and inverse temperature are chosen such that we are precisely in this regime of ferromagnetic ordering. See \cref{thm:MSregime} and \cref{thm:free_energy} for the connection between the finite-volume Gibbs states via the partition function, the finite-volume exponential tilting functions and their uniform limit, and the regime of ferromagnetic ordering given here.


We emphasize that the column of the table corresponding to $d=1$ are results that are obtained in \cite{Koskinen2023}. The methods used to prove the results in this manuscript for $d \geq 2$ can be almost directly applied to the case $d=1$, noting the lack of continuous symmetry breaking, to obtain the same results with weaker assumptions. Let us now explain our results which are summarized in the Table~\ref{tab:results} line-by line.  \medskip

\rowcolors{2}{gray!25}{white}
\begin{table}

    \centering
    \begin{tabular}{c|c|c|c}
    \rowcolor{gray!50}
        \multicolumn{1}{c|}{}& $d=1$  & $d=2$  & $d\geq 3 $\\
        \rule{0pt}{5mm}
        Cluster points, lattice RF & $\overline{\{\bar\nu^h_z, z \in \Z\}}$&$\overline{\{\bar\nu^h_z, z \in \Z^2\}} $ & $\{\nu^h_{\Omega}, \Omega \in \mathbb{S}^{d-1}  \}$ \\
        \rule{0pt}{5mm}
  Cluster points, cont. RF & $\overline{\{\bar\nu^h_z, z \in \R\}}$&$\overline{\{\bar\nu^h_z, z \in \R^2\}}$  & same as for lattice RF \\
  \rule{0pt}{4mm}
   
        AW-metastate & $\frac{1}{2}(\delta_{\nu^h_+} +\delta_{\nu^h_-})$ & \multicolumn{2}{c}{$\int_{\mathbb{S}^{d-1}}d\Omega\, \rho_\P(\Omega)\delta_{\nu^h_{\Omega}}$ }\\
        \rule{0pt}{4mm}
        Law of NS-metastate& $n_+\delta_{\nu^h_+} +(1-n_+)\delta_{\nu^h_-}$ &  \multicolumn{2}{c}{$\int_0^1 \delta_{\nu^h_{\hat B_t}}dt, \,\,B_\cdot$ indep.} \\
        \rule{0pt}{4mm}
         Cluster points NS-metast. &$p\delta_{\nu^h_+} +(1-p)\delta_{\nu^h_-}, p\in [0,1]$& \multicolumn{2}{c}{$\int_{\mathbb{S}^{d-1}}\eta(d\Omega)\delta_{\nu^h_{\Omega}}, 
         \eta \in \mathcal{M}_1(\mathbb{S}^{d-1})$}     \\
    \end{tabular}\medskip
    \caption{Overview of the results on random symmetry breaking in the ferromagnetic regime, depending on spin-dimension $d$.  The measures $\nu^h_{\Omega}$ denote pure states magnetized in direction $\Omega$, on which 
    the metastates are supported. $d=1,2$ cluster points also contain mixtures of exponential tilts $\bar\nu^h_z$.}
    \label{tab:results}
\end{table}
%

\subsection{Cluster points: pure states and mixtures of exponential tilts}
In the first line of the table the set of 
the cluster points of the sequence of Gibbs measures $\mu^h_n$ on the increasing volumes $\{1,\dots,n\}$
w.r.t. the topology that metrizes weak convergence on the state space $(\mathbb{R}^d)^\mathbb{N}$ given in \cref{sec:topologyMeasures}  
for $\P$-a.e. random field realization $h=(h(i))_{\in \N}$  are displayed, depending on spin-dimension $d$.  

\textit{Pure States}. In the ferromagnetic regime, the role of the pure infinite-volume Gibbs states is played by 
a collection of product measures $\nu^h_{\Omega}$, where $\Omega$ 
is a vector which runs over the unit sphere in local spin space, i.e. $\Omega\in \mathbb{S}^{d-1}\subset \R^d$. 
These measures $\nu^h_{\Omega}$ turn out to be Gaussian measures which are independent over the spins. 
Moreover,  they keep their dependence 
on the infinite-volume collection of fields $h$ in a strictly local way: This means that the field $h(i)$ acting 
on the $i$-th spin variable in the Hamiltonian enters the state only 
through the factor for the local spin $\phi(i)$, and we have 
$$ 
\nu^h_{\Omega}(d\phi)=\prod_{i}\nu^{h(i)}_{\Omega;i}(d\phi(i))
$$
All local factors feel $\Omega$ in the same way. Physically speaking 
 $\Omega$ describes an overall orientation of the spin distribution 
which carries over to an appearance of a macroscopic total magnetization in the direction $\Omega$.

One can summarize the structure of the pure states $\nu_\Omega^h$ by stating that the single-dimensional $(i,j)$-marginal for the field component $\phi_j(i)$
of this random probability distribution is almost surely distributed like
\begin{align*}
\nu_\Omega^h|_{(i,j)} \sim \frac{1}{\sqrt{\beta}} G_j(i) + \sqrt{1 - \frac{d}{\beta} - || s ||^2}\,\, \Omega_j + h_j(i)   
\end{align*}
where $G := \{ G_j (i)\}_{j \in \{ 1,...,d\}, i \in \mathbb{N}}$ is collection of independent identically distributed standard Gaussians. Note that the Gaussian $G$ and the random field $h$ exist on different probability spaces. For more details and precise definitions, see \cref{def:PS} for the definition of the probability measure $\nu^{x,y,h}$, see \cref{sec:relabelling} for the connection of $\nu^{x,y,h}$ to the pure state $\nu^h_\Omega$, and see \cref{thm:CSD} for the pure state cluster point result in dimension $d \geq 3$.

Looking at the first line of the table, we see that the collection of measures  
$\{\nu^h_{\Omega}, \Omega \in \mathbb{S}^{d-1}       \}$ 
equals the \textit{full set} of  cluster points of the sequence of Gibbs measures, \textit{only} 
in spin dimension $d\geq 3$. The full set of cluster points turns out to be $\P$-almost surely strictly 
bigger in dimensions $d=1,2$. In these lower dimensions it contains also mixtures of a particular type, which  
we are going to describe now. \medskip

\textit{Spherical mixtures of $z$-tilts}. 
In lower dimensions $d=1,2$ the a.s. cluster points of the Gibbs measures also contain non-product measures which are of the type 
$$ 
\bar\nu^h_z := \left( \int_{\mathbb{S}^{d-1}}d\Omega \,\, e^{\beta r^* \langle \Omega, z \rangle} \right)^{-1}\int_{\mathbb{S}^{d-1}}d\Omega \,\, e^{\beta r^* \langle \Omega, z \rangle}\nu^h_{\Omega}, 
$$
where $z\in \R^d$ is a parameter for an exponential tilt, and 
$d\Omega$ is the non-normalized surface measure on $\mathbb{S}^{d-1}$. These measures thus describe $\mathbb{S}^{d-1}$-averages of exponentially 
$\langle \Omega, z \rangle$-tilted pure states in total magnetization direction $\Omega$. 

It is simple but important to realize that, in any dimension, the topological closure of the set of the mixtures 
$\overline{\{\bar\nu^h_z, z \in \Z^d\}}$ equals the union of the countable set 
$\{\bar\nu^h_z, z \in \Z^d\}$ together with the set of product states 
$\{\nu^h_{\Omega}, \Omega \in \mathbb{S}^{d-1}\}$. Indeed, the latter product states reappear 
as \textit{fully concentrated tilt-averages}, which are  obtained as limit points of sequences of tilted states, 
for which the tilt size  $|z|$ goes to infinity.   
Because of this, in $d=1,2$ the set of cluster points  (which we recall is always a closed set, by elementary topological generalities) 
contains mixtures, but necessarily also the pure states,  while the product states are the only cluster points 
$d\geq 3$. See \cref{thm:CSD} for the full proof of these results.

Comparing the  first two lines of the table we find differences only with regard to the nature of the random field 
distribution to be of \textit{lattice nature} or \textit{continuum nature}. 
By lattice nature we mean that sums of random fields a.s. lie in a multiple of $\Z^d$. 
(This can be generalized to different lattices, see \cref{thm:CSD} and \cref{rmk:possibleValues} concerning possible and recurrent values of random walks below.)

\medskip

\subsection{Aizenman-Wehr metastate}
The Aizenman-Wehr metastate whose behavior is displayed 
in the third line of the table 
describes intuitively the probability to  which infinite-volume Gibbs
measure a very large volume will be asymptotically close to, if the choice of the large volume is independent 
of the realization of the quenched randomness.  

The expression we see, namely  
 $$\int_{\mathbb{S}^{d-1}}d\Omega\, \rho_\P(\Omega)\delta_{\nu^h_{\Omega}}$$
 is a (possibly distorted) sphere-average of the pure states, and only of the pure states. (Note that there 
 are models for which also mixtures appear in the metastates, see \cite{Kuelske1997}). 
 The density function describing the distortion $\rho_{\P}$ turns out to depend on the random field distribution only via its 
 covariance matrix $\Sigma$, and it is proportional to $\left( \left< \Omega, \Sigma^{-1} \Omega \right> \right)^{-\frac{d}{2}}$.

 Hence it has to equal a constant 
 in the case of an isotropic random field distribution, but also 
 when $h(i)$ are the increments of a simple symmetric random walk. 
 It is in particular absent in the case $d=1$, and so the symmetric expression $\frac{1}{2}(\nu^h_+ +\nu^h_-)$ known 
 from \cite{Koskinen2023} appears.
 Note that the non-product cluster points in $d=1,2$ obtain zero mass  i.e. they are 
 \textit{invisible under the AW-metastate}. 
 This means that for a typical realization of randomness, the system will be typically very close to a pure state, 
 even in $d=1,2$.  This is to say, mixtures are atypical in the AW-metastate sense. 
 Note that in particular the AW-metastate does not distinguish between a continuous and a discrete random field distribution,
as opposed to set of the cluster points which makes this distinction. See \cref{thm:AWMS} for the construction of the Aizenman-Wehr metastate. \medskip

\subsection{Newman-Stein metastate}
The NS-metastate describes the 
limit of the empirical means $\overline{\kappa}_N^h$ which is typically observed through (possibly non-linear) but 
local observables on the space of measures, which we can think to be of the form 
$g(\mu)=\tilde g(\mu(f_1), \dots, \mu(f_l))$, with local spin observables $f_i:(\mathbb{R}^d)^\mathbb{N} \to \mathbb{R}$. In this article, due to being able to prove results directly for generic bounded Lipschitz functions on $(\mathbb{R}^d)^\mathbb{N}$, see \cref{def:PS} for an example, we are able to proceed more directly. Limit results for this object remember the dependence 
on the fields $h$ only in a local way, where the distribution over the product measures $\nu^h_{\Omega}$ turns out to be 
governed by the total field sum $\sum_{i=1}^N h(i)$ which becomes asymptotically independent 
of the behavior of $h(i)$ in a window of finitely many $i$'s. 
This (by strong Gaussian 
approximation on the above random walk, and rescaling) 
gives rise to an independent $\R^d$-valued Brownian motion $B_t$ which just remembers the covariances of the 
$h(i)$'s. For details compare the proof of \cref{thm:NSdistconv} below. 
This Brownian motion then enters the limiting expression 
via its projection to the sphere $\mathbb{S}^{d-1}$, which is denoted 
by $\hat B_t$. This explains the limiting result of the table as the random 
empirical mean $\int_0^1 \delta_{\nu^h_{\hat B_t}}dt$, where $\hat B_t$ is independent of 
the collection of random fields $h$, see \cref{thm:NS_asymptotic} and \cref{thm:NSdistconv}. 

The form of the result is the same in \textit{all} spin dimensions $d\geq 1$, but 
the limiting formula is richer in higher dimensions. 
One notes that the corresponding expression 
becomes very explicit in $d=1$, where it simplifies and an arcsine-distributed variable $n_+$ pops up 
which governs the random weight on the two atoms $\nu_+^h$ and $\nu_+^h$. 

The last line of the table states that the \textit{a.s. cluster points of the empirical averages} of the form 
$\kappa_N$ are equal to the full set of \textit{possible mixtures}  
obtained by choosing arbitrary mixing measures on the sphere, i.d. 
$\{\int_{\mathbb{S}^{d-1}}\eta(d\Omega)\delta_{\nu^h_{\Omega}}, 
         \eta \in \mathcal{M}_1(\mathbb{S}^{d-1})\}$. 
         This formula can again be read in all dimensions $d\geq 1$. It is a huge set in $d\geq 2$, but   
it also holds in $d=1$ where it simplifies to be the set convex combinations between just  two measures.  
We note that the bold statement that really all Borel measures on the sphere are obtained, 
 is naturally explained  by the corresponding result on the sphere. Namely,  
 the empirical measures of the sphere-projected Brownian motions  $\int_0^1 \delta_{\hat B_t}dt$ 
have all Borel measures on the sphere as a.s. limit points, see \cref{thm:NSclusterpoints}.

\subsection{Overlap distribution for non-scaled random fields}

The study of overlaps between two replicas of a disordered system with the same disorder configuration 
is a central part of many works in disordered systems, in particular spin glasses \cite{Mezard1986, Talagrand2011, chatterjee2024featuresspinglassrandom}. We mention that overlaps were constructively used to analyze the breakup of the free state into a continuum of non-translation invariant glassy pure states for  Potts and clock models at low temperatures on trees \cite{JiLa26,  Coquille2026}.

For two replicas $a$ and $b$, we consider the overlap distribution defined as the following pushforward measure
\begin{align*}
{R_n^{a,b}}_* (\mu_n^h \otimes \mu_n^h), \ R_n^{a,b} (\phi^a, \phi^b) := \frac{1}{n} \sum_{i=1}^n \left< \phi^a (i), \phi^b (i)\right> .    
\end{align*}
To see heuristically what could be expected, note that the finite-volume Gibbs state can be asymptotically approximated in law, in the sense of the proof of \cref{thm:AWMS}, by $\mu_n^h \approx \nu^h_\frac{S_n}{|| S_n||}$, and one can readily informally compute the overlap distribution in the large-$n$ limit since the $\nu_\Omega^h$ probability measures are factorized. However, to give a proof of this, we will proceed with a more direct approach as follows. By using the representation given for the finite-volume Gibbs states in \cref{def:mixture_representation_2}, it follows that
\begin{align*}
{R_n^{a,b}}_* (\mu_n^h \otimes \mu_n^h) = \int_{B_{2d} (0,1)} \alpha_n^h (dx^a, dy^a)  \int_{B_{2d} (0,1)} \alpha_n^h (d x^b,dy ^b) \ {R_n^{a,b}}_* (\nu_n^{x^a, y^a} \otimes \nu_n^{x^b, y^b}) ,     
\end{align*}
where the form of the mixing probability measure $\alpha_n^h$ is given in \cref{def:mixingProbabilityMeasure}. By direct estimation, we show that ${R_n^{a,b}}_* (\nu_n^{x^a, y^a} \otimes \nu_n^{x^b, y^b}) \approx \delta_{\left< x^a, x^b \right> + \left< y^a, y^b \right>}$ in the large $n$-limit uniformly in the variables $(x^a,y^a,x^a,x^b)$ for expectations of bounded Lipschitz functions. The precise statement and proof are given in \cref{thm:overlapMC}. The analysis of the asymptotics of the mixing measure proceed in precisely the same way as for the non-overlap case, and the almost sure limiting points of the overlap distribution show an anomalous structure in dimension $d = 2$, just as in the case without the overlaps. This result, along with the limiting points for $d \geq 3$, are given and proved in \cref{thm:overlapCSD}. For dimensions $d \geq 3$, the random walk is always transient, and one finds that $\alpha_n^h \approx \delta_{(r^* \frac{S_n}{|| S_n||}, y^*)}$ in the large-$n$ limit, where $r^* = \sqrt{1 - \frac{d}{\beta} - || s||^2}$ and $y^* = s$. It follows then that ${R_n^{a,b}}_* (\mu_n^h \otimes \mu_n^h) \approx \delta_{{(r^*)}^2 + || y^*||^2}$, and we see that the limit of the overlap distributions is $\mathbb{P}$-almost surely trivial.
This behavior carries over to the convergence in law of the overlap distribution, and we conclude this subsection with the following triviality result.
\begin{theorem}
Suppose that $h$ satisfies (A).

For any $d$, it follows that
\begin{align*}
\lim_{n \to \infty} {R_n^{a,b}}_* (\mu_n^h \otimes \mu_n^h) = \delta_{{(r^*)^2 + || y^*||^2}} = \delta_{1 - \frac{d}{\beta}}
\end{align*}
in law, and hence also in probability.
\end{theorem}
\noindent
We see that although the finite-volume Gibbs states can converge along subsequences to an uncountable number of pure states indexed by the sphere, which can be regarded as a spin-glass feature, the limit of the overlap distribution is trivial.

By a similar (and easier) computation we 
see that in the paramagnetic regime 
$1-d /\beta \leq \Vert s\Vert ^2$
the overlap takes the $\beta$-independent 
deterministic value 

\begin{align*}
\lim_{n \to \infty} {R_n^{a,b}}_* (\mu_n^h \otimes \mu_n^h) = \delta_{\Vert s\Vert^2}
\end{align*}
This ensures continuity of the overlap distribution 
at the second-order 
phase transition curve $1-d/\beta = \Vert s\Vert ^2$.

\medskip
\subsection{Overlap distributions for scaled random fields}
In the spirit of \cite{chatterjee2024featuresspinglassrandom}, we modify the Hamiltonian by scaling the random fields by a volume dependent term $\frac{1}{\sqrt{n}}$, so that we redefine the Hamiltonian as
\begin{align*}
H_n^{\frac{h}{\sqrt{n}}} (\phi) := - \frac{1}{2n} \sum_{i,j=1}^n \left< \phi(i), \phi(j) \right> - \frac{1}{\sqrt{n}} \sum_{i=1}^n \left< h(i), \phi(i) \right>
\end{align*}
for the rest of this section, and we will henceforth refer to this variation of the model as the model with scaled random fields.

For the overlap distribution of the scaled random fields, we proceed as we did for the non-scaled case. The critical difference appears in the asymptotics of the mixing measures, and we obtain the following approximation
\begin{align*}
{R_n^{a,b}}_* (\mu_n^{\frac{h}{\sqrt{n}}} \otimes \mu_n^{\frac{h}{\sqrt{n}}}) \approx \int_{\mathbb{S}^{d - 1}} \gamma^{\frac{S_n}{\sqrt{n}}} (d \Omega^a) \int_{\mathbb{S}^{d-1}} \gamma^{\frac{S_n}{\sqrt{n}}} (d \Omega^b) \ \delta_{{(r^*)}^2 \left< \Omega^a, \Omega^b \right>}  
\end{align*}
valid for all dimensions, where now $r^* = \sqrt{1 - \frac{d}{\beta}} > 0$, the probability measure $\gamma^z$ is given by
\begin{align*}
\gamma^z(d \Omega) =  \frac{d \Omega \ e^{\beta r^* \left< z, \Omega \right>}}{\int_{\mathbb{S}^{d-1}}d \Omega \ e^{\beta r^* \left< z, \Omega \right>}} ,    
\end{align*}
and we denote $\mu_n^{\frac{h}{\sqrt{n}}}$ to be the corresponding scaled random field finite-volume Gibbs states to differentiate it from the non-scaled case. The precise statement and proof of this approximation is given in \cref{thm:overlapApproximation}.

To describe the asymptotic behavior of the overlap 
distribution we denote by $\rho^R(dq)$
the family of distributions on 
the interval $[-1,1]$
depending on a radial variable $R>0$
as follows 
\begin{equation}
 \begin{split}   
&\int\rho^R(dq)f(q):=
\int_{\mathbb{S}^{d-1}}\gamma^{R e_1}(d\Omega^a)
\int_{\mathbb{S}^{d-1}}\gamma^{R e_1}(d\Omega^b)
f\bigl((r^*)^2\langle \Omega^a,\Omega^b \rangle\bigr) 
 \end{split}
\end{equation}
where $e_1$ is an arbitrary unit
vector on the $d-1$-dimensional sphere.  We have replica-symmetry breaking 
and non-self averaging of the overlap distribution. More precisely, the following theorem holds.

\begin{theorem} Suppose that $h$ satisfies (A).

It follows that
\begin{align*}
\lim_{n \to \infty} d_{\operatorname{BL}_1} \left( {R_n^{a,b}}_* (\mu_n^{\frac{h}{\sqrt{n}}} \otimes \mu_n^{\frac{h}{\sqrt{n}}}),  \rho^{\frac{|| S_n ||}{\sqrt{n}}} \right) = 0 \text{ a.s}.    
\end{align*}
\end{theorem}

While the functional dependence of the measure
$\rho^R(dq)$ involves only $\beta, d$, 
the total strength of the random fields in volume $n$
enters the formula by means of the random quantity $\Vert S_n\Vert/\sqrt{n}$.

Let us explain these results and compare with the findings in \cite{chatterjee2024featuresspinglassrandom} on the random field Ising model. 

\subsection{$d=1$: Discrete overlap distribution, RSB, NSA, RFIM-like behavior}

For $d=1$ the situation is as follows. 
The overlap distribution 
$\rho^{\frac{|S_n|}{\sqrt{n}}}$ is supported on the values $\pm (r^*)^2$. 
The quenched expectation 
takes the value 

$$\int_{[-1,1]}q\rho^{\frac{|S_n|}{\sqrt{n}}}(dq)=(r^*)^2\tanh^2(\beta r^* |S_n|/\sqrt{n}) $$
This follows 
from an inspection 
of the $\gamma$-measures 
in the case $d=1$. This formula can be compared to \cite[Theorem 2.1]{chatterjee2024featuresspinglassrandom}

\subsection{$d\geq 2$: Continuous (atomless) overlap distribution, many states, RSB, NSA}

Now the distribution  
$\rho^{\frac{|| S_n||}{\sqrt{n}}}$ is 
a Lebesgue-continuous distribution which is supported on the whole interval $(r^*)^2[-1,1]$, 
for a.e. random field realization. The continuous symmetry breaking 
(many pure states) which happens 
only for $d\geq 2$ is responsible 
for this continuity (atomless 
property) of the overlap distribution.

\medskip

\subsection{ A short remark on ultrametricity}

Following \cite{chatterjee2024featuresspinglassrandom}, we say that ultrametricity holds if for any replicas $a$, $b$, and $c$ which are pairwise different, we have 
\begin{align*}
&(\mu_n^{\frac{h}{\sqrt{n}}} \otimes \mu_n^{\frac{h}{\sqrt{n}}} \otimes \mu_n^{\frac{h}{\sqrt{n}}}) (R_n^{a,c} \geq \operatorname{min} \{ R_n^{a,b}, R_n^{b,c}\}) \approx 1
\end{align*}
with large probability w.r.t. the random 
field distribution $\P$. 

Using the approximation given in \cref{thm:overlapApproximation} we see the following. In the case $d=1$ there is ultrametricity 
for trivial reasons, which are the same 
as in \cite[Theorem 2.2]{chatterjee2024featuresspinglassrandom}. 
Indeed, the sphere $\mathbb{S}^0=\{-1,1\}$
appears in the above formula, 
and it is elementary to check 
that the condition in the indicator 
holds automatically, 
for all values of $(\Omega^a,\Omega^b,\Omega^c)\in 
\{-1,1\}^3$, 
without any 
assumption on the $\gamma$-measure. 

This is no longer the case for $d\geq 2$. Take 
$d=2$ for instance. First of all, all triples 
$(\Omega^a,\Omega^b,\Omega^c)\in (\mathbb{S}^1)^3$
appear with positive Lebesgue density 
with respect to the threefold product 
of the gamma-measure. 
Second we note that clearly the sphere $\mathbb{S}^1$ is not an ultrametric space. 
This is seen e.g. with the choice 
$(\Omega^a,\Omega^b,\Omega^c)=((1,0),(1,1)/\sqrt{2},(0,1)
)$ which falsifies 
the condition in the indicator. 
Hence we have proved the following 
theorem. 

\begin{theorem} Suppose that $h$ satisfies (A). 
For the finite-volume Gibbs states $\mu_n^{\frac{h}{\sqrt{n}}}$, the overlap distributions satisfy ultrametricity if $d = 1$, and the overlap distributions do not satisfy ultrametricity if $d \geq 2$.
\end{theorem}  

\medskip

\subsection{Metastates for scaled random fields}
Using the same asymptotic representation as for the overlaps, we have the following approximation
\begin{align*}
\mu_n^{\frac{h}{\sqrt{n}}} \approx \int_{\mathbb{S}^{d-1}} \gamma^{\frac{S_n}{ \sqrt{n}}} (d \Omega) \nu_\Omega^0 ,    
\end{align*}
which is valid in the bounded Lipschitz metric. Since the finite-volume Gibbs states depend only on the scaled sums of the external fields, one can deduce the metastates directly by referencing the appropriate limit theorems for random walks. The metastates are given and proved in \cref{thm:metastatesScaled}. The results are presented in \cref{tab:results2}.

Since the probability measures $\nu_\Omega^0$ do not depend parametrically on $h$, and the scaled sums of random walks are asymptotically independent of local field configurations of $h$, the Aizenman-Wehr metastate takes the $h$-independent form 
\begin{align*}
\kappa^h (d \mu) := \mathbb{E} \delta_{\overline{\nu}_{B_1}^0} (d \mu) .
\end{align*}
\begin{remark}Note that we see 
field-independent non-product measures  
which are mixed with a continuous 
mixing measure that involves a Gaussian 
variable. This is reminiscent of the 
situation in the finite-pattern Hopfield model.
In both cases the metastate is continuously 
distributed (even for $d=1$) over a set 
of mixtures of a particular form. This 
continuous distribution is driven by a Gaussian.
\end{remark}
\noindent
The Newman-Stein metastate
\begin{align*}
\overline{\kappa}^h := \int_0^1 dt \ \delta_{\overline{\nu}^0_{\frac{B_t}{\sqrt{t}}}}    
\end{align*}
is a consequence of the functional central limit theorem. 
\rowcolors{2}{gray!25}{white}
\begin{table}
    \centering
    \begin{tabular}{c|c|c}
    \rowcolor{gray!50}
        \multicolumn{1}{c|}{Scaled random fields}& $d=1$   & $d\geq 2 $\\
        \rule{0pt}{5mm}
        Cluster points, lattice/cont. RF & \multicolumn{2}{c}{$\overline{\{\overline{\nu}^0_{z}, z \in \mathbb{R}^{d}  \}}$} \\
        \rule{0pt}{4mm}
        AW-metastate &  \multicolumn{2}{c}{$E \delta_{ \bar\nu^0_{B_1}  }$}
        \\
        \rule{0pt}{4mm}
        Law of NS-metastate&   \multicolumn{2}{c}{$\int_{[0,1]} \delta_{\bar\nu^0_{B_t/\sqrt{t}}} dt, \,\,B_\cdot$ indep.} \\
        \rule{0pt}{5mm}
          Overlap distribution & \multicolumn{2}{c}{$
          \rho^{\frac{\|S_n\|}{\sqrt{n}}}
         $}
          \\
          \rule{0pt}{4mm}
    RSB/NSA  &
           \multicolumn{2}{c}{yes/yes}\\
           \rule{0pt}{4mm}
          Many states /Ultrametricity  &
          no/yes & yes/no    \\   
    \end{tabular}\medskip

    \caption{Overview of the results on random symmetry breaking in the ferromagnetic regime, depending on spin-dimension $d$, for the model with scaled random fields.  The measures $\bar\nu^0_{z}$ which are mixtures of pure states 
    of the non-random model, play a bigger role than for the model with size-independent RF scaling. 
    }
    \label{tab:results2}
\end{table}

\medskip 

\noindent {\bf Reading guide.}
The remainder of the paper is organized as follows. \cref{sec:FVGS} contains a rigorous formulation of the finite-volume Gibbs states of this particular model, and a subsection on the modes of convergence of the finite-volume Gibbs states. In \cref{sec:AsympAnalysis}, we develop the fundamental results concerning the asymptotic analysis of the finite-volume Gibbs states, and, finally, in \cref{sec:random} and \cref{sec:scaledRandomOverlaps}, we apply these asymptotics to construct and analyze the metastates and overlap distributions of the random field case. Note that the results derived in \cref{sec:FVGS} and \cref{sec:AsympAnalysis} hold for deterministic external fields $h$ with additional assumptions, and the proofs are written to emphasize this. The application to random fields $h$ then involves translating the results from the "deterministic" case to the new random case with a.s. properties being inherited from the random field. The appendices contain lengthy calculations or technical observations with references or proofs. The statements of the results are always given in full notation, typically including all relevant dependencies. In the proofs, we will usually omit the dependencies by dropping the sub or superscripted parameters, unless there is a possibility of confusion or the parametric dependence is relevant to the result itself.
\newpage
\section{Finite-volume Gibbs states} \label{sec:FVGS}
\subsection{Definitions and construction of states} \label{sec:constructionOfStates}
\noindent
We begin with the explicit construction of the finite-volume Gibbs states via functionals. The Hamiltonian function $H_n^h : \left( \mathbb{R}^d \right)^n \to \mathbb{R}$ is given by
\begin{align*}
H_n^h (\phi) := - \frac{1}{2 n} \sum_{i,j=1}^n \left< \phi(i), \phi(j) \right> - \sum_{i=1}^n \left< h(i), \phi(i) \right> ,
\end{align*}
where $h$ is $\left( \mathbb{R}^{d} \right)^n$-valued, and $\left< \cdot, \cdot \right>$ is the $d$-dimensional Euclidean inner-product. The magnetization vector $M_n : (\mathbb{R}^d)^n \to \mathbb{R}^d$ is given by
\begin{align*}
M_n (\phi) := \sum_{i=1}^n \phi(i),    
\end{align*}
and the Hamiltonian can be written to include the magnetization vector as follows
\begin{align*}
H_n^h (\phi) =  - \frac{1}{2 n} || M_n (\phi) ||^2 - \sum_{i=1}^n \left< h(i), \phi(i) \right> ,
\end{align*}
where $||\cdot ||$ is the norm corresponding to the $d$-dimensional Euclidean inner-product. The particle number function $N_n : \left( \mathbb{R}^d \right)^n \to \mathbb{R}$ is given by
\begin{align*}
N_n (\phi) := \sum_{i=1}^n \left< \phi(i), \phi(i) \right> .
\end{align*}
The finite-volume Gibbs states $\{ \mu_n^{\beta,h} \}_{n \in \mathbb{N}}$ are, at first, formally given by their actions
\begin{align} \label{def:FVGS_formal}
\mu^{\beta,h}_n [f] := \frac{1}{Z_n (\beta, h)} \int_{ \left( \mathbb{R}^d \right)^n} d \phi \ \delta (N_n (\phi) - n) e^{- \beta H^h_n (\phi)}  f (\phi) 
\end{align}
on $f \in C_b ( \left( \mathbb{R}^d \right)^n)$, where $C_b ( \left( \mathbb{R}^d \right)^n)$ is the space of bounded continuous functions on $\left( \mathbb{R}^d \right)^n$, $d \phi$ is the Lebesgue measure on $\left( \mathbb{R}^d \right)^n$, $\beta > 0$ is the inverse temperature, and $Z_n (\beta, h)$ is the partition function which normalizes the finite-volume Gibbs states into probability measures. This type of formal definition given by an action uniquely determines a Radon probability measure by the Riesz-Markov-Kakutani theorem, see \cite[Chapter 6]{Rudin1987}. Rigorously, we can define the finite-volume Gibbs states by using hyperspherical coordinates on $\mathbb{S}^{n d - 1}$ as follows
\begin{align} \label{def:delta_rigorous}
 \int_{ \left( \mathbb{R}^d \right)^n} d \phi \ \delta (N_n (\phi) - n) e^{- \beta H_n^h (\phi)}  f (\phi) = \frac{n^{\frac{nd-2}{2}}}{2} \int_{\mathbb{S}^{nd - 1}} d \Omega \  e^{- \beta H_n^h (\sqrt{n} \Omega)} f (\sqrt{n} \Omega),
\end{align}
where we formally identify
\begin{align*} 
 \phi_j (i) = \sqrt{n}  \Omega_j (i) ,
\end{align*}
where $(i,j) \in [n] \times [d]$, where we use the notation $[n] := \{1,2,...,n\}$. The rigorous definition can then be written explicitly using this formula.  

To proceed, we linearize and orthogonalize the Hamiltonian and the particle number function, after which we preform a series of rigorous delta function computations to arrive at the following mixture representation of the finite-volume Gibbs states
\begin{align} \label{def:mixture_representation}
\mu_n^{\beta,h} [f] =  \frac{1}{Z_n (\beta,h)}  \int_{B_{2d} (0,1)} \frac{dx dy}{(1 - || x ||^2 - || y ||^2)^{d+1}}\  e^{n \psi_n^{\beta, h} (x,y)} \nu_n^{x,y,h} [f] ,
\end{align}
where, $B_{2d}(0,1)$ is the $2d$-dimensional unit ball, 
$x,y\in \mathbb{R}^d$. 
For convenience, we have redefined the normalization constant $Z_n (\beta, h)$ to the following
\begin{align*}
Z_n (\beta,h) :=  \int_{B_{2d} (0,1)} \frac{dx dy}{(1 - || x ||^2 - || y ||^2)^{d+1}}\  e^{n \psi_n^{\beta, h} (x,y)} .
\end{align*}
In addition, we can identify the distribution $\alpha_n^{\beta,h}$ on $B_{2d}(0,1)$ given by its action
\begin{align} \label{def:mixingProbabilityMeasure}
\alpha_n^{\beta,h} [g]  = \frac{1}{Z_n (\beta,h)}  \int_{B_{2d} (0,1)} \frac{dx dy}{(1 - || x ||^2 - || y ||^2)^{d+1}}\  e^{n \psi_n^{\beta, h} (x,y)} g(x,y),   
\end{align}
where $g \in C_b (\mathbb{R}^{2d})$. From which we can rewrite the finite-volume Gibbs states in the following way
\begin{align} \label{def:mixture_representation_2}
\mu_n^{\beta,h} [f] = \int_{B_{2d} (0,1)} \alpha_n^{\beta,h} (dx,dy) \  \nu_n^{x,y,h} [f] = \alpha_n^{\beta,h} [\nu_n^{\cdot, \cdot,h} [f]].    
\end{align}
Note that in the marginal distribution of $\alpha_n^{\beta,h}$ the first $d$-components corresponds to the distribution of the magnetization density $\frac{M_n}{n}$ with respect to the finite-volume Gibbs states. The computations leading to this representation are presented in \cref{sec:deltaComputations}, and we will now explain each of these objects in detail.

In the process of linearization and orthogonalization, we construct a basis $\{e^h_{i',j',n}\}_{(i',j') \in [n] \times [d]}$ of the space $\left( \mathbb{R}^d \right)^n $
\begin{align*}
(e^h_{1,j',n})_j (i) := \frac{\delta_{j,j'}}{\sqrt{n}} , \ (e^h_{2,j',n})_j (i)  := \frac{\delta_{j,j'}(h_j (i) - (m_n)_j)}{\sqrt{n} (s_n)_j}
\end{align*}
where
\begin{align} \label{def:sampleMeanStandardDeviation}
m_n := \frac{1}{n} \sum_{i=1}^n h(i) \in \mathbb{R}^d, \ (s_n)_j := \sqrt{\frac{1}{n} \sum_{i=1}^n h_j(i)^2 - \left((m_n)_j \right)^2} \in [0, \infty) , 
\end{align}
and $\delta_{j,j'}$ is a function that is $1$ when $j = j' \in [d]$ and $0$ otherwise. We call the sequences of vectors $(m_n)$ and $(s_n)$ the sample means and sample standard deviations respectively. We then complete the orthonormal basis by constructing the remaining $\{e^h_{i',j',n}\}_{(i',j') \in ([n] \setminus \{1,2\}) \times [d]}$ basis vectors by whatever orthonormalization process one desires.

We denote $\psi_n^{\beta,h} : B_{2d} (0,1) \to \mathbb{R}$ to be the finite-volume exponential tilting function given by
\begin{align} \label{def:FVETF}
\psi_n^{\beta,h} (x,y) := \frac{\beta}{2} || x ||^2 + \beta \left< m_n, x \right> + \beta \left<s_n, y  \right> + \frac{d}{2} \ln (1 - || x ||^2 - || y ||^2) .
\end{align}
We denote $\nu_n^{x,y,h}$ to be the \textit{shifted  microcanonical probability measure} corresponding to the probability measure on $\left( \mathbb{R}^{d} \right)^n$ given by its action
\begin{align} \label{def:FVMCM}
\nu_n^{x,y,h} [f]   := &\frac{1}{|\mathbb{S}^{(n-2)d - 1}|} \int_{\mathbb{S}^{(n-2)d - 1}}d \Omega \\ &\times f \left( \sum_{j=1}^d (x_j \sqrt{n} e^h_{1,j,n} + y_j  \sqrt{n} e^h_{2,j,n}) + \sqrt{n} \sqrt{1 - || x ||^2 - || y ||^2} \sum_{3 \leq i \leq n, 1 \leq j \leq d} \Omega_j (i) e^h_{i,j,n} \right) \notag ,
\end{align}
where $(x,y) \in B_{2d} (0,1)$, $f \in C_b ((\mathbb{R}^d)^n)$, and $|\mathbb{S}^{(n-2)d - 1}|$ is the surface measure of the $(n-2)d-1$-dimensional unit sphere. 
We will use the following probabilistic representation of the surface 
measure on the sphere in terms of auxiliary standard Gaussians, which together with the appropriate normalization leads to the form 
\begin{align} \label{def:FVMCMprobabilistic}
\nu_n^{x,y,h} \sim \sum_{j=1}^d (x_j \sqrt{n} e^h_{1,j,n} + y_j  \sqrt{n} e^h_{2,j,n})  + \sqrt{n} \sqrt{1 - || x ||^2 - || y ||^2} \sum_{3 \leq i \leq n, 1 \leq j \leq d} \frac{G_{j} (i)}{|| \pi_{([n] \setminus \{1,2\} )\times [d]} (G)||} e^h_{i,j,n} ,     
\end{align}
where $G$ is a standard $\left( \mathbb{R}^d \right)^n$-valued Gaussian random variable, $\pi_{I \times J} : \left( \mathbb{R}^d \right)^n \to  \left( \mathbb{R}^J \right)^I$ is the canonical coordinate projection, where $(I,J) \subset [n] \times [d]$, and we use the notation $\sim$ to imply that that the probability distribution corresponding to the law of the random variable on the right is the same as the probability distribution on the left. We will use the notation $\nu_n^{x,y,h} |_{(i,j)} := ({\pi_{\{ i \} \times \{ j\}}})_* \nu_n^{x,y,h}$, where $({\pi_{\{ i \} \times \{ j\}}})_* \nu_n^{x,y,h}$ is the pushforward measure of $\nu_n^{x,y,h}$ by the mapping $\pi_{\{i\} \times \{j\}}$. We will call $\nu_n^{x,y,h} |_{(i,j)}$ the single-site single-component marginal distribution of $\nu_n^{x,y,h}$. 
Here and throughout the paper 
we will use the norm symbol 
$|| \cdot ||$ 
to denote the Euclidean norm 
on $\mathbb{R}^k$ for various values of $k$ 
which are clear from the context, 
as in the above formula for $k=(n-2)d$. 

Initially such probability measures are defined on $\left( \mathbb{R}^d \right)^n$, but these probability measures can be trivially extended to $\left( \mathbb{R}^d \right)^{\mathbb{N}}$ by tensoring on the Dirac measure at $0$ for the remaining components. To be more precise, we redefine $\nu_n^{x,y,h} := (\pi_{[n] \times [d]}, \pi_{(\mathbb{N} \setminus [n]) \times [d]} (0))_* \nu_n^{x,y,h}$. The redefined measure is now given on  $\left( \mathbb{R}^d \right)^\mathbb{N}$. For the rest of this manuscript, we will assume that all probability measures, unless otherwise stated, have been extended in this way if they are defined on $\left( \mathbb{R}^d \right)^n$.
\subsection{Convergence, topology of states, and cluster points} \label{sec:topologyMeasures}
\noindent
Given the construction of the previous subsection, the finite-volume Gibbs states are probability measures on $(\mathbb{R}^d)^{\mathbb{N}}$, and we will denote this space by $\mathcal{M}_1 ((\mathbb{R}^d)^\mathbb{N})$. To construct a suitable notion of convergence on the space of probability measures, we first metrize the space $(\mathbb{R}^d)^\mathbb{N} \simeq \prod_{i \in \mathbb{N}} \mathbb{R}^d \simeq \prod_{(i,j) \in \mathbb{N} \times [d]} \mathbb{R}$ using the given metric
\begin{align*}
d(\phi, \phi') := \frac{1}{C(d)} \sum_{(i,j) \in \mathbb{N} \times [d]} \frac{\min \{ |\phi_j (i) - \phi'_j (i)|, 1 \}}{2^{i + j}} , \ C(d) := \sum_{(i,j) \in \mathbb{N} \times [d]} \frac{1}{2^{i + j}} .
\end{align*}
This metric describes the product topology on $(\mathbb{R}^d)^\mathbb{N}$, and turns it into a separable complete metric space, see \cite[Chapter 3]{Ethier1986}.  As usual, we call topological spaces that are separable and completely metrizable \textit{Polish spaces}.

Note that the space of vector-valued sequences $(\mathbb{R}^d)^\mathbb{N}$, 
is a Polish space, w.r.t. product topology. The space of probability measure on a Polish space, in this case $\mathcal{M}_1 ((\mathbb{R}^d)^\mathbb{N})$ together 
with the notion of weak convergence, is a Polish space again. 
The  metric which metrizes weak 
convergence can be chosen to convenience, 
we will for our work restrict to the bounded Lipschitz metric. 

To be precise, for $X$ a Polish space, we say that a sequence of probability measures $\{\mu_n \}_{n=1}^\infty$ on $X$ converges weakly to another probability measure $\mu$ on $X$ if
\begin{align*}
\lim_{n \to \infty} \mu_n [f] = \mu [f]
\end{align*}
for any $f \in C_b (X)$. 
The bounded Lipschitz metric $d_{\operatorname{BL}_1}$ is given by
\begin{align} \label{eq:boundedLipschitzMetric}
d_{\operatorname{BL_1}} (\mu, \nu) := \sup_{f \in \operatorname{BL}_1 (\mathbb{R}^d)^\mathbb{N}} |\mu[f] - \nu [f]| ,
\end{align}
where $\operatorname{BL}_1 ((\mathbb{R}^d)^\mathbb{N})$ is the space of $1$-bounded $1$-Lipschitz functions with respect to the metric $d$.


In this article, we will also need the definition of the cluster points $\operatorname{clust} (\mu_n)$ of a sequence of probability measures $(\mu_n)$.
We define it as the collection of probability measures that can be obtained as convergent subsequences of $(\mu_n)$, explicitly, we have
\begin{align} \label{def:clusterPoints}
\operatorname{clust} (\mu_n) := \{\mu : \exists (n_k), \ \lim_{k \to \infty} \mu_{n_k} = \mu \} 
.   
\end{align}
\section{Asymptotic analysis of finite-volume Gibbs states} \label{sec:AsympAnalysis}
\subsection{Shifted microcanonical probability measures}
In this subsection, we will give a detailed proof of the uniform (in $x,y$) convergence of the shifted microcanonical probability measures. 
For the purpose of the approximations used, 
we consider the sequence $h$ to be a fixed deterministic object. The approximations which 
we want to obtain, lead us to the natural assumptions 
appearing in the hypothesis below. We will turn 
to consequences for i.i.d. sequences in a later part of the paper. The present treatment allows 
us to bypass problems which may  arise from possibly uncountably many sets of zero in the space of random fields.

\begin{lemma} \label{def:PS}
Suppose that
\begin{align*}
\lim_{n \to \infty} m_n = m \in \mathbb{R}^d, \ \lim_{n \to \infty} s_n = s \in (0, \infty)^d , \ \sum_{(i,j) \in \mathbb{N} \times [d]} \frac{|h_j(i)|}{2^{i+j}} < \infty .
\end{align*}
 
It follows that
\begin{align*}
\lim_{n \to \infty} \sup_{(x,y) \in B_{2d}(0,1)} d_{\operatorname{BL}_1} (\nu_n^{x,y,h}, \nu^{x,y,h}) = 0 ,
\end{align*}
where $\nu^{x,y,h}$ is a product measure with single-site single-component marginal distributions given by
\begin{align*}
\nu^{x,y,h}|_{(i,j)} \sim  \sqrt{\frac{1 - || x ||^2 - || y ||^2}{d}} G_j (i) + x_j + y_j \frac{h_j (i) - m_j}{s_j}  , 
\end{align*} 
where $G := \{ G_j (i)\}_{(i,j) \in \mathbb{N} \times [d]}$ is a standard $(\mathbb{R}^d)^\mathbb{N}$-valued Gaussian random variable. 

Furthermore, the mapping $(x,y) \mapsto \nu^{x,y,h}$ is continuous, and the mapping $\Omega \mapsto \nu^{r \Omega, y, h}$ is bounded Lipschitz with respect to the angular variables $\Omega \in \mathbb{S}^{d-1}$, where we understand the metric of the target space of probability measures to be given by the bounded Lipschitz metric $d_{\operatorname{BL}_1}$. 
\end{lemma}
\begin{proof}
Consider the coupling $\Gamma_n^{x,y,h}$ of $\nu_n^{x,y,h}|_{[n] \times [d]}$ and $\nu^{x,y,h}|_{[n] \times [d]}$ given by the probabilistic representation $\Gamma_n^{x,y,h} \sim T_n^{x,y,h} (G)$, where 
\begin{align*}
(T^{x,y,h}_{n}(G))_1 &= \sum_{j=1}^d (x_j \sqrt{n} e^h_{1,j,n} + y_j  \sqrt{n} e^h_{2,j,n}) + \frac{\sqrt{n}}{|| \pi_{([n] \setminus \{1,2\} )\times [d]} (G)|| } \sqrt{1 - || x ||^2 - || y ||^2} \sum_{3 \leq i \leq n, 1 \leq j \leq d} G_{j} (i) e^h_{i,j,n}
\end{align*}
and
\begin{align*}
(T^{x,y,h}_{n}(G))_2 &= \sum_{j=1}^d \delta_{j} \left(x_j  + y_j  \frac{h - m_j}{s_j} \right)  + \sqrt{\frac{1 - || x ||^2 - || y ||^2}{d}} \sum_{1 \leq i \leq n, 1 \leq j \leq d} G_j (i) e^h_{i,j,n}, 
\end{align*}
where we have used the notation $\delta_j  \in \{0,1\}^{[n]\times [d]}$ for the vector with components $(\delta_j)_{j'} (i') := \delta_{j,j'}$, and $G$ is a standard $(\mathbb{R}^d)^\mathbb{N}$-valued Gaussian random variable. Using the properties of the standard Gaussian $G$ and orthonormality of $\{e_{i,j,n}^h\}_{(i,j) \in [n] \times [d]}$, one can check that $(T_{n}^{x,y,h} (G))_1 \sim \nu_n^{x,y}|_{[n] \times [d]}$, and that $(T_{n}^{x,y,h} (G))_2 \sim \nu^{x,y}|_{[n] \times [d]}$.  

There are two main differences between the two expressions: The difference between 
the mean values, and 
the appearance of the normalization involving the Gaussians, in front of the second sum, for the first expression, but not the second expression.  
Our aim is now to show that both differences 
become small, in the limit of large $n$. 

Let us now denote $\Gamma_{i,j,n}^{x,y,h} := \Gamma^{x,y,h}_n |_{(i,j) \times (i,j)}$. Using this coupling, and the trivial bounds $|| x||, || y || \leq 1$, we have the following inequality
\begin{align} \label{eq:gammaInequality}
\Gamma^{x,y,h}_{i,j,n} [|\phi_j (i) - \phi_j' (i) |]  &\leq \left| \frac{h_j (i) - m_j}{s_j} - \frac{h_j (i) - (m_n)_j}{(s_n)_j} 
\right| \\  &+ \frac{1}{\sqrt{d}} \mathbb{E}_G\left| \sum_{ 1 \leq j' \leq d} (G_{j'} (1) (e_{1,j',n}^h)_j (i) + G_{j'} (2) (e_{2,j',n}^h)_j (i))\right| \nonumber \\&+ \mathbb{E}_G \left| \left( \frac{\sqrt{n}}{|| G_{[3,n]}||} - \frac{1}{\sqrt{d}} \right)   \sum_{3 \leq i' \leq n, 1 \leq j' \leq d} G_{j'} (i') (e^h_{i',j',n})_j (i) \right| \nonumber
\end{align} 
where we have denoted $G_{[3,n]} := \pi_{([n] \setminus \{1,2\} )\times [d]} (G)$. 

Now, we proceed one term at a time. First, we have
\begin{align*}
\left| \frac{h_j (i) - m_j}{s_j} - \frac{h_j (i) - (m_n)_j}{(s_n)_j} 
\right| \leq |h_j(i)| \max_{j \in [d]} \left\{\left| \frac{1}{(s_n)_j} - \frac{1}{s_j}\right| \right\} +  \max_{j \in [d]} \left\{\left| \frac{(m_n)_j}{(s_n)_j} - \frac{m_j}{s_j}\right| \right\}  .   
\end{align*}
For the second term on the r.h.s.  
of \cref{eq:gammaInequality}, by direct computation and estimation, we arrive at an upper bound of the form 
\begin{align*}
     \frac{1}{\sqrt{d}} \mathbb{E}_G\left| \sum_{ 1 \leq j' \leq d} (G_{j'} (1) (e_{1,j',n}^h)_j (i) + G_{j'} (2) (e_{2,j',n}^h)_j (i))\right| &= \frac{1}{\sqrt{n d}} \mathbb{E}_G \left| G_j (1) + G_j (2) \frac{h_j(i) - (m_n)_j}{(s_n)_j}\right| \\
     &\leq \frac{1}{\sqrt{n d}} + \frac{|h_j (i)|}{\sqrt{n d}} \max_{j \in [d]} \left\{ \frac{1}{(s_n)_j} \right\} + \frac{1}{\sqrt{n d}} \max_{j \in [d]} \left\{ \frac{|(m_n)_j| }{(s_n)_j} \right\} .
\end{align*}
Next, using the orthonormality of $\{ e_{i,j,n}^h \}_{(i,j) \in [n] \times [d]}$, it follows that
\begin{align*}
    \mathbb{E}_G \left|   \sum_{3 \leq i' \leq n, 1 \leq j' \leq d} G_{j'} (i') (e^h_{i',j',n})_j (i) \right|^2 \leq 1 ,
\end{align*}
from which, by the Cauchy-Schwartz inequality, we have
\begin{align*}
 \mathbb{E}_G \left| \left( \frac{\sqrt{n}}{|| G_{[3,n]}||} - \frac{1}{\sqrt{d}} \right)   \sum_{3 \leq i' \leq n, 1 \leq j' \leq d} G_{j'} (i') (e^h_{i',j',n})_j (i) \right| \leq  \left( \mathbb{E}_G \left| \ \frac{\sqrt{n}}{|| G_{[3,n]}||} - \frac{1}{\sqrt{d}} \right|^2 \right)^{\frac{1}{2}}
\end{align*}
In totality, we find that
\begin{align} \label{eq:localBound}
\sup_{(x,y) \in B_{2d}(0,1)}\Gamma^{x,y,h}_{i,j,n} [|\phi_j (i) - \phi_j' (i) |] &\leq| h_j(i)| \max_{j \in [d]} \left\{\left| \frac{1}{(s_n)_j} - \frac{1}{s_j}\right| \right\} +  \max_{j \in [d]} \left\{\left| \frac{(m_n)_j}{(s_n)_j} - \frac{m_j}{s_j}\right| \right\}  \\
&+ \frac{1}{\sqrt{n d}} + \frac{|h_j (i)|}{\sqrt{n d}} \max_{j \in [d]} \left\{ \frac{1}{(s_n)_j} \right\} + \frac{1}{\sqrt{n d}} \max_{j \in [d]} \left\{ \frac{|(m_n)_j| }{(s_n)_j} \right\} \nonumber \\ &+ \left( \mathbb{E}_G \left| \ \frac{\sqrt{n}}{|| G_{[3,n]}||} - \frac{1}{\sqrt{d}} \right|^2 \right)^{\frac{1}{2}} . \nonumber  
\end{align}

To compute the asymptotics of the last term on the right hand side of the inequality, we use hyperspherical coordinates to obtain the following
\begin{align*}
\mathbb{E}_G\left| \frac{\sqrt{n}}{|| G_{[3,n]}||} - \frac{1}{\sqrt{d}} \right|^2  &= \left( \int_0^\infty dr \ r^{(n - 2)d - 1} e^{- \frac{1}{2} r^2} \right)^{-1} \int_0^\infty dr \ r^{(n - 2)d - 1} e^{- \frac{1}{2} r^2} \left( \frac{\sqrt{n}}{r} - \frac{1}{\sqrt{d}} \right)^2 \\
&= \frac{1}{d} \left( \int_0^\infty dr \ e^{- \frac{2d + 3}{2} r^2} e^{((n - 2)d - 3) ( \ln r - \frac{1}{2} r^2)} r^2 \right)^{-1} \int_0^\infty dr \ e^{- \frac{2d + 3}{2} r^2} e^{((n - 2)d - 2) ( \ln r - \frac{1}{2} r^2)}  \left( r - 1 \right)^2 .
\end{align*}
For the above integral, since the mapping $r \mapsto \ln r - \frac{1}{2} r^2$ is smooth and attains its unique global maximum at the point $r^* = 1$, by the Laplace method, it follows that 
\begin{align} \label{eq:laplaceConvergence}
\lim_{n \to \infty} \mathbb{E}_G\left| \frac{\sqrt{n}}{|| G_{[3,n]}||} - \frac{1}{\sqrt{d}} \right|^2 = 0 .
\end{align}
By combining \cref{eq:localBound} and \cref{eq:laplaceConvergence}, it follows that
\begin{align*}
    \lim_{n \to \infty} \sup_{(x,y) \in B_{2d}(0,1)}\Gamma^{x,y,h}_{i,j,n} [|\phi_j (i) - \phi_j' (i) |] = 0 
\end{align*}
for any fixed $(i,j) \in \mathbb{N} \times [d]$. 

Now, we need to elevate the componentwise bound in $(i,j)$ to a global-type bound in accordance with \cref{thm:couplingBound}, which we state here for convenience
\begin{align*}
d_{\operatorname{BL}_1} (\nu_n^{x,y,h}, \nu^{x,y,h}) \leq \frac{1}{C(d)}\sum_{(i,j) \in [n] \times [d]} \frac{\Gamma_{i,j,n}^{x,y,h} [|\phi_j(i) - \phi'_j(i)|]}{2^{i + j}} + \frac{2}{C(d)} \sum_{(i,j) \in (\mathbb{N} \setminus [n]) \times [d]} \frac{1}{2^{i + j}} .
\end{align*}
Using \cref{eq:localBound} along with the above inequality, and some trivial upper-bounds like extending partial sums to full series with positive terms, it follows that
\begin{align*}
\sup_{(x,y) \in B_{2d}(0,1)}d_{\operatorname{BL}_1} (\nu_n^{x,y,h}, \nu^{x,y,h})    &\leq \max_{j \in [d]}  \left\{\left| \frac{1}{(s_n)_j} - \frac{1}{s_j}\right| \right\} \frac{1}{C(d)} \sum_{(i,j) \in \mathbb{N} \times [d]} \frac{|h_j(i)|}{2^{i + j}}  + \max_{j \in [d]} \left\{\left| \frac{(m_n)_j}{(s_n)_j} - \frac{m_j}{s_j}\right| \right\} \\ &+  \frac{1}{\sqrt{n d}} + \frac{1}{\sqrt{n d}} \max_{j \in [d]} \left\{ \frac{1}{(s_n)_j} \right\} \sum_{(i,j) \in \mathbb{N} \times [d]} \frac{|h_{j}(i)|}{2^{i+ j}} + \frac{1}{\sqrt{n d}} \max_{j \in [d]} \left\{ \frac{|(m_n)_j| }{(s_n)_j} \right\}\\ &+  \left( \mathbb{E}_G \left| \ \frac{\sqrt{n}}{|| G_{[3,n]}||} - \frac{1}{\sqrt{d}} \right|^2 \right)^{\frac{1}{2}} + \frac{2}{C(d)} \sum_{(i,j) \in (\mathbb{N} \setminus [n]) \times [d]} \frac{1}{2^{i + j}} .
\end{align*}
Recalling that the series over $|h_j(i)|$ is finite by assumption, it follows that
\begin{align*}
    \lim_{n \to \infty} \sup_{(x,y) \in B_{2d}(0,1)}d_{\operatorname{BL}_1} (\nu_n^{x,y,h}, \nu^{x,y,h})  = 0 ,
\end{align*}
as desired.

For the continuity properties, we construct another coupling $\Gamma \sim S(G)$, where $G$ is a standard $(\mathbb{R}^d)^\mathbb{N}$-valued Gaussian random variable, between $\nu^{x,y,h}$ and $\nu^{x',y',h}$ by setting
\begin{align*}
    (S (G)_1)_{j}(i) &= \sqrt{\frac{1 - || x ||^2 - || y ||^2}{d}} G_j (i) + x_j + y_j \frac{h_j (i) - m_j}{s_j} \\  (S (G)_2)_{j}(i) &= \sqrt{\frac{1 - || x' ||^2 - || y' ||^2}{d}} G_j (i) + x'_j + y'_j \frac{h_j (i) - m_j}{s_j}
\end{align*}
for all $(i,j) \in \mathbb{N} \times [d]$. One can verify that $\nu^{x,y,h} \sim S(G)_1$ and $\nu^{x',y',h} \sim S(G)_2$, and we have the following componentwise inequality
\begin{align*}
\Gamma [|\phi_{j}(i) - \phi'_{j} (i)|]  &\leq \left|\sqrt{\frac{1 - || x ||^2 - || y||^2}{d}} -  \sqrt{\frac{1 - || x' ||^2 - || y'||^2}{d}}\right| \mathbb{E}_G |G_j (i)|  \\&+ |x_j - x_j'| + \frac{|h_j(i)|}{|s_j|} |y_j - y_j'| + \frac{|m_j|}{|s_j|} |y_j - y_j'| .   
\end{align*}
It follows that
\begin{align*}
d_{\operatorname{BL}_1} (\nu^{x,y,h}, \nu^{x',y',h}) &\leq \frac{1}{C(d)} \sum_{(i,j) \in \mathbb{N} \times [d]} \frac{\Gamma [|\phi_{j}(i) - \phi'_{j} (i)|]}{2^{i + j}} \\
&\leq \left|\sqrt{\frac{1 - || x ||^2 - || y ||^2}{d}} - \sqrt{\frac{1 - || x' ||^2 - || y' ||^2}{d}} \right| + || x - x'|| \\
&+ \left( \max_{j \in [d]} \left\{ \frac{1}{|s_j|} \right\} \sum_{(i,j) \in \mathbb{N}} \frac{|h_j(i)|}{2^{i+j}} + \max_{j \in [d]} \left\{ \frac{|m_j|}{|s_j|}\right\} \right) || y - y'|| .
\end{align*} 
Both the continuity and angular Lipschitz property follow from this inequality.
\end{proof}
\subsection{Exponential tilting functions}
\noindent
In this subsection, we analyze the exponential tilting functions and deduce the parameters that we say are in the ferromagnetic regime. We also clarify the connection between the concentration properties of the mixing probability measure of the finite-volume Gibbs states, and the limiting exponential tilting function.

This first result concerns the uniform convergence of the finite-volume exponential tilting functions.
\begin{lemma} \label{thm:uniconvFVETF} Suppose that
\begin{align*}
\lim_{n \to \infty} m_n = m \in \mathbb{R}^d, \ \lim_{n \to \infty} s_n = s \in (0, \infty)^d  .    
\end{align*}
It follows that
\begin{align*}
\lim_{n \to \infty} \sup_{(x,y) \in B_{2d} (0,1)} |\psi_n^{\beta,h} (x,y) - \psi^{\beta,h} (x,y)| = 0 ,
\end{align*}
where $\psi^{\beta, h} : B_{2d} (0,1) \to \mathbb{R}$ is given by
\begin{align*}
\psi^{\beta,h} (x,y) := \frac{\beta || x ||^2}{2} + \beta \left<m, x \right> + \beta \left<y, s \right> + \frac{d}{2} \ln (1 - || x ||^2 - || y ||^2) .
\end{align*}
\end{lemma}
\begin{proof}
We have
\begin{align*}
\psi_n (x,y) - \psi (x,y) = \beta \left< x, m_n - m \right> + \beta \left< y, s_n - s \right>
\end{align*}
so that
\begin{align*}
\left| \psi_n (x,y) - \psi (x,y) \right| \leq || x || \ || m_n - m || + || y ||  \ || s_n - s || .
\end{align*}
The result follows.
\end{proof}
\noindent
Next, we present the relevant details concerning the set of global maximizing points of the limiting exponential tilting function, and exactly specify it for the ferromagnetic regime.
\begin{lemma} \label{thm:MSregime}
If $|| s || \geq 1$ and $\beta > 0$, or if $|| s || < 1$ and $0 < \beta \leq \frac{d}{1 - ||s ||^2}$, it follows that $\psi^{\beta,h}$ has a single unique global maximizing point contained in the unit ball $B_{2d} (0,1)$.

If $||s || < 1$, $\beta > \frac{d}{1 - || s ||^2}$, and $m = 0$, it follows that the collection of global maximizing points of $\psi^{\beta, h}$ denoted by $M^*(\beta,h)$ is given by
\begin{align*}
M^* (\beta, h) = \sqrt{1 - \frac{d}{ \beta} - ||s||^2}\,\, \mathbb{S}^{d - 1} \times \left\{ s \right\} .
\end{align*}
\end{lemma}
\begin{proof}
It is instructive to first deal with the case $m \not = 0$, since in this case there is always a unique global maximizing point. Our aim is to deduce the global maximizing points of the mapping
\begin{align*}
(x,y) \mapsto \psi(x,y) =    \frac{\beta || x ||^2}{2} + \beta \left<m, x \right> + \beta \left<y, s \right> + \frac{d}{2} \ln (1 - || x ||^2 - || y ||^2) . 
\end{align*}
Using the change of variables $(O^h, U^h)$ detailed in \cref{sec:coordinateTransforms}, along with hyperspherical coordinates, it follows that we can investigate the equivalent problem of finding the global maximizing points of the  mapping
\begin{align*}
 (r_1, \theta_1, r_2, \theta_2) \mapsto  \frac{\beta}{2} r_1^2 + \beta ||m|| r_1 \cos \theta_1 + \beta ||s|| r_2 \cos \theta_2 + \frac{d}{2} \ln (1 - r_1^2 - r_2^2) ,    
\end{align*}
where $\theta_1, \theta_2 \in [0, \pi]$ and $(r_1,r_2) \in B_2 (0,1) \cap [0, \infty)^2$. It is trivial that the global maximizing point for the angular variables must be given by $\theta_1 = \theta_2 = 0$. We are then left with the mapping
\begin{align*}
(r_1,r_2) \mapsto d \left( \frac{\beta(d)}{2} r_1^2 + \beta(d)  || m || r_1 + \beta(d) || s|| r_2 + \frac{1}{2} \ln (1 - r_1^2 - r_2^2) \right) , \ \beta(d) := \frac{\beta}{d} ,
\end{align*}
where we have introduced the re-scaled $\beta(d)$. The mapping inside the parentheses on the right hand side corresponds exactly to the limiting exponential tilting function of the $1$-dimensional random field mean-field spherical model, see \cite{Koskinen2023}, restricted to $B_2 (0,1) \cap [0, \infty)^2$ with choice of parameters $J := 1$, $\beta := \beta(d)$, $m^\parallel = || m||$, and $m^\perp := || s ||$. It is known that this mapping has a unique global maximizing point for any $\beta(d)$, see \cite[Lemma 3.4.1]{Koskinen2023}, and for our particular choice of parameters, this unique global maximizing point belongs to $B_2(0,1) \cap [0, \infty)^2$.

Now, we deal with the case $m = 0$, where there can either be a unique global maximizing point or a symmetric collection of non-unique global maximizing points depending on $\beta$ and $|| s ||$. Proceeding as we did for the previous case, we first consider the equivalent maximization problem for the mapping
\begin{align*}
(r_1,  \theta_1, r_2, \theta_2) \mapsto \frac{\beta}{2} r_1^2 + \beta ||s|| r_2 \cos \theta_2 + \frac{d}{2} \ln (1 - r_1^2 - r_2^2) .
\end{align*}
This time the angular variable $\theta_1$ is absent, and thus any $\theta_1 \in [0, \pi]$ is a valid global maximizer which reflects the rotation symmetry of the original maximization problem. Again, the global maximizing point for the angular variable $\theta_2$ must be $\theta_2 = 0$, and we again consider the mapping
\begin{align*}
(r_1,r_2) \mapsto d \left( \frac{\beta(d)}{2} r_1^2 + \beta(d) || s|| r_2 + \frac{1}{2} \ln (1 - r_1^2 - r_2^2) \right)  .   
\end{align*}
We are in the same situation as before, namely, the mapping inside the parentheses corresponds to the restricted limiting exponential tilting function of the $1$-dimensional random field mean-field spherical model, with the same choice of parameters as before. By \cite[Lemma 3.4.2]{Koskinen2023}, we have the following possibilities for the global maximizing points.
\begin{itemize}
\item If $|| s || \geq 1$ and $\beta(d) > 0$, there exists a unique global maximizing point $(r_1^*, r_2^*)$ of the given mapping, where
\begin{align*}
r_1^* = 0, \ r_2^* = \sqrt{1 + \left(\frac{1}{2 \beta(d) || s||} \right)^2} - \frac{1}{2 \beta(d) || s ||} .
\end{align*}
\item If $|| s || < 1$ and $\beta(d) \leq \frac{1}{1 - || s||^2}$, there exists a unique global maximizing point $(r_1^*, r_2^*)$ of the given mapping, where
\begin{align*}
r_1^* = 0, \ r_2^* = \sqrt{1 + \left(\frac{1}{2 \beta(d) || s||} \right)^2} - \frac{1}{2 \beta(d) || s ||} .
\end{align*}
\item If $|| s || < 1$ and  $\beta(d) > \frac{1}{1 - || s ||^2}$, there exists a unique global maximizing point $(r_1^*, r_2^*)$ of the given mapping, where 
\begin{align*}
r_1^* =  \sqrt{1 - \frac{1}{\beta(d)} - || s||^2}, \ r_2^* = || s || .   
\end{align*}
\end{itemize}
Note that in the first two parameter regimes, since $r_1^* = 0$, there is no rotational symmetry. However, in the third parameter regime, since $r_1^* > 0$, we have rotational symmetry of the original maximization problem. The result follows by applying the inverse change of coordinates to the solutions given. 
\end{proof}
\noindent
The following concentration result is a standard application of the Laplace method.
\begin{lemma} \label{thm:free_energy} Suppose that
\begin{align*}
\lim_{n \to \infty} m_n = m \in \mathbb{R}^d, \ \lim_{n \to \infty} s_n = s \in (0, \infty)^d .    
\end{align*}
It follows that
\begin{align*}
\limsup_{n \to \infty} \frac{1}{n} \ln  \int_{A} \frac{dx dy}{(1 - || x ||^2 - || y ||^2)^{d+1}} \  e^{n \psi_n^{\beta, h} (x,y)} \leq \sup_{(x,y) \in A} \psi^{\beta,h} (x,y) 
\end{align*}
for any set $A \subset B_{2d} (0,1)$ with positive Lebesgue measure, and
\begin{align*}
\liminf_{n \to \infty} \frac{1}{n} \ln  \int_{B_{2d}(0,1)} \frac{dx dy}{(1 - || x ||^2 - || y ||^2)^{d+1}} \  e^{n \psi_n^{\beta, h} (x,y)} \geq \sup_{(x,y) \in B_{2d}(0,1)} \psi^{\beta,h} (x,y) .
\end{align*}
As a consequence, it follows that
\begin{align*}
\lim_{n \to \infty} \alpha_n^{\beta,h} (B) = 0 ,    
\end{align*}
for any $B \subset B_{2d}(0,1)$ with positive finite Lebesgue measure such that 
\begin{align*}
\sup_{(x,y) \in B} \psi^{\beta,h} (x,y) < \sup_{(x,y) \in B_{2d}(0,1)} \psi^{\beta,h} (x,y) .
\end{align*}
\end{lemma}
\begin{proof}
For the upper bound, we rewrite the integral 
by expressing the $n$-independent part of 
integrand in exponential form 
as follows 
\begin{align*}
\int_{A} \frac{dx dy}{(1 - || x ||^2 - || y ||^2)^{d+1}} \  e^{n \psi_n (x,y)} &= \int_{A} dx dy \  \frac{e^{\frac{2 (d+1)}{d} \psi_n (x,y)} }{(1 - ||x ||^2 - || y ||^2)^{d+1}} e^{\left(n - \frac{2 (d+1)}{d} \right) \psi_n (x,y)} \\
&= \int_{A} dx dy \  e^{\frac{2 (d + 1)}{d} \left( \frac{\beta}{2} || x ||^2 + \beta \left< m_n, x \right> + \beta \left< s_n, y \right>\right)} e^{\left(n - \frac{2 (d+1)}{d} \right) \psi_n (x,y)} \\
&= \int_{A} dx dy \  e^{\frac{2 (d + 1)}{d} \left( \frac{\beta}{2} || x ||^2 + \beta \left< m_n, x \right> + \beta \left< s_n, y \right>\right)} \\&\times e^{\left(n - \frac{2 (d+1)}{d} \right) (\psi_n (x,y) - \psi (x,y))} e^{\left(n - \frac{2 (d+1)}{d} \right) \psi(x,y)} \\
&= \int_{A} dx dy \  e^{\frac{2 (d + 1)}{d} \left( \frac{\beta}{2} || x ||^2 + \beta \left< m_n, x \right> + \beta \left< s_n, y \right>\right)} \\ &\times e^{\left(n - \frac{2 (d+1)}{d} \right) (\beta \left< m_n - m,x \right> + \beta \left< s_n - s, y \right>)} e^{\left(n - \frac{2 (d+1)}{d} \right) \psi(x,y)} .
\end{align*}
We compute directly
\begin{align*}
\int_{A} \frac{dx dy}{(1 - || x ||^2 - || y ||^2)^{d+1}} \  e^{n \psi_n (x,y)} &\leq \left( \int_A dx dy \right) e^{ \frac{2(d+1)}{d} \left( \frac{\beta}{2} + \beta || m_n || + \beta || s_n||\right)}  e^{ \left( n - \frac{2 (d + 1)}{d} \right) \beta ( ||m_n - m|| + || s_n - s||)} \\&\times e^{\left( n - \frac{2 (d + 1)}{d} \right) \sup_{(x,y) \in A} \psi (x,y)} .
\end{align*}
The limsup results after taking the limit of the scaled logarithm. For the lower bound, let $M^*$ be the set of global maximizing points of $\psi$, and denote by $\psi^*$ the value of $\psi$ at any global maximizing point. By continuity, the set $\psi^{-1} (\psi^* - \varepsilon, \psi^* + \varepsilon)$ is open, it is non-empty since $M^*$ belongs to it, and, for small enough $\varepsilon > 0$, the set does not intersect with the boundary $\mathbb{S}^{2d - 1}$, which can be seen from the explicit determination of the global maximizing points in  \cref{thm:MSregime}. It follows that
\begin{align*}
\int_{B_{2d}(0,1)} \frac{dx dy}{(1 - || x ||^2 - || y ||^2)^{d+1}} \  e^{n \psi_n (x,y)} &\geq  e^{n (\psi^* - \varepsilon)}e^{- n \sup_{(x,y) \in B_{2d}(0,1)} |\psi_n(x,y) - \psi(x,y)|} \\
&\times \int_{\psi^{-1} (\psi^* - \varepsilon, \psi^* + \varepsilon)} \frac{dx dy}{(1 - ||x ||^2 - || y||^2)^{d+1}} .
\end{align*}
The liminf result now follows by first taking the liminf of the scaled logarithm, 
thereby using \cref{thm:uniconvFVETF} to control the sup over $x,y$, 
and then letting $\varepsilon \to 0$. 

The concentration result follows easily. 
\end{proof}
\subsection{Finite-volume Gibbs states in the non-ferromagnetic regime}
In this subsection, we characterize the infinite-volume Gibbs states in the non-ferromagnetic regime. The main result is given below.
\begin{theorem} \label{thm:nonFerromagneticIVGS}
Suppose that 
\begin{align*}
\lim_{n \to \infty} m_n = m \in \mathbb{R}^d , \ \lim_{n \to \infty} s_n = s \in (0, \infty)^d, \ \sum_{(i,j) \in \mathbb{N} \times [d]} \frac{|h_j(i)|}{2^{i+j}} < \infty .    
\end{align*}
In addition, suppose that one of the following holds:
\begin{enumerate}
\item 
\begin{align*}
m \not = 0 .
\end{align*}
\item 
\begin{align*}
m = 0, \ || s || \geq 1 .    
\end{align*}
\item 
\begin{align*}
m = 0, \ || s || < 1, \ \beta \leq \frac{d}{1 - || s ||^2} .    
\end{align*}
\end{enumerate}
It follows that
\begin{align*}
\lim_{n \to \infty} d_{\operatorname{BL}_1} (\mu_n^{\beta,h}, \nu^{x^*,y^*,h}) = 0 ,    
\end{align*}
where $(x^*,y^*) \in B_{2d}(0,1)$ is the unique global maximizing point of the limiting exponential tilting function $\psi^{\beta,h}$.
\end{theorem}
\begin{proof}
Using \cref{def:PS}, it follows that
\begin{align*}
\lim_{n \to \infty} d_{\operatorname{BL}_1} \left( \mu_n, \alpha_n [\nu^{\cdot,\cdot}] \right) \leq \lim_{n \to \infty}  \sup_{(x,y) \in B_{2d} (0,1)} d_{\operatorname{BL}_1} (\nu_n^{x,y}, \nu^{x,y}) = 0 .   
\end{align*}
Using \cref{thm:free_energy}, with the given non-ferromagnetic parameter regime, it follows that
\begin{align*}
\lim_{n \to \infty} \alpha_n (B) = 0 ,    
\end{align*}
for any closed set $B \subset B_{2d}(0,1)$ which does not contain the unique global maximizing point $(x^*, y^*) \in B_{2d}(0,1)$ of the limiting exponential tilting function $\psi$. This implies that
\begin{align*}
|\alpha_n [g] - g(x^*,y^*)| \leq \alpha_n [ || (x,y) - (x^*,y^*)||] 
&\leq \alpha_n [ \mathbbm{1}((x,y) \in B((x^*,y^*), \varepsilon))|| (x,y) - (x^*,y^*)||] \\ &+  \alpha_n [ \mathbbm{1}((x,y) \not \in B((x^*,y^*), \varepsilon))|| (x,y) - (x^*,y^*)||] \\
&\leq \varepsilon + 2 \alpha_n ( (B((x^*,y^*), \varepsilon)^c) 
\end{align*}
for any $g \in \operatorname{BL}_1 (\mathbb{R}^d \times \mathbb{R}^d)$.  It follows that
\begin{align*}
\limsup_{n \to \infty} d_{\operatorname{BL}_1} (\alpha_n, \delta_{(x^*,y^*)}) \leq \varepsilon , 
\end{align*}
and letting $\varepsilon \to 0^+$, this implies that 
\begin{align*}
\lim_{n \to \infty} d_{\operatorname{BL}_1} (\alpha_n, \delta_{(x^*,y^*)})   = 0 ,
\end{align*}
which implies that $\alpha_n$ converges to $\delta_{(x^*,y^*)}$ weakly. It follows that
\begin{align*}
d_{\operatorname{BL}_1} (\mu_n, \nu^{x^*,y^*}) \leq d_{\operatorname{BL}_1} \left( \mu_n, \alpha_n [\nu^{\cdot,\cdot}] \right) + d_{\operatorname{BL}_1} (\alpha_n [\nu^{\cdot, \cdot}], \nu^{x^*,y^*}) 
\leq d_{\operatorname{BL}_1} \left( \mu_n, \alpha_n [\nu^{\cdot,\cdot}] \right) + \alpha_n \left[d_{\operatorname{BL}_1} (\nu^{\cdot, \cdot}, \nu^{x^*,y^*})\right] .
\end{align*}
Now, since $(x,y) \mapsto d_{\operatorname{BL}_1} (\nu^{x, y}, \nu^{x^*,y^*})$ is continuous and vanishing at $(x^*,y^*)$ by \cref{def:PS}, it follows that
\begin{align*}
\lim_{n \to \infty}  d_{\operatorname{BL}_1} (\mu_n, \nu^{x^*,y^*}) = 0 , 
\end{align*}
and the result follows.
\end{proof}
\subsection{Finite-volume Gibbs states with bounded sums of external fields in the ferromagnetic regime}
\noindent
A posteriori, for the ferromagnetic regime, we know that the asymptotic analysis of the finite-volume Gibbs states is split into two distinct regimes depending on the boundedness or unboundedness of the sums over sites of the external fields. In this subsection, we will give the asymtptotics for the bounded case. This case is presented in the following result. 
\begin{lemma} \label{thm:bounded_convergence}
Suppose that 
\begin{align*}
\lim_{n \to \infty} m_n = 0 \in \mathbb{R}^d , \ \lim_{n \to \infty} s_n = s \in (0, \infty)^d , \ \sum_{(i,j) \in \mathbb{N} \times [d]} \frac{|h_j(i)|}{2^{i+j}} < \infty, \  || s || < 1, \ \beta > \frac{d}{1 - || s ||^2} .   
\end{align*}
In addition, suppose that
\begin{align*}
\lim_{n \to \infty} nm_n = z \in \mathbb{R}^d .    
\end{align*}
It follows that
\begin{align*}
\lim_{n \to \infty} d_{\operatorname{BL}_1} (\mu_n^{\beta,h}, \overline{\nu}^{z , r^*, y^*, h}) = 0 , 
\end{align*}
where
\begin{align*}
\overline{\nu}^{z, r^*, y^*, h} := \frac{1}{\int_{\mathbb{S}^{d - 1}} d \Omega \ e^{\beta r^* \left< \Omega, z \right>}} \int_{\mathbb{S}^{d - 1}} d \Omega \ e^{\beta r^* \left< \Omega, z \right>} \nu^{r^* \Omega,y^*, h} ,   
\end{align*}
and
\begin{align*}
r^* := \sqrt{1 - \frac{d}{\beta} - ||s ||^2}, \ y^* = s .
\end{align*}
\end{lemma}
\begin{proof} 
Starting from the representation for the finite-volume Gibbs states given in \cref{def:mixture_representation}, for any $f \in \operatorname{BL}_1 ((\mathbb{R}^d)^\mathbb{N})$, we have
\begin{align*}
\left| \mu_n [f] - \frac{1}{Z_n} \int_{B_{2d}(0,1)} \frac{dxdy}{(1 - ||x||^2 - || y||^2)^{d+1}} e^{n \psi_n(x,y)} \nu^{x,y}[f] \right| \leq \sup_{(x,y) \in B_{2d}(0,1)} d_{\operatorname{BL}_1} (\nu_n^{x,y}, \nu^{x,y}) .
\end{align*}
For the resulting integral, we change the order of integration and rewrite the integrand as follows
\begin{align*}
&\frac{1}{Z_n} \int_{B_{2d}(0,1)} \frac{dxdy}{(1 - ||x||^2 - || y||^2)^{d+1}} e^{n \psi_n(x,y)} \nu^{x,y}[f] \\
&= \frac{1}{Z_n} \int_{B_{1+d} (0,1)} \frac{dr dy \ r^{d - 1} \mathbbm{1}(r > 0)}{(1 - r^2 - || y ||^2)^{d+1}} e^{n \left( \frac{1}{2} r^2 +  \beta \left< s_n, y  \right> + \frac{d}{2} \ln (1 - r^2 - || y ||^2)\right)} \\ &\times \int_{\mathbb{S}^{d-1}} d \Omega \ e^{\beta r \left< n m_n, \Omega \right>}  \nu^{r \Omega, y} [f] .
\end{align*}
Let us now denote $\rho_n$ to be the probability measure on $B_{1 +d}(0,1)$ given by
\begin{align*}
\rho_n (dr,dy) = \frac{1}{Q_n}  \frac{dr dy \ r^{d - 1} \mathbbm{1}(r > 0)}{(1 - r^2 - || y ||^2)^{d+1}} e^{n \left( \frac{1}{2} r^2 +  \beta \left< s_n, y  \right> + \frac{d}{2} \ln (1 - r^2 - || y ||^2) \right)} ,
\end{align*}
where $Q_n$ is a normalization constant. We then have
\begin{align*}
&\frac{1}{Z_n} \int_{B_{2d}(0,1)} \frac{dxdy}{(1 - ||x||^2 - || y||^2)^{d+1}} e^{n \psi_n(x,y)} \nu^{x,y}[f] = \frac{\rho_n \left[ \int_{\mathbb{S}^{d-1}} d \Omega \ e^{\beta r \left< n m_n, \Omega \right>} \nu^{r \Omega, y} [f]\right]}{\rho_n \left[ \int_{\mathbb{S}^{d-1}} d \Omega \ e^{\beta r \left< n m_n, \Omega \right>} \right]}   
\end{align*}
One can repeat the proof of \cref{thm:free_energy} for the probability measure $\rho_n$, and show that it too satisfies the concentration property
\begin{align*}
\limsup_{n \to \infty} \frac{1}{n} \ln \rho_n (A') \leq \sup_{(r,y) \in A'} \psi(r \Omega,y) - \sup_{(r \Omega,y) \in B^{+}_{1+d}(0,1)} \psi(r \Omega,y)       
\end{align*}
for any $A' \subset B^{+}_{1+d} (0,1)$ with finite positive Lebesgue measure, where $B^{+}_{1 + d} (0,1) := B_{1+d}(0,1) \cap ((0, \infty) \times \mathbb{R}^d)$, and $\Omega \in \mathbb{S}^{d-1}$ does not actually appear in the supremum due to the rotational invariance of $\psi$ for this particular case $m = 0$. The concentration property of $\rho_n$ implies that
\begin{align*}
\lim_{n \to \infty} \rho_n = \delta_{(r^*, y^*)}
\end{align*}
weakly. For the integrand, by \cref{def:PS}, we observe that the mapping
\begin{align*}
(r,y) \mapsto \sup_{\Omega \in \mathbb{S}^{d-1}} \sup_{f \in \operatorname{BL}_1 ((\mathbb{R}^d)^\mathbb{N})}|\nu^{r \Omega,y}[f] - \nu^{r^* \Omega, y^*} [f]|  
\end{align*}
is uniformly bounded in $(r,y)$ and continuous. The sequence of mappings 
\begin{align*}
(r, \Omega, y) \mapsto e^{\beta r \left< n m_n, \Omega \right>}     
\end{align*}
is uniformly bounded in $(r,y,\Omega)$, continuous, and uniformly convergent to its pointwise limit. We thus have
\begin{align*}
&\left| \rho_n \left[ \int_{\mathbb{S}^{d-1}} d \Omega \ e^{\beta r \left< n m_n, \Omega \right>} \nu^{r \Omega, y} [f]\right] - \int_{\mathbb{S}^{d-1}} d \Omega \ e^{\beta r^* \left< z, \Omega \right>} \nu^{r^* \Omega, y^*} [f]  \right| \\
&\leq |\mathbb{S}^{d-1}| e^{\beta || n m_n||} \rho_n \left[ \sup_{\Omega \in \mathbb{S}^{d-1}} \sup_{f \in \operatorname{BL}_1 ((\mathbb{R}^d)^\mathbb{N})}|\nu^{r \Omega,y}[f] - \nu^{r^* \Omega, y^*} [f]| \right] \\
&+ |\mathbb{S}^{d-1}| \sup_{(r,y,\Omega) \in B_{1+d}^{+} (0,1) \times \mathbb{S}^{d-1}} |e^{\beta r \left< n m_n, \Omega \right>} - e^{\beta r^* \left< z, \Omega \right>}| .
\end{align*}
Since the right hand side does not depend on the chosen $f$, it follows that
\begin{align*}
\lim_{n\to \infty}\sup_{f \in \operatorname{BL}_1 ((\mathbb{R}^d)^\mathbb{N})}\left| \rho_n \left[ \int_{\mathbb{S}^{d-1}} d \Omega \ e^{\beta r \left< n m_n, \Omega \right>} \nu^{r \Omega, y} [f]\right] - \int_{\mathbb{S}^{d-1}} d \Omega \ e^{\beta r^* \left< z, \Omega \right>} \nu^{r^* \Omega, y^*} [f]  \right|  = 0 ,
\end{align*}
where we make use of the weak convergence of $\rho_n$ to $\delta_{(r^*,y^*)}$. By the same argument
\begin{align*}
\lim_{n \to \infty} \rho_n \left[ \int_{\mathbb{S}^{d-1}} d \Omega \ e^{\beta r \left< n m_n, \Omega \right>} \right] = \int_{\mathbb{S}^{d-1}} d \Omega \ e^{\beta r^* \left< z, \Omega \right>} ,
\end{align*}
and the result follows.
\end{proof}
\begin{remark} \label{rmk:bounded_convergence_density}
Using the same conditions and assumptions as in \cref{thm:bounded_convergence}, it would follow that
\begin{align*}
\lim_{n \to \infty} d_{\operatorname{BL}_1} \left( \alpha_n^{\beta,h}, \ \left( \int_{\mathbb{S}^{d-1} } d \Omega \ e^{\beta r^* \left< z, \Omega \right>}\right)^{-1} \int_{\mathbb{S}^{d-1} } d \Omega \ e^{\beta r^* \left< z, \Omega \right>} \delta_{r^* \Omega, y^*} \right)   = 0 . 
\end{align*}

\end{remark}
\subsection{Finite-volume Gibbs states with unbounded sums of external fields in the ferromagnetic regime}
\noindent
In this subsection, we consider the other asymptotic regime which involves the case where the sum over sites of the external fields are unbounded. This regime is more involved, and we begin with a result concerning the asymptotic representation of the finite-volume Gibbs states which captures the idea that the symmetry breaking of the model happens in the direction of the sum over sites of the external field.
\begin{lemma} \label{thm:mixturerep2} Suppose that
\begin{align*}
\lim_{n \to \infty} m_n = 0 \in \mathbb{R}^d , \ \lim_{n \to \infty} s_n = s \in (0, \infty)^d , \ \sum_{(i,j) \in \mathbb{N} \times [d]} \frac{|h_j(i)|}{2^{i+j}} < \infty, \  || s || < 1, \ \beta > \frac{d}{1 - || s ||^2} .   
\end{align*}
It follows that
\begin{align*}
\lim_{n \to \infty} \sup_{f \in \operatorname{BL}_1 ((\mathbb{R}^d)^\mathbb{N})} \left| \mu_n^{\beta,h} [f] -  \frac{1}{Q_n(\beta,h)} \int_{\mathbb{S}^{d-1}} d \Omega (\theta, \varphi_2, ..., \varphi_{d-1})\int_{B_{2}(0,1)} dr dy_1 \ \mathbbm{1}(r > 0) e^{n(d) \Psi_n^{\beta,h} (r, \theta, y_1)} \nu^{r^* {(O_n^h)}^{-1} (\Omega), y^*, h} [f] \right| = 0 , 
\end{align*}
where $Q_n(\beta,h)$ is a normalization constant, $n(d) := n - \frac{2(d+1)}{d} + \frac{d-1}{d}$, $O_n^h$ is the orthogonal transformation given in \cref{sec:coordinateTransforms}, the mapping $\Psi_n^{\beta,h} : B_{2} (0,1) \times [0, \pi] \to \mathbb{R}$ is given by
\begin{align*}
\Psi_n^{\beta,h} (r, \theta, y_1) := \frac{\beta}{2} r^2 + \beta r || m_n || \cos \theta + \beta || s_n|| y_1 + \frac{d}{2} \ln (1 - r^2 - y_1^2) ,    
\end{align*}
where we understand that $(r,y_1) \in B_{2} (0,1)$ and $\theta \in [0, \pi]$, and
\begin{align*}
r^* = \sqrt{1 - \frac{d}{\beta} - ||s ||^2}, \ y^* = s .    
\end{align*}
\end{lemma}
\begin{proof}
First, as in the previous result, we use the following bound
\begin{align*}
\left| \mu_n [f] - \frac{1}{Z_n} \int_{B_{2d}(0,1)} \frac{dxdy}{(1 - ||x||^2 - || y||^2)^{d+1}} e^{n \psi_n(x,y)} \nu^{x,y}[f] \right| \leq \sup_{(x,y) \in B_{2d}(0,1)} d_{\operatorname{BL}_1} (\nu_n^{x,y}, \nu^{x,y}) 
\end{align*}
for any $f \in \operatorname{BL}_1 ((\mathbb{R}^d)^\mathbb{N})$. Next, we rewrite the subsequent representation of the finite-volume Gibbs states as follows
\begin{align*}
&\frac{1}{Z_n} \int_{B_{2d}(0,1)} \frac{dxdy}{(1 - ||x||^2 - || y||^2)^{d+1}} e^{n \psi_n(x,y)} \nu^{x,y}[f]  \\ &=
\frac{1}{Z_n} \int_{B_{1 + d} (0,1)} dr dy \ r^{d-1} \mathbbm{1} (r > 0)  \int_{\mathbb{S}^{d-1}} d \Omega e^{\frac{2(d+1)}{d} \left( \frac{\beta}{2} r^2 + \beta r \left< m_n, \Omega \right> + \beta \left<s_n, y \right>\right)}  e^{\left(n - \frac{2(d+1)}{d} \right) \psi_n (r \Omega,y) } \nu^{r \Omega,y}[f] .
\end{align*}
Denote by $\rho_n$ the probability measure on $B_{1+d} (0,1) \times \mathbb{S}^{d-1}$ given by
\begin{align*}
\rho_n (dr, d\Omega, d y) := \frac{dr d \Omega dy  \ \mathbbm{1} (r > 0) \   e^{\left(n - \frac{2(d+1)}{d} \right) \psi_n (r \Omega,y) }}{ \int_{\mathbb{S}^{d-1}} d \Omega \int_{B_{1+d} (0,1)} dr dy \ \mathbbm{1}(r > 0)  e^{\left(n - \frac{2(d+1)}{d} \right) \psi_n (r \Omega,y)}}  , 
\end{align*}
where we emphasize that $(r,y) \in B_{1+d} (0,1)$ and $\Omega \in \mathbb{S}^{d-1}$. Denote by $g_n$ the function $B_{1+d} (0,1) \times \mathbb{S}^{d-1}$ given by
\begin{align*}
g_n(r,\Omega, y) :=  r^{d-1} e^{\frac{2(d+1)}{d} \left( \frac{\beta}{2} r^2 + \beta r \left< m_n, \Omega \right> + \beta \left<s_n, y \right>\right)} .
\end{align*}
We can rewrite the given representation of the finite-volume Gibbs states as
\begin{align*}
&\frac{1}{Z_n} \int_{B_{1 + d} (0,1)} dr dy \ r^{d-1} \mathbbm{1} (r > 0)  \int_{\mathbb{S}^{d-1}} d \Omega e^{\frac{2(d+1)}{d} \left( \frac{\beta}{2} r^2 + \beta r \left< m_n, \Omega \right> + \beta \left<s_n, y \right>\right)}  e^{\left(n - \frac{2(d+1)}{d} \right) \psi_n (r \Omega,y) } \nu^{r \Omega,y}[f]  \\&= \frac{ \rho_n \left[ g_n (r,\Omega,y) \nu^{r \Omega,y} [f] \right]}{\rho_n \left[ g_n (r,\Omega, y) \right] } ,  
\end{align*}
where are writing out each integrating variable $(r, \Omega, y)$ for clarity. 

Now, we again use the same technique as in the previous result, namely that
\begin{align*}
(r,y) \mapsto \sup_{\Omega \in \mathbb{S}^{d-1}} \sup_{f \in \operatorname{BL}_1 ((\mathbb{R}^d)^\mathbb{N})}|\nu^{r \Omega,y}[f] - \nu^{r^* \Omega, y^*} [f]|  
\end{align*}
is uniformly bounded and continuous in $(r,y)$, and the quantity
\begin{align*}
(r,y) \mapsto \sup_{\Omega \in \mathbb{S}^{d-1}}|g_n(r,\Omega, y) - g (r^*, \Omega, y^*)|   
\end{align*}
is uniformly bounded and continuous in $(r,y)$, where $g(r, \Omega,y) := r^{d-1} e^{\frac{2(d+1)}{d} \left( \frac{\beta}{2} r^2 + \beta \left<s, y \right>\right)}$ is the pointwise limit of $g_n (r, \Omega,y)$. We have
\begin{align*}
&\left| \rho_n \left[ g_n (r,\Omega,y) \nu^{r \Omega,y} [f] \right] - \rho_n \left[ g (r^*,\Omega,y^*) \nu^{r^* \Omega,y^*} [f] \right]\right|  \\ &\leq \rho_n \left[\sup_{\Omega \in \mathbb{S}^{d-1}}|g_n(r,\Omega, y) - g (r^*, \Omega, y^*)|  \right]  
+ || g ||_\infty \rho_n \left[\sup_{\Omega \in \mathbb{S}^{d-1}} \sup_{f \in \operatorname{BL}_1 ((\mathbb{R}^d)^\mathbb{N})}|\nu^{r \Omega,y}[f] - \nu^{r^* \Omega, y^*} [f]|  \right] ,
\end{align*}
where we have denoted $|| g ||_\infty := \sup_{ \Omega \in \mathbb{S}^{d-1}} |g(r^*, \Omega, y^*)| < \infty$. Although the weak convergence of $\rho_n$ cannot be deduced when considered with all variables present, since the right hand side of the above inequality does not depend on $\Omega$, the integration with respect to $\rho_n$ occurs only over the variables $(r,y)$.  If we denote this marginal distribution by $\rho_n'$, then, using the same concentration technique as in \cref{thm:free_energy} and \cref{thm:bounded_convergence}, it follows that
\begin{align*}
\lim_{n \to \infty} \rho_n'  = \delta_{(r^*, y^*)}
\end{align*}
weakly, which in turn implies that
\begin{align*}
\lim_{n \to \infty} \sup_{f \in \operatorname{BL}_1 ((\mathbb{R}^d))^\mathbb{N}}\left| \rho_n \left[ g_n (r,\Omega,y) \nu^{r \Omega,y} [f] \right] - \rho_n \left[ g (r^*,\Omega,y^*) \nu^{r^* \Omega,y^*} [f] \right]\right| = 0 .    
\end{align*}
Combining together the numerator and denominator, noting that $g(r, \Omega, y)$ does not depend on $\Omega$, and cancelling like terms, it follows that
\begin{align*}
\lim_{n \to \infty} \sup_{f \in \operatorname{BL}_1 ((\mathbb{R}^d))^\mathbb{N}} \left| \mu_n [f] - \rho_n [\nu^{r^* \Omega, y^*} [f]] \right| = 0 .   
\end{align*}

To continue, we use the coordinate transformation $(O_n^h, U_n^h)$ given in \cref{sec:coordinateTransforms} on the numerator part $\rho_n [\nu^{r^* \Omega, y^*} [f]]$ of to obtain the following
\begin{align*}
&\int_{\mathbb{S}^{d-1}} d \Omega \ \int_{B_{1+d} (0,1)} dr dy \ \mathbbm{1}(r > 0) \ e^{\left(n - \frac{2(d+1)}{d} \right) \psi_n (r \Omega,y)} \nu^{r^* \Omega, y^*} [f] \\
&= \int_{\mathbb{S}^{d-1}} d \Omega (\theta, \varphi_2,..., \varphi_{d-1}) \ \int_{B_{1+d} (0,1)} dr dy_1 dy_{> 1} \ \mathbbm{1}(r > 0) \\ &\times e^{\left(n - \frac{2(d+1)}{d} \right) \left( \frac{\beta}{2} r^2 + \beta r ||m_n || \cos \theta + \beta || s_n || y_1 + \frac{d}{2} \ln (1 - r^2 - || y_{>1}||^2 - y_1^2)\right)} \nu^{r^* O_n^{-1} (\Omega (\theta, \varphi_2,..., \varphi_{d-1})), y^*} ,
\end{align*}
where the original $y$ variable in the integrand is first transformed via the orthogonal map $U_n$, and then split into a variable $y_1 \in \mathbb{R}$, and a remaining portition $y_{> 1} \in \mathbb{R}^{d-1}$, where $d y_{> 1}$ is the corresponding Lebesgue measure. We also emphasize that the hyperspherical coordinates $\Omega$ are not the same coordinates, and that they correspond to the transformed variables by the map $O_n$. We isolate the integral of $y_{> 1}$ as
\begin{align*}
\int_{\mathbb{R}^{d-1}} dy_{>1} \mathbbm{1}(0 < r^2 + y_1^2 + y_{>1}^2 < 1) \left( 1 - r^2 - y_1^2 - y_{>1}^2 \right)^{\frac{d}{2} \left(n - \frac{2(d+1)}{d} \right)} .   
\end{align*}
Using homogeneity, it follows that
\begin{align*}
&\int_{\mathbb{R}^{d-1}} dy_{>1} \mathbbm{1}(0 < r^2 + y_1^2 + y_{>1}^2 < 1) \left( 1 - r^2 - y_1^2 - y_{>1}^2 \right)^{\frac{d}{2} \left(n - \frac{2(d+1)}{d} \right)}  \\
&= \mathbbm{1}(r^2 + y_1^2 < 1)\left(1 - r^2 - y_1^2 \right)^{\frac{d - 1}{2}} \left( 1 - r^2 - y_1^2 \right)^{\frac{d}{2} \left( n - \frac{2 (d+1)}{d}\right)} \\ &\times\int_{\mathbb{R}^{d-1}} dy_{>1} \mathbbm{1}(0 <  y_{>1}^2 < 1) \left( 1 - y_{>1}^2 \right)^{\frac{d}{2} \left(n - \frac{2(d+1)}{d} \right)} \\
&= \mathbbm{1}(r^2 + y_1^2 < 1) \left( 1 - r^2 - y_1^2 \right)^{\frac{d}{2} \left( n - \frac{2 (d+1)}{d} + \frac{d-1}{d}\right)} \\ &\times\int_{\mathbb{R}^{d-1}} dy_{>1} \mathbbm{1}(0 <  y_{>1}^2 < 1) \left( 1 - y_{>1}^2 \right)^{\frac{d}{2} \left(n - \frac{2(d+1)}{d} \right)} .
\end{align*}
Returning to the non-isolated integral, it follows that
\begin{align*}
&\int_{\mathbb{S}^{d-1}} d \Omega (\theta, \varphi_2,..., \varphi_{d-1}) \ \int_{B_{1+d} (0,1)} dr dy_1 dy_{> 1} \ \mathbbm{1}(r > 0) \\ &\times e^{\left(n - \frac{2(d+1)}{d} \right) \left( \frac{\beta}{2} r^2 + \beta r ||m_n || \cos \theta + \beta || s_n || y_1 + \frac{d}{2} \ln (1 - r^2 - || y_{>1}||^2 - y_1^2)\right)} \nu^{r^* O_n^{-1} (\Omega (\theta, \varphi_2,..., \varphi_{d-1})), y^*}  \\
&= A_n(d) \int_{\mathbb{S}^{d-1}} d \Omega (\theta, \varphi_2,..., \varphi_{d-1}) \ \int_{B_{2} (0,1)} dr dy_1 \ \mathbbm{1}(r > 0) \\ &\times \left( 1 - r^2 - y_1^2 \right)^{\frac{d}{2} \left( n - \frac{2 (d+1)}{d} + \frac{d-1}{d}\right)}  e^{\left(n - \frac{2(d+1)}{d} \right) \left( \frac{\beta}{2} r^2 + \beta r ||m_n || \cos \theta + \beta || s_n || y_1 \right)} \nu^{r^* O_n^{-1} (\Omega (\theta, \varphi_2,..., \varphi_{d-1})), y^*} ,
\end{align*}
where
\begin{align*}
A_n(d) :=  \int_{\mathbb{R}^{d-1}} dy_{>1} \mathbbm{1}(0 <  y_{>1}^2 < 1) \left( 1 - y_{>1}^2 \right)^{\frac{d}{2} \left(n - \frac{2(d+1)}{d} \right)} .   
\end{align*}

Now, we again rewrite part of the integrand as follows
\begin{align*}
&\left( 1 - r^2 - y_1^2 \right)^{\frac{d}{2} \left( n - \frac{2 (d+1)}{d} + \frac{d-1}{d}\right)}  e^{\left(n - \frac{2(d+1)}{d} \right) \left( \frac{\beta}{2} r^2 + \beta r ||m_n || \cos \theta + \beta || s_n || y_1 \right)}  \\
&= e^{- \frac{d-1}{d} \left( \frac{\beta}{2} r^2 + \beta r || m_n || \cos \theta + \beta || s_n || y_1 \right)} e^{ \left( n - \frac{2 (d+1)}{d} + \frac{d-1}{d}\right) \left(  \frac{\beta}{2} r^2 + \beta r || m_n || \cos \theta + \beta || s_n || y_1 + \frac{d}{2} \ln (1 - r^2 - y_1^2)\right)} .
\end{align*}
Now we repeat the same argument as before to simplify the integrand one last time, namely, we consider a probability measure $\rho_n''$ on $  B_2 (0,1) \times \mathbb{S}^{d-1}$ given by
\begin{align*}
\rho_n'' (dr, d \Omega, dy_1) := \frac{ dr dy_1 d \Omega \ \mathbbm{1}(r > 0) e^{ \left( n - \frac{2 (d+1)}{d} + \frac{d-1}{d}\right) \left(  \frac{\beta}{2} r^2 + \beta r || m_n || \cos \theta + \beta || s_n || y_1 + \frac{d}{2} \ln (1 - r^2 - y_1^2)\right)} }{Z_n''} ,
\end{align*}
and the function $g_n''$ on $ B_2 (0,1) \times \mathbb{S}^{d-1}$ given by
\begin{align*}
g_n'' (r, \Omega, y_1) = e^{- \frac{d-1}{d} \left( \frac{\beta}{2} r^2 + \beta r || m_n || \cos \theta + \beta || s_n || y_1 \right)} ,    
\end{align*}
so that
\begin{align*}
\rho_n [\nu^{r^* \Omega, y^*} [f]] = \frac{\rho_n'' [g_n'' (r, \Omega, y_1)] \nu^{r^* O_n^{-1} (\Omega), y^*} [f]]}{\rho_n'' [g_n''(r, \Omega, y_1)]} .   
\end{align*}
Again, one shows that the sequence $\{g_n'' \}_{n = 1}^\infty$ is uniformly convergent in  $(r, \Omega, y_1)$ to its pointwise limit $g''$, and the probability measure $\rho_n''$ satisfies the same type of concentration property as the previous probability measure $\rho_n'$ ultimately leading to its weak convergence to $\delta_{(r^*, y_1^*)}$. It follows that
\begin{align*}
\lim_{n \to \infty} \sup_{f \in \operatorname{BL}_1 ((\mathbb{R}^d)^\mathbb{N})}\left| \rho_n'' [g_n'' (r, \Omega, y_1) \nu^{r^* O_n^{-1} (\Omega), y^*}[f]] - g'' (r^*, \Omega, y_1^*) \rho_n'' [ \nu^{r^* O_n^{-1} (\Omega), y^*}[f]] \right| = 0 ,   
\end{align*}
and one should note that $g''$ does not depend on the angular variable $\Omega$. Cancelling like terms the final result follows.
\end{proof}
\noindent
The purpose of this asymptotic representation is that we will next prove a concentration result for $\theta \in [0, \delta]$ for small $\delta > 0$. In this way, we capture the idea that the magnetization of this model points strongly in the direction of the sum over sites of the external field. We will then apply this concentration result by applying the concentration inequality given in \cref{sec:analyticConcentration}. To that end, we present the following result which gives a necessary lower bound for our problem.
\begin{lemma} \label{thm:angular_concentration}
Suppose that
\begin{align*}
\lim_{n \to \infty} m_n = 0 \in \mathbb{R}^d , \ \lim_{n \to \infty} s_n = s \in (0, \infty)^d , || s || < 1, \ \beta > \frac{d}{1 - || s ||^2} .   
\end{align*}   
In addition, suppose that
\begin{align*}
\lim_{n \to \infty} n || m_n || = \infty .
\end{align*}
It follows that for small but fixed $\delta > 0$  there exists a constant $C (\delta) > 0$ such that for 
large enough $n \in \mathbb{N}$
\begin{align*}
&\frac{\int_{\mathbb{S}^{d - 1}} d \Omega (\theta, \varphi_2,..., \varphi_{d-1}) \ \int_{B_{2} (0,1)} dr dy_1 \ \mathbbm{1}(r > 0) \mathbbm{1}(0 \leq \theta \leq \delta) \ e^{n(d) \Psi_n^{\beta,h} (r, \theta,y_1)}}{\int_{\mathbb{S}^{d - 1}} d \Omega (\theta, \varphi_2,..., \varphi_{d-1}) \ \int_{B_{2} (0,1)} dr dy_1 \ \mathbbm{1}(r > 0) \mathbbm{1}(\delta \leq \theta \leq \pi) \ e^{n(d) \Psi_n^{\beta,h} (r, \theta,y_1)}} \\& \geq \frac{C (\delta) e^{n(d) (\Psi_n^{\beta,h} (r_n^*,0,y_n^*) - \Psi_n^{\beta,h} (r_n^* (\delta), \delta, y_n^* (\delta)))}}{\sqrt{n(d) || m_n ||}^{d - 1}} ,
\end{align*}
where $n(d) = n - \frac{2(d+1)}{d} + \frac{d-1}{d}$,  where we denote $(r_n^*(\delta), \delta, y_n^*(\delta))$ to be the unique global maximizing point of the mapping 
\begin{align*}
\Psi_n^{\beta,h} (r, \theta, y_1) := \frac{\beta}{2} r^2 + \beta r || m_n || \cos \theta + \beta || s_n|| y_1 + \frac{d}{2} \ln (1 - r^2 - y_1^2) ,    
\end{align*}
and where we understand that $(r,y_1) \in B_{2} (0,1)$ and $\theta \in [\delta, \pi]$, we use a convention where $(r_n^*, y_n^*) := (r_n^*(0), y_n^*(0))$,  and
\begin{align*}
r^* = \sqrt{1 - \frac{d}{\beta} - ||s ||^2}, \ y^* = s .    
\end{align*}
\end{lemma}
\begin{proof} 
To ease notation, we will write $y$ in the formulas below for the scalar variable $y_1$. The first step is the following upper-bound
\begin{align*}
\int_{\mathbb{S}^{d - 1}} d \Omega \ \int_{B_{2} (0,1)} dr dy \ \mathbbm{1}(r > 0) \mathbbm{1}(\delta \leq \theta \leq \pi) \ e^{n(d) \Psi_n (r, \theta,y)} \leq |\mathbb{S}^{d-1}| \int_{B_{2} (0,1)} dr dy \ \mathbbm{1}(r > 0)  \ e^{n(d) \Psi_n (r, \delta,y)} ,
\end{align*}
where we have used the monotonicity in $\theta$ of the map $(r, \theta,y) \mapsto \Psi_n (r, \theta, y)$. For the purpose of this proof, when the other integration variables contained with $d \Omega$ are not explicitly written out, it is because they are not explicitly needed. Using the result for the $1$-dimensional model, see \cite[Lemma 3.5.3]{Koskinen2023} , it follows that
\begin{align*}
\lim_{n \to \infty} \frac{n(d) \int_{B_{2} (0,1)} dr dy \ \mathbbm{1}(r > 0)  \ e^{n(d) \Psi_n (r, \delta,y)}}{e^{n(d) \Psi_n (r_n^* (\delta), \delta, y_n^* (\delta))}} = \int_{\mathbb{R}^d} dr dy \ e^{\frac{1}{2} \left< (r,y), H_{(r,y)}[\Psi](r^*,y^*) (r,y) \right>} ,
\end{align*}
where $(r_n^* (\delta), \delta, y_n^* (\delta))$ is the unique global maximizing point of the mapping $(r, \theta, y) \mapsto \Psi_n (r,\theta, y)$, and $H_{(r,y)} [\Psi]$ is the Hessian of the limiting function of $\Psi_n$ considered as a function of two variables $(r,y)$ since the limit does not depend on $\theta$. Note that the uniqueness of the global maximizing point can be deduced from \cite[Lemma 3.4.1]{Koskinen2023} and the monotonicity of the map with respect to $\theta$. It follows that there exists a constant $C(\delta) > 0$ such that for large enough $n$, we have
\begin{align*}
\int_{\mathbb{S}^{d - 1}} d \Omega \ \int_{B_{2} (0,1)} dr dy \ \mathbbm{1}(r > 0) \mathbbm{1}(\delta \leq \theta \leq \pi) \ e^{n(d) \Psi_n (r, \theta,y)} \leq \frac{C(\delta) e^{n(d) \Psi_n (r_n^* (\delta), \delta, y_n^* (\delta))}}{n(d)}
\end{align*}

For the lower bound, we go through the standard process of obtaining a Laplace-type asymptotic for a non-degenerate quadratic global maximizing point. To that end, we have
\begin{align*}
&\int_{\mathbb{S}^{d - 1}} d \Omega \ \int_{B_{2} (0,1)} dr dy \ \mathbbm{1}(r > 0) \mathbbm{1}(0 \leq \theta \leq \delta) \ e^{n(d) \Psi_n (r, \theta,y)} \\ &\geq e^{n(d) \Psi_n (r_n^*, 0, y_n^*)} \int_{\mathbb{S}^{d-1}} d \Omega  \int_{B_2(0, \delta)} dr dy \ \mathbbm{1}(0 \leq \theta \leq  \delta) e^{n(d) \left( \Psi_n (r_n^* + r, \theta, y_n^* + y) - \Psi_n (r_n^*, 0, y_n^*)\right)} .
\end{align*}
Now, we wish to consider the Taylor approximation of $\Psi_n$ at the global maximizing point $(r_n^*, 0, y_n^*)$. First, we note that
\begin{align*}
    \Psi_n (r, \theta, y) = \Psi (r,y) + \beta || m_n || r \cos (\theta) + \beta (|| s_n|| - || s ||) y ,
\end{align*}
where $H[\Psi]$ is the Hessian of the limiting function $\Psi$ in only the coordinates $(r,y)$, followed by the second-order expansion of $\Psi_n$ at the critical point $(r_n^*, 0, y_n^*)$
\begin{align*}
    \Psi_n (r, \theta,y) - \Psi_n (r_n^*, 0, y_n^*) &= \int_0^1 dt \ (1 - t) \left< (r - r_n^*, y - y_n^*), H [\Psi] (r_n^* (1 - t) + r t, y_n^* (1 - t) + y t) (r - r_n^*, y - y_n^*) \right> \\
    &- 2 \beta ||m_n || (r - r_n^*) \theta \int_0^1 dt \ (1 - t) \sin (\theta t) \\
    & - \beta || m_n|| \theta^2 \int_0^1 dt \ (1 - t) (r_n^* (1 - t) + r t) \cos (\theta t)
\end{align*}
Plugging in this expansion, we can directly compute the following limit
\begin{align*}
&\lim_{n \to \infty} n(d) \left( \Psi_n \left(r_n^* + \frac{r}{\sqrt{n(d)}}, \frac{\theta}{\sqrt{n(d) || m_n ||}}, y_n^* + \frac{y}{\sqrt{n(d)}} \right) - \Psi_n (r_n^*, 0, y_n^*) \right)  \\ &= - \frac{\beta r^*}{2} \theta^2 + \frac{1}{2} \left< (r,y), H[\Psi] (r^*, y^*) (r,y)\right>  .
\end{align*}
Prior to taking the limit, one should recall that the volume element $d \Omega$ is explicitly given by
\begin{align*}
d \Omega (\theta, \varphi_2,..., \varphi_{d-1}) = \sin (\theta)^{d-2} \sin (\varphi_2)^{d-3} ... \sin(\varphi_{d-2}) d \theta d \varphi_2... d \varphi_{d-1} .
\end{align*}
In the change of variables $\theta \mapsto \frac{\theta}{\sqrt{n(d) || m_n ||}}$, we isolate the singularity of the $\sin$ as follows
\begin{align*}
&d \Omega \left( \frac{\theta}{\sqrt{n(d) || m_n ||}}, \varphi_2,..., \varphi_{d-1}\right) \\&= \frac{1}{\sqrt{n(d) || m_n ||}^{d - 2}} \theta^{d-2} \left( \frac{\sin \left( \frac{\theta}{\sqrt{n(d) || m_n||}}\right)}{\frac{\theta}{\sqrt{n(d) || m_n ||}}} \right)^{d - 2}  \sin (\varphi_2)^{d-3} ... \sin(\varphi_{d-2}) \frac{d \theta}{\sqrt{n(d) || m_n ||}} d \varphi_2... d \varphi_{d-1}
\end{align*}
Combining these observations, by Fatou's lemma, for large enough $n$, it follows that
\begin{align*}
&\liminf_{n \to \infty} \sqrt{n(d) || m_n ||}^{d-1} \int_{\mathbb{S}^{d-1}} d \Omega \left(\frac{\theta}{\sqrt{n(d) || m_n ||}}, \varphi_2,..., \varphi_{d-1} \right) \int_{B_2 (0, \sqrt{n(d)}\delta)} dr dy \ \mathbbm{1}(0 \leq \theta \leq \sqrt{n(d) || m_n||} \delta) \\ &\times e^{n(d) \left( \Psi_n (r_n^* + \frac{r}{\sqrt{n(d)}}, \frac{\theta}{\sqrt{n(d) || m_n ||}}, y_n^* + \frac{y}{\sqrt{n(d)}}) - \Psi_n (r_n^*, 0, y_n^*) \right)} \\
&\geq  |\mathbb{S}^{d-2}| \int_0^\infty d \theta \ \theta^{d-2} e^{- \frac{\beta r^*}{2} \theta^2} \int_{\mathbb{R}^{2}} dr dy \ e^{\frac{1}{2}\left< (r,y), H[\Psi] (r^*, y^*) (r,y) \right>} . 
\end{align*}
Returning the original integral, 
using the above result, it follows that there exists a positive constant $C'(\delta) > 0$ such that
\begin{align*}
\int_{\mathbb{S}^{d - 1}} d \Omega \ \int_{B_{2} (0,1)} dr dy \ \mathbbm{1}(r > 0) \mathbbm{1}(0 \leq \theta \leq \delta) \ e^{n(d) \Psi_n (r, \theta,y)} \geq C' (\delta) \frac{e^{n(d) \Psi_n (r_n^*, 0, y_n^*)}}{n(d) \sqrt{n(d) || m_n ||}^{d-1}}
\end{align*}
for large enough $n$. 
This can be seen by using the integral form of the remainder term in the Taylor expansion in which the third derivatives appear. 
Combining together the two estimates, we find that
\begin{align*}
\frac{\int_{\mathbb{S}^{d - 1}} d \Omega \ \int_{B_{2} (0,1)} dr dy \ \mathbbm{1}(r > 0) \mathbbm{1}(0 \leq \theta \leq \delta) \ e^{n(d) \Psi_n (r, \theta,y)}}{\int_{\mathbb{S}^{d - 1}} d \Omega \ \int_{B_{2} (0,1)} dr dy \ \mathbbm{1}(r > 0) \mathbbm{1}(\delta \leq \theta \leq \pi) \ e^{n(d) \Psi_n (r, \theta,y)}} \geq C (\delta)\frac{e^{n(d) (\Psi_n (r_n^*,0,y_n^*) - \Psi_n (r_n^* (\delta), \delta, y_n^* (\delta)))}}{\sqrt{n(d) || m_n ||}^{d - 1}} .
\end{align*}
with suitably redefined 
$C(\delta)$.
\end{proof}
\noindent
Using this concentration bound, we have the following main result of this subsection. It expresses how the random field sums select the overall direction of the pure states in the large volume approximation of the finite-volume Gibbs measures.  
\begin{lemma} \label{thm:unbounded_convergence} Suppose that
\begin{align*}
\lim_{n \to \infty} m_n = 0 \in \mathbb{R}^d , \ \lim_{n \to \infty} s_n = s \in (0, \infty)^d , \ \sum_{(i,j) \in \mathbb{N} \times [d]} \frac{|h_j(i)|}{2^{i+j}} < \infty , \  || s || < 1, \ \beta > \frac{d}{1 - || s ||^2} .   
\end{align*}
In addition, suppose that
\begin{align*}
\lim_{n \to \infty} n || m_n || = \infty .
\end{align*}
It follows that
\begin{align*}
\lim_{n \to \infty} d_{\operatorname{BL}_1} (\mu_n^{\beta,h}, \nu^{r^* \widehat{m}_n, y^*, h}) = 0 ,
\end{align*}
where
\begin{align*}
\widehat{m}_n := \frac{m_n}{|| m_n||} , \ r^* = \sqrt{1 - \frac{d}{\beta} - || s ||^2}, \ y^* = s .
\end{align*}
\end{lemma}
\begin{proof}
We will proceed in steps. To begin, we fix arbitrary but small $\delta > 0$. In the first step, we start from the asymptotic representation of the finite-volume Gibbs state given in \cref{thm:mixturerep2}, we condition the angular variable $\theta$ to the set $\theta \in [0, \delta]$, and we use the standard concentration bound in \cref{sec:analyticConcentration}, to obtain the following
\begin{align*}
&\left| \mu_n [f] - \frac{1}{Q^\delta_n} \int_{\mathbb{S}^{d-1}} d \Omega \int_{B_{2}(0,1)} dr dy \ \mathbbm{1}(r > 0) e^{n(d) \Psi_n (r, \theta, y)} \mathbbm{1}(\theta \in [0, \delta])\nu^{r^* O_n^{-1} (\Omega), y^*} [f] \right| \\ &\leq \varepsilon_n +  \frac{2}{1 + \frac{C (\delta) e^{n(d) (\Psi_n (r_n^*,0,y_n^*) - \Psi_n (r_n^* (\delta), \delta, y_n^* (\delta)))}}{\sqrt{n(d) || m_n ||}^{d - 1}}} ,    
\end{align*}
where $\varepsilon_n \to 0, \ n \to \infty$ is the approximation error appearing from \cref{thm:mixturerep2}, we are again easing the notation by dropping the subindex $1$ from $y_1$ so that $y$ and $y_1$ agree in the formulas, $Q_n^\delta$ is a normalization constant, $f \in \operatorname{BL}_1 ((\mathbb{R}^d)^\mathbb{N})$ is arbitrary, and we have already applied the lower bound given in \cref{thm:angular_concentration} which yields the positive fixed constant $C(\delta) > 0$. Now, by combining the upper bounds for the continuity proof in \cref{def:PS} and the form of $O_n$ given in \cref{sec:coordinateTransforms}, it follows that
\begin{align*}
\mathbbm{1}(\theta \in [0, \delta]) \left| \nu^{r^* O_n^{-1}(\Omega), y^*} [f] - \nu^{r^* \frac{n m_n}{|| n m_n||}, y^*} [f] \right| &\leq r^* \mathbbm{1}(\theta \in [0, \delta]) \left| \left| O_n^{-1} (\Omega) - \frac{m_n}{|| m_n||} \right| \right| \\
&\leq r^* \mathbbm{1}(\theta \in [0, \delta]) \sqrt{(1 - \cos \theta)^2 + (d-1) \sin^2 \theta} \\
&\leq r^* \sup_{\theta \in [0, \delta]}  \sqrt{(1 - \cos \theta)^2 + (d-1) \sin^2 \theta} .
\end{align*}
The right hand side is a constant that is clearly vanishing as $\delta \to 0$. It then follows that
\begin{align*}
&\left|  \frac{1}{Q^\delta_n} \int_{\mathbb{S}^{d-1}} d \Omega \int_{B_{2}(0,1)} dr dy \ \mathbbm{1}(r > 0) e^{n(d) \Psi_n (r, \theta, y)} \mathbbm{1}(\theta \in [0, \delta])\nu^{r^* O_n^{-1} (\Omega), y^*} [f]  - \nu^{\frac{m_n}{||m_n||}, y^*} [f] \right| \\ &\leq  r^* \sup_{\theta \in [0, \delta]}  \sqrt{(1 - \cos \theta)^2 + (d-1) \sin^2 \theta} . 
\end{align*}
Combining together the given inequalities, it follows that
\begin{align*}
|\mu_n [f] - \nu^{r^* \frac{m_n}{|| m_n||}, y^*} [f]| &\leq \varepsilon_n +  r^* \sup_{\theta \in [0, \delta]}  \sqrt{(1 - \cos \theta)^2 + (d-1) \sin^2 \theta}   +   \frac{2}{1 + \frac{C (\delta) e^{n(d) (\Psi_n (r_n^*,0,y_n^*) - \Psi_n (r_n^* (\delta), \delta, y_n^* (\delta)))}}{\sqrt{n(d) || m_n ||}^{d - 1}}} .
\end{align*}

For the second step, we will prove that the second term on the right hand side inequality is exponentially  decreasing. To begin with, we rewrite $\Psi_n$ as follows 
\begin{align*}
\Psi_n (r, \delta, y) = \Psi (r, y) + \beta r || m_n||  \cos (\delta) + \beta y (|| s_n|| - || s ||)  .
\end{align*}
Since $(r_n^* (\delta), y_n^* (\delta))$ is the unique critical point of $\Psi_n (\cdot, \delta, \cdot)$, it follows that
\begin{align*}
\nabla_{(r,y)} [\Psi_n] (r_n^* (\delta), y_n^* (\delta)) = 0 \iff \nabla [\Psi] (r_n^* (\delta), y_n^* (\delta)) = (- \beta || m_n|| \cos(\delta), - \beta (|| s_n|| - || s ||)) .
\end{align*}
Since $\Psi$ has a negative definite Hessian which is continuous in the variables $(r,y)$ at the point $(r^*, y^*)$, it follows that $\nabla [\Psi]$ is locally continuously differentiable and invertible around the point $(r^*,y^*)$, and we have
\begin{align*}
(r_n^* (\delta), y_n^* (\delta)) = (\nabla [\Psi])^{-1} (- \beta || m_n|| \cos (\delta), - \beta (|| s_n|| - || s ||)) .
\end{align*}
Since $\nabla[\Psi]$ is locally continuously differentiable and invertible, it follows that $(\nabla [\Psi])^{-1}$ is also locally continuously differentiable around the origin  $(0,0)$. For a small neighbourhood of the origin $U \subset \mathbb{R}^2$, we can thus consider the continuously differentiable function $g : U \to \mathbb{R}$ given by
\begin{align*}
g (x,y) := (\Psi \circ (\nabla[\Psi])^{-1})(x,y) - \left< (\nabla [\Psi])^{-1} (x,y), (x,y) \right> .
\end{align*}
Using this function, it follows that
\begin{align*}
\Psi_n(r_n^*,0, y_n^*) = g( - \beta || m_n|| , - \beta (|| s_n|| - || s||)), \ \Psi_n (r_n^* (\delta), \delta, y_n^* (\delta)) = g(\cos (\delta) (- \beta || m_n|| ), - \beta (|| s_n|| - || s||)) .
\end{align*}
By developing the derivative of $g$ in the first component variable, it follows that
\begin{align*}
\Psi_n (r_n^*, 0, y_n^*) - \Psi_n (r_n^* (\delta), \delta, y_n^* (\delta)) 
&=  g( - \beta || m_n|| , - \beta (|| s_n|| - || s||)) - g(\cos (\delta) (- \beta || m_n|| ), - \beta (|| s_n|| - || s||)) \\
&= - \beta || m_n || (1 - \cos (\delta)) \\&\times \int_0^1 dt \ (\partial^{(1,0)} g)( - \beta ||m_n|| (1 - t) + \cos (\delta) (- \beta || m_n||) t, - \beta ( || s_n|| - || s||))
 .
\end{align*}
By direct computation, it follows that
\begin{align*}
    \nabla [g] (x,y) = - (\nabla [\Psi])^{-1} (x,y) \implies (\partial^{(1,0)} g) (0,0) = - r^* < 0 . 
\end{align*}
By using the continuous differentiability of $g$, it follows that 
\begin{align*}
\lim_{n \to \infty}  \frac{\Psi_n (r_n^*, 0, y_n^*) - \Psi_n (r_n^* (\delta), \delta, y_n^* (\delta)) }{|| m_n||} = - \beta (1 - \cos (\delta)) (\partial^{(1,0)} g)(0,0) =  \beta r^* (1 - \cos (\delta)) > 0 .  
\end{align*}
Returning to the exponential, we rewrite it as follows 
\begin{align*}
\frac{e^{n(d) (\Psi_n (r_n^*,0,y_n^*) - \Psi_n (r_n^* (\delta), \delta, y_n^* (\delta)))}}{\sqrt{n(d) || m_n ||}^{d - 1}} = e^{n (d) || m_n|| \left( \frac{\Psi_n (r_n^*, 0, y_n^*) - \Psi_n (r_n^* (\delta), \delta, y_n^* (\delta)) }{|| m_n||} - \frac{d-1}{2(n(d) || m_n||)} \ln (n(d) || m_n||) \right)} .  
\end{align*}
Since $\lim_{n \to \infty} n(d) || m_n || = \infty$, it follows that
\begin{align*}
\lim_{n \to \infty} \left( \frac{\Psi_n (r_n^*, 0, y_n^*) - \Psi_n (r_n^* (\delta), \delta, y_n^* (\delta)) }{|| m_n||} - \frac{d-1}{2(n(d) || m_n||)} \ln (n(d) || m_n||) \right) = \beta r^* (1 - \cos (\delta)) > 0 ,     
\end{align*}
which implies that
\begin{align*}
\lim_{n \to \infty}  \frac{e^{n(d) (\Psi_n (r_n^*,0,y_n^*) - \Psi_n (r_n^* (\delta), \delta, y_n^* (\delta)))}}{\sqrt{n(d) || m_n ||}^{d - 1}} = \infty ,    
\end{align*}
from which we obtain
\begin{align*}
\lim_{n \to \infty}  \frac{2}{1 + \frac{C (\delta) e^{n(d) (\Psi_n (r_n^*,0,y_n^*) - \Psi_n (r_n^* (\delta), \delta, y_n^* (\delta)))}}{\sqrt{n(d) || m_n ||}^{d - 1}}} = 0 .   
\end{align*}
Returning now to the original inequality, it follows that
\begin{align*}
d_{\operatorname{BL}_1} (\mu_n, \nu^{r^*\frac{m_n}{||m_n||}, y^*}) \leq  \varepsilon_n + r^* \sup_{\theta \in [0, \delta]}  \sqrt{(1 - \cos \theta)^2 + (d-1) \sin^2 \theta}  &+ \frac{2}{1 + \frac{C (\delta) e^{n(d) (\Psi_n (r_n^*,0,y_n^*) - \Psi_n (r_n^* (\delta), \delta, y_n^* (\delta)))}}{\sqrt{n(d) || m_n ||}^{d - 1}}}   
\end{align*}
for any arbitrary but small $\delta > 0$. In particular, we have
\begin{align*}
\limsup_{n \to \infty} d_{\operatorname{BL}_1} (\mu_n, \nu^{r^*\frac{m_n}{||m_n||}, y^*}) \leq  r^* \sup_{\theta \in [0, \delta]}  \sqrt{(1 - \cos \theta)^2 + (d-1) \sin^2 \theta} ,   
\end{align*}
and since the left hand side does not depend on $\delta > 0$, we can take the limit as $\delta \to 0$ and the result follows.
\end{proof}
\begin{remark} \label{rmk:unbounded_convergence_density}
Using the same assumptions and conditions as in \cref{thm:unbounded_convergence}, it would follow that
\begin{align*}
\lim_{n \to \infty} d_{\operatorname{BL}_1} (\alpha_n^{\beta,h}, \delta_{r^* \widehat{m}_n, y^*}) = 0 .    
\end{align*}
\end{remark}
\subsection{Limit points}
\noindent
We conclude this section with a result concerning the limit points of the finite-volume Gibbs states. It is a direct application of both \cref{thm:bounded_convergence} and \cref{thm:unbounded_convergence}.
\begin{theorem} \label{thm:clusterpointsdeterministic}
Suppose that
\begin{align*}
\lim_{n \to \infty} m_n = 0 \in \mathbb{R}^d , \ \lim_{n \to \infty} s_n = s \in (0, \infty)^d , \ \sum_{(i,j) \in \mathbb{N} \times [d]} \frac{|h_j(i)|}{2^{i+j}} < \infty , \  || s || < 1, \ \beta > \frac{d}{1 - || s ||^2} .   \end{align*}
It follows that
\begin{align*}
\operatorname{clust} \left( \mu_n^{\beta,h} \right) \subset \{ \overline{\nu}^{z, r^*, y^*, h} : z \in \mathbb{R}^d\} \cup \{ \nu^{r^* \Omega, y^*, h} : \Omega \in \mathbb{S}^{d-1}\} ,
\end{align*}
where
\begin{align*}
r^* = \sqrt{1 - \frac{d}{\beta} - || s ||^2}, \ y^* = s .   
\end{align*}
\end{theorem}
\begin{proof}
Let $(\mu_{n_k})$ be a convergent subsequence. When investigating the corresponding subsequence $(n_k || m_{n_k} ||)$, there are only two possibilities. The first is that this subsequence is bounded, in which case, by compactness, there exists $z \in \mathbb{R}^d$ such that $n_{k_j} m_{n_{k_j}} \to z$ as $j \to \infty$ from which we obtain $\mu_{n_{k_j}} \to \overline{\nu}^z$ as $j \to \infty$, by  \cref{thm:bounded_convergence}. The other possibility is that this subsequence is unbounded, which means 
that there is a subsubsequence 
such that $n_{k_j} || m_{n_{k_j}} ||\uparrow \infty$ with $j\uparrow \infty$. 
In this case, by compactness, there exists $\Omega \in \mathbb{S}^{d-1}$ such that along possibly 
another subsubsubsequence we have that 
$\frac{n_{k_{j_l}} m_{n_{k_{j_l}}}}{|| n_{k_{j_l}} m_{n_{k_{j_l}}} ||} \to \Omega$ as $l \to \infty$ from which we obtain 
correspondingly $\mu_{n_{k_{j_l}}} \to \nu^{r^* \Omega, y^*}$ as $l \to \infty$, using this time \cref{thm:unbounded_convergence}.
\end{proof}
\begin{remark} \label{rmk:limit_points_density}
Using the same assumptions and conditions of \cref{thm:clusterpointsdeterministic}, it would follow that
\begin{align*}
\operatorname{clust} \left( \alpha_n^{\beta,h} \right) \subset &\left\{ \left( \int_{\mathbb{S}^{d-1}} d \Omega \ e^{\beta r^* \left< z, \Omega \right>} \right)^{-1} \int_{\mathbb{S}^{d-1}} d \Omega \ e^{\beta r^* \left< z, \Omega \right>} \delta_{r^* \Omega, y^*} : z \in \mathbb{R}^d \right\} \cup \{ \delta_{r^* \Omega, y^*} : \Omega \in \mathbb{S}^{d-1}\} .
\end{align*}
One should note the lack of direct $h$-dependence in this result.
\end{remark}
\section{Random fields and metastates} \label{sec:random}
\subsection{Preliminaries and definitions for random finite-volume Gibbs states} \label{sec:randomFVGS}
\noindent
We now consider the primary application to the case where $h$ is a random external field. First, we show that under mild conditions the conditions under which the results of the previous section hold are satisfied almost surely.
\begin{lemma} \label{thm:simplestronglaw} Suppose that $h : (\Omega, \mathcal{F}, \mathbb{P}) \to \left( \mathbb{R}^d \right)^\mathbb{N}$ is a random variable with independent identically distributed components $\{ h(i)\}_{i \in \mathbb{N}}$ such that each component of each site $\{ h_j(i) \}_{(i,j) \in \mathbb{N} \times [d]}$ has a finite second moment $\mathbb{E} h_j(i)^2 < \infty$.

If we denote
\begin{align*}
m_n^h := \frac{1}{n} \sum_{i=1}^n h(i), \ (s_n^h)_j := \sqrt{\frac{1}{n} \sum_{i=1}^n h_j (i)^2 - ((m_n)_j)^2} ,    
\end{align*}
then it follows that
\begin{align*}
\lim_{n \to \infty} m_n^h = \mathbb{E} h(0), \ \lim_{n \to \infty} (s_n^h)_j = \sqrt{\mathbb{E} h_j(0)^2 - (\mathbb{E} h_j(0))^2}, \ \sum_{(i,j) \in \mathbb{N} \times [d]} \frac{|h_j(i)|}{2^{i + j}} < \infty   
\end{align*}
$\mathbb{P}-$almost surely. 
\end{lemma}
\begin{proof}
The first two limits follow by the strong law of large numbers. For the finiteness of the sum, we have
\begin{align*}
\sum_{(i,j) \in [n] \times [d]} \frac{|h_j(i)|}{2^{i+j}} \leq C(d) \max_{j \in [d]} \mathbb{E} |h_j(0)| + \sum_{(i,j) \in [n]] \times [d]} \frac{|h_j(i)| - \mathbb{E} |h_j(i)|}{2^{i+j}} . 
\end{align*}
Now note that
\begin{align*}
\mathbb{E} \left(\frac{|h_j(i)| - \mathbb{E} |h_j(i)|}{2^{i+j}} \right)^2 = \frac{\mathbb{E} |h_j(0)|^2 - (\mathbb{E} |h_j(0)|)^2}{4^{i+j}} .
\end{align*}
If we sum over the $(i,j) \in \mathbb{N} \times [d]$, it is clear that the corresponding series is finite. By the Kolmogorov-Khintchin theorem, see \cite[Theorem 2.5.6]{Durrett2019}, it follows that
\begin{align*}
\sum_{(i,j) \in \mathbb{N} \times [d]} \frac{|h_j(i)| - \mathbb{E} |h_j(i)|}{2^{i+j}}   < \infty
\end{align*}
$\mathbb{P}$-almost surely, from which the $\mathbb{P}$-almost sure finiteness of the sum in this result follows. 
\end{proof}
\noindent
Since $h \mapsto \mu_n^{\beta,h}$ is a continuous mapping, it follows that if $h$ is a random external field then we can identify $\mu_n^{\beta,h}$ as a random probability measure by the mapping $\omega \mapsto h (\omega) \mapsto \mu_n^{\beta, h (\omega)}$. When we now refer to $\mathbb{P}$-almost sure properties, we mean with respect to the underlying probability space that $h$ is built on. 

From \cref{thm:simplestronglaw}, combined together with \cref{thm:MSregime}, we see that the interesting random external fields which should generate non-trivial behavior in the large-volume limits of the finite-volume Gibbs states almost surely should at least satisfy
\begin{align*}
\mathbb{E} h(0) = 0, \ \mathbb{E} ||h(0) ||^2 < 1    
\end{align*}
and the inverse temperature $\beta > 0$ should be chosen such that
\begin{align*}
\beta > \frac{d}{1 -\mathbb{E} ||h(0) ||^2} . 
\end{align*}
With reference to \cref{thm:bounded_convergence}, \cref{thm:unbounded_convergence}, and \cref{thm:clusterpointsdeterministic}, we also see that the behaviour of the sum of the random fields at the sites 
\begin{align*}
S_n := \sum_{i=1}^n h(i) = n m_n^h      
\end{align*}
determines the possible limit points of the finite-volume Gibbs states almost surely or otherwise.

To that end, we have the following result detailing the behaviour of the $d$-dimensional random walk $\{ S_n \}_{n=1}^\infty$ as it relates to our model.
\begin{lemma} \label{thm:standard_information} Suppose that $h : (\Omega, \mathcal{F}, \mathbb{P}) \to \left( \mathbb{R}^d \right)^\mathbb{N}$ is a random variable with independent identically distributed components $\{ h(i)\}_{i \in \mathbb{N}}$ such that each component of each site $\{ h_j(i) \}_{(i,j) \in \mathbb{N} \times [d]}$ has a finite second moment $\mathbb{E} h_j(i)^2 < \infty$, is centered $\mathbb{E} h(i) = 0$, and the covariance matrix $\Sigma_{j,k} :=  \mathbb{E} h_j(i) h_k(i)$ has full rank $d$.

For any $d \geq 2$, the following properties hold:
\begin{enumerate}
\item We have
\begin{align*}
\operatorname{clust} \left( \frac{S_n}{\sqrt{n}} \right) = \mathbb{R}^d
\end{align*}
almost surely.
\end{enumerate}
For $d \geq 3$, the following properties hold:
\begin{enumerate}
\item We have
\begin{align*}
\lim_{n \to \infty} || S_n || = \infty
\end{align*}
almost surely.
\end{enumerate}
For $d = 2$, the following properties hold:
\begin{enumerate}
\item We have
\begin{align*}
P = \operatorname{clust} (S_n)
\end{align*}
almost surely, where $P$ is the set of recurrent points of the random walk $(S_n)$ given by
\begin{align*}
P := \{z \in \mathbb{R}^2 : \forall \varepsilon > 0, \ \mathbb{P} (S_n\in B(z, \varepsilon) \text{ infinitely often}) = 1\} .    
\end{align*}
\end{enumerate}
\end{lemma}
\begin{proof}
We will prove things in the order they appear. For the first point, by the reverse Fatou lemma, it follows that
\begin{align*}
\mathbb{P} \left( \left(\frac{S_n}{\sqrt{n}} \right) \in B(z, \varepsilon) \text{ infinitely often}\right) \geq \limsup_{N \to \infty} \mathbb{P}\left( \frac{S_n}{\sqrt{n}} \in B(z, \varepsilon)\right) = \mathbb{P}(B_1 \in B(z, \varepsilon) > 0)    
\end{align*}
for any $z \in \mathbb{R}^d$, and $\varepsilon > 0$, where $B_1$ is a possibly correlated Gaussian random variable on $\mathbb{R}^d$. The set
\begin{align*}
\left(\frac{S_n}{\sqrt{n}} \right) \in B(z, \varepsilon) \text{ infinitely often}    
\end{align*}
belongs to the exchangeable sigma-algebra, and thus, by the Hewitt-Savage theorem, see \cite[Chapter 12]{Klenke2020},  it follows that it is either has probability $1$ or $0$. Since this set has positive probability, it follows that 
\begin{align*}
 \mathbb{P} \left( \left(\frac{S_n}{\sqrt{n}} \right) \in B(z, \varepsilon) \text{ infinitely often}\right) = 1 .  
\end{align*}
Denote
\begin{align*}
\Omega' := \bigcap_{q \in \mathbb{Q}^d, \ k \in \mathbb{N}} \left\{  \left(\frac{S_n}{\sqrt{n}} \right) \in B\left(q, \frac{1}{k} \right) \text{ infinitely often}\right\} .   
\end{align*}
This is countable intersection of sets of probability $1$, and thus it follows that $\mathbb{P}(\Omega') = 1$. Now, given an arbitrary $z \in \mathbb{R}^d$, then by density, it follows that there exists a sequence $(q_j)$ in $\mathbb{Q}^d$ such that
\begin{align*}
\lim_{j \to \infty} q_j = z .    
\end{align*}
In the set $\Omega'$, for any $j$, choose $l_j$ such that $l_j > l_{j-1}$ and
\begin{align*}
\left| \left| \frac{S_{l_j}}{\sqrt{l_j}} - q_j \right| \right| \leq \frac{1}{j} .    
\end{align*}
It follows that
\begin{align*}
\lim_{j \to \infty} \frac{S_{l_j}}{\sqrt{l_j}} = z  
\end{align*}
almost surely, and since this holds for any $z \in \mathbb{R}^d$ the claim holds.

For the next statement, the random walk $(S_n)$ is transient in dimension $d \geq 3$, which implies that
\begin{align*}
\mathbb{P} \left( S_n\in B(0,M) \text{ infinitely often}\right) = 0 \iff  \mathbb{P} \left(  S_n \in B(0,M) \text{ finitely often} \right) = 1 .  
\end{align*}
for any $M \in \mathbb{N}$. In the same way as before, we have
\begin{align*}
\mathbb{P} \left( \bigcap_{M \in \mathbb{N}} \left( S_n\right) \in B(0,M) \text{ finitely often}\right) = 1 .    
\end{align*}
Now, in particular given a $M \in \mathbb{N}$, there exists $N \in \mathbb{N}$ such that for all $n \geq N$, it follows that 
\begin{align*}
|| S_n || \geq M .    
\end{align*}
This implies that
\begin{align*}
\lim_{n \to \infty} ||S_n || = \infty   
\end{align*}
almost surely.

For the final statement, we will need to construct a new set of probability $1$. We will largely mirror the ideas of the proof of a similar statement for normalized one-dimensional random walks given in \cite[Theorem 1]{Kesten1970}. First, we will use $I$ to denote half-open rational rectangles which are sets of the form
\begin{align*}
I = [a_1,b_1) \times [a_2,b_2),   
\end{align*}
where $a_1 < b_1$, $a_2 < b_2$, and $a_1,b_1,a_2,b_2 \in \mathbb{Q}$. By the Hewitt-Savage theorem, we have
\begin{align*}
\mathbb{P}\left( S_n \in I \text{ infinitely often} \right) = 0,1 .    
\end{align*}
Note that if the probability above is $0$, then it follows that
\begin{align*}
\mathbb{P}(S_n \in I \text{ finitely often}) = 1 .    
\end{align*}
Denote the countable collections $A$ and $B$ of half-open rational rectangles by
\begin{align*}
A := \{ I : \mathbb{P}\left( S_n \in I \text{ infinitely often} \right) = 1 \} , \ B := \{ I : \mathbb{P}\left( S_n \in I \text{ finitely often} \right) = 1 \}  .
\end{align*}
The following set
\begin{align*}
\Omega'' := \bigcap_{I \in A} \{  S_n \in I \text{ infinitely often} \} \cap \bigcap_{I \in B}  \left\{S_n \in I \text{ finitely often}\right\}  
\end{align*}
is a set of probability $1$, and we will prove our results almost surely with respect to this set. We will first prove that
\begin{align*}
P \subset \operatorname{clust} (S_n)    
\end{align*}
almost surely. To that end, by recurrence, it follows that at least every half-open rational interval that contains $p$ in its interior belongs to the set $A$. Let $(q_n)$ be a sequence in $\mathbb{Q}^2$ such that
\begin{align*}
\lim_{n \to \infty} q_n = p .    
\end{align*}
For large enough $n \geq N$, it follows that 
\begin{align*}
p \in \operatorname{int} \left( \left[ (q_n)_1 - \frac{1}{n}, (q_n)_1 + \frac{1}{n}\right) \times \left[ (q_n)_2 - \frac{1}{n}, (q_n)_2 + \frac{1}{n}\right)\right) .  
\end{align*}
If we denote
\begin{align*}
I_n :=  \left[ (q_n)_1 - \frac{1}{n}, (q_n)_1 + \frac{1}{n}\right) \times \left[ (q_n)_2 - \frac{1}{n}, (q_n)_2 + \frac{1}{n}\right) ,    
\end{align*}
then since $\mathbb{P}(S_n \in I \text{ infinitely often}) = 1$, by the same subsequence construction as for the first claim in this result, we can construct a random subsequence $S_{n_k}$ converging to $p$ almost surely. It follows that
\begin{align*}
P \subset \operatorname{clust} (S_n) .    
\end{align*}
For the other inclusion, choose $p \not \in P$. It follows that there must exist a half-open rational rectangle $I$ containing $p$ such that $\mathbb{P}(S_n \in \operatorname{I} \text{ finitely often}) = 1$. Towards a contradiction, suppose that $p \in \operatorname{clust} (S_n)$ almost surely. This is equivalent to stating that $\mathbb{P}(S_n \in I' \text{ infinitely often for every half-open rational rectangle containing } p) = 1$. We have
\begin{align*}
&\mathbb{P}(S_n \in I' \text{ infinitely often for every half-open rational rectangle containing } p) \\
&= 1 - \mathbb{P}(S_n \in I' \text{ finitely often for some half-open rational rectangle containing } p) \\
&\leq 1 - \mathbb{P}(S_n \in I \text{ finitely often}) = 0 .
\end{align*}
This is a contradiction, and thus we must have $p \not \in \operatorname{clust} (S_n)$ almost surely, which implies that
\begin{align*}
P^c \subset \operatorname{clust}(S_n)^c \implies \operatorname{clust}(S_n) \subset P .    
\end{align*}
In totality, we see that
\begin{align*}
P = \operatorname{clust}(S_n)    
\end{align*}
almost surely, as desired.
\end{proof}
\noindent
Motivated by this collection of results and observations, we bundle together these various conditions into a single list of assumptions.
\begin{assumptionA} \label{assump:A} We say that a measurable random external field $h : (\Omega, \mathcal{F}, \mathbb{P}) \to (\mathbb{R}^d)^\mathbb{N}$ satisfies the conditions (A) if:
\begin{enumerate}
\item The components $h(i)$ are independent identically distributed $\mathbb{R}^d$-valued random variables. 
\item The components $h(i)$ satisfy $\mathbb{E} |h_{j} (i)|^{2} < \infty$ for any $j \in [d]$.
\item The components $h(i)$ are centered $\mathbb{E} h(i) = 0$.
\item The covariance matrix $ \Sigma := (\mathbb{E} h_j(i) h_k (i))_{j,k=1}^d$ has full rank $d$.    
\item The second moments satisfy $\mathbb{E} || h(i) ||^2 < 1$.
\item The inverse temperature $\beta > 0$ is chosen so that $\beta > \frac{d}{1 - \mathbb{E} ||h(i)||^2}$.    

\end{enumerate}
\end{assumptionA}
\noindent
\subsection{Relabelling of finite-volume Gibbs states, pure states, and $z$-tilted probability measures} \label{sec:relabelling}
Given \cref{assump:A}, we also change the notation of the finite-volume Gibbs states to emphasize the relevant details. First, we define the constants
\begin{align*}
r^* = \sqrt{1 - \frac{d}{\beta} - \mathbb{E} || h(i)||^2}, \ (y^*)_j = \sqrt{\mathbb{E} h_j(i)^2} .     
\end{align*}
The finite-volume Gibbs states $\mu_n^h$ are defined by $\mu_n^h := \mu_n^{\beta,h}$, and the mixing probability measures $\alpha_n^{\beta,h}$ are defined by $\alpha_n^h := \alpha_n^{\beta,h}$.  The pure states $\nu_\Omega^h$ for $\Omega \in \mathbb{S}^{d-1}$ are defined by by $\nu_\Omega^h := \nu^{r^* \Omega, y^*, h}$. Explicitly, we see that $\nu_\Omega^h$ is a factorized probability measure with single-site single-component marginal distributions given by
\begin{align*}
\nu_\Omega^h |_{(i,j)} \sim \frac{G_j (i)}{\sqrt{\beta}} + \sqrt{1 - \frac{d}{\beta} - \mathbb{E} || h(i) ||^2} \Omega + h_j(i) ,
\end{align*}
where $\{G_j(i) \}_{(i,j) \in \mathbb{N} \times [d]}$ is a standard Gaussian random variable. Note that $h$ and $G$ live in two different probability spaces. The $z$-tilted probability measures $\overline{\nu}_z^h$ for $z \in \mathbb{R}^d$ are defined by
\begin{align*}
\overline{\nu}_z^h := \left( \int_{\mathbb{S}^{d-1}} d \Omega \ e^{\beta r^* \left< z, \Omega \right>} \right)^{-1} \int_{\mathbb{S}^{d-1}} d \Omega \ e^{\beta r^* \left< z, \Omega \right>} \nu_\Omega^h .   
\end{align*}
\noindent
\subsection{Chaotic size dependence}
\noindent
Using the limit theorems for the random walk, we have the following chaotic size dependence result.
\begin{theorem} \label{thm:CSD}
Suppose that $h$ satisfies (A).

For dimension $d = 2$, it follows that 
\begin{align*}
\operatorname{clust} (\mu_n^h) = \{ \overline{\nu}_z^h : z \in P \} \cup \{ \nu_\Omega^h : \Omega \in \mathbb{S}^{d-1}\}
\end{align*}
almost surely, where $P \subset \mathbb{R}^d$ is the set of recurrent values of $(S_n)$, and, for dimensions $d \geq 3$, it follows that
\begin{align*}
\operatorname{clust} (\mu_n^h) = \{ \nu_\Omega^h : \Omega \in \mathbb{S}^{d-1}\}
\end{align*}
almost surely.
\end{theorem}
\begin{proof}
For all dimensions $d \geq 2$, it follows that if $\frac{S_{n_k}}{\sqrt{n_k}} \to q$, where $z \in \mathbb{R}^d \setminus \{0 \}$, as $k \to \infty$, then $\ || S_{n_k} || \to \infty$ and $\widehat{S}_{n_k}\to \widehat{z}$ as $k \to \infty$. By \cref{thm:unbounded_convergence}, it follows that
\begin{align*}
\lim_{k\to \infty} d_{\operatorname{BL}_1} (\mu_{n_k}^h, \nu^h_{\widehat{z}}) = 0 .     
\end{align*}
Since $\operatorname{clust}\left( \frac{S_n}{\sqrt{n}} \right) = \mathbb{R}^d$ almost surely, it follows that
\begin{align*}
\{ \nu_\Omega^h : \Omega \in \mathbb{S}^{d-1} \} \subset \operatorname{clust} (\mu_n^h)
\end{align*}
almost surely. 

For dimensions $d \geq 3$, since
\begin{align*}
\lim_{n \to \infty} || S_n || = \infty    
\end{align*}
almost surely, it follows that if $(\mu^h_{n_k})$ is a convergent subsequence, then by compactness of $\mathbb{S}^{d-1}$, there must exist a further convergent subsubsequence $(\mu^h_{n_{k_j}})$ such that
\begin{align*}
\lim_{j \to \infty} \widehat{S}_{n_{k_j}} = \Omega .  
\end{align*}
It follows that
\begin{align*}
\lim_{j \to \infty} d_{\operatorname{BL}_1} (\mu_{n_{k_j}}^h, \nu_\Omega^h) = 0 ,    
\end{align*}
which implies that
\begin{align*}
 \lim_{k \to \infty} d_{\operatorname{BL}_1} (\mu_{n_{k}}^h, \nu_\Omega^h) = 0,     
\end{align*}
from which we obtain
\begin{align*}
\operatorname{clust} (\mu_n^h) \subset  \{ \nu_\Omega^h : \Omega \in \mathbb{S}^{d-1}\}   
\end{align*}
almost surely, which implies that for dimension $d \geq 3$, we must have
\begin{align*}
 \operatorname{clust} (\mu_n^h) = \{ \nu_\Omega^h : \Omega \in \mathbb{S}^{d-1}\} .   
\end{align*}
For the case of dimension $d = 2$, by recurrence of the random walk, it follows that for every $z \in P$ the set of recurrent values, there exists a subsequence $(S_{n_k})$ such that
\begin{align*}
 \lim_{n \to \infty} S_{n_k} = z .   
\end{align*}
Along this subsequence, it follows that
\begin{align*}
\lim_{k \to \infty} d_{\operatorname{BL}_1} (\mu_{n_k}^h, \overline{\nu}_z^h) = 0    ,
\end{align*}
which implies that
\begin{align*}
\{ \overline{\nu}_z^h : z \in P \} \cup \{ \nu_\Omega^h : \Omega \in \mathbb{S}^{d-1}\} \subset \operatorname{clust} (\mu_n^h) .    
\end{align*}
In the other direction, we repeat the argument of \cref{thm:clusterpointsdeterministic}. In particular, if we have a convergent subsequence $(\mu_{n_k}^h)$ then the corresponding subsequence of the random walk $(S_{n_k})$ is either bounded or contains a further unbounded subsubsequence. If the subsequence is bounded then, by compactness, there exists a further subsubsequence $(S_{n_{k_j}})$ such that
\begin{align*}
\lim_{j \to \infty} S_{n_{k_j}} = z \in P    
\end{align*}
in which case we have
\begin{align*}
\lim_{j \to \infty} d_{\operatorname{BL}_1} (\mu_{n_{k_j}}^h, \overline{\nu}_z^h) = 0 ,  
\end{align*}
and hence the same limit also for the original subsequence. If there is an unbounded subsubsequence $(S_{n_{k_j}})$, then we repeat the argument used for $d = 3$, and it follows that 
\begin{align*}
\lim_{k \to \infty} d_{\operatorname{BL}_1} (\mu_{n_{k}}^h, \nu_\Omega^h)    
\end{align*}
for some $\Omega \in \mathbb{S}^{d-1}$. Since these are the only two cases, it follows
\begin{align*}
 \operatorname{clust} (\mu_n^h) \subset \{ \overline{\nu}_z^h : z \in P \} \cup \{ \nu_\Omega^h : \Omega \in \mathbb{S}^{d-1}\}    
\end{align*}
from which the result follows. 
\end{proof}
\noindent
\begin{remark} \label{rmk:mixing_probability_measure_cluster}
Using the same assumptions and conditions of \cref{thm:CSD}, for dimension $d = 2$, it follows that 
\begin{align*}
\operatorname{clust} (\alpha_n^h) = \overline{\left\{ \left( \int_{\mathbb{S}^1} d \Omega \ e^{ \beta r^* \left< z, \Omega \right>} \right)^{-1} \int_{\mathbb{S}^1} d \Omega \ e^{ \beta r^* \left< z, \Omega \right>} \delta_{r^* \Omega, y^*} : z \in P \} \cup \{ \delta_{r^* \Omega, y^*} : \Omega \in \mathbb{S}^{d-1} \right\}}
\end{align*}
almost surely, where $P \subset \mathbb{R}^d$ is the set of recurrent values of $(S_n)$, and, for dimensions $d \geq 3$, it follows that
\begin{align*}
\operatorname{clust} (\alpha_n^h) = \{ \delta_{r^* \Omega, y^*} : \Omega \in \mathbb{S}^{d-1}\}
\end{align*}
almost surely. We once again note the explicit lack of parametric $h$-dependence in these results.
\end{remark}
\begin{remark} \label{rmk:possibleValues}
In this remark, we present some additional details concerning the set of recurrent values $P \subset \mathbb{R}^d$ of the random walk $(S_n)$ appearing in \cref{thm:CSD}. Most of what is presented here is contained in \cite[Chapter 5, Section 5.4]{Durrett2019}. We say that $x \in \mathbb{R}^d$ is a possible value of the random walk $(S_n)$ if for every $\varepsilon > 0$ there exists $n$ such that $\mathbb{P}(|| S_n - x || < \varepsilon) > 0$. The set of recurrent points of a random walk is either the empty set, or it is a closed (additive) subgroup of $\mathbb{R}^d$ which coincides exactly with the set of possible values. In our case, since the random field satisfies \cref{assump:A}, it follows that the corresponding random walk in dimension $d = 2$ has a non-empty set of recurrent values, and thus they coincide with the possible values. In the context of this work, we present two archetypical examples of random fields. The first is the case where the random field components are distributed as standard $2$-dimensional Gaussians. In this case, since the support of the Gaussian itself is the whole space, we see that the set of recurrent values must be the entirety of $\mathbb{R}^2$. For the second case, we consider the random field components distributed uniformly on the set $\{-1,1 \}^2$. Here instead the possible values are given by all of $\mathbb{Z}^2$. The first case is what would consider a continuous random field, and the second case constitutes a lattice random field. By modifying the distribution of the components of the random field, we can obtain different examples of sets of possible values.  
\end{remark}
\noindent
This cluster point result ensures that the collection $(\mu_n^h)$ is tight almost surely. 
\subsection{Aizenman-Wehr metastate}
\noindent
For the investigation of the Aizenman-Wehr metastate, we need to prove the uniform tightness of the so-called intensity measure, see \cref{sec:tightnessConvergenceRandomMeasures}. We have the following  result.
\begin{lemma} Suppose that $h$ satisfies (A).

It follows that the collection of intensity measures $(\mathbb{E} \mu_n^h)$ is uniformly tight.
\end{lemma}
\begin{proof}
To prove the uniform tightness of the intensity measure, it is enough to prove the uniform tightness of its marginals. For this, observe that
\begin{align*}
\mathbb{E} \mu_n^h [\phi_j (i)^2] =  \frac{1}{n} \mathbb{E} \mu_n^h \left[ \sum_{i=1}^n \phi_j (i)^2 \right] \leq \mathbb{E} \mu_n^h \left[ \frac{1}{n} \sum_{j=1}^d\sum_{i=1}^n \phi_j (i)^2 \right] = 1 .
\end{align*}
Using Chebyshev's inequality, it follows that
\begin{align*}
1 - \frac{1}{C^2} \leq \mathbb{E} \mu_n^h (|\phi_j (i)| \leq C) , 
\end{align*}
which shows that when $C \to \infty$, then $\mathbb{E} \mu_n^h (|\phi_j (i)| \leq C) \to 1$, which proves the uniform tightness of this particular marginal. Since $(i,j)$ was arbitrary, it follows that all the marginals are uniformly tight which implies that intensity measure is uniformly tight.
\end{proof}
\noindent
We can now construct the(a) Aizenman-Wehr metastate of the given model. We will apply the results given in \cref{thm:unbounded_convergence}. We have the following result.
\begin{theorem} \label{thm:AWMS}
Suppose that $h$ satisfies (A).

It follows that
\begin{align*}
\lim_{n \to \infty} \mu_n^h = \nu^h_{\widehat{B}_1}
\end{align*}
in law, where $B_1$ is a possibly correlated $d$-dimensional Gaussian random variable with covariance matrix $\Sigma$ independent of $h$.

The(an) Aizenman-Wehr metastate $\kappa^h$ is the(a) measurable map $\kappa^{\cdot} : (\mathbb{R}^d)^\mathbb{N} \to \mathcal{M}_1 (\mathcal{M}_1 ((\mathbb{R}^d)^\mathbb{N}))$ that satisfies
\begin{align*}
\mathbb{E} f \left( h, \nu^h_{\widehat{B}_1}\right) = \mathbb{E} \kappa^h [f (h, \cdot)]  
\end{align*} 
for any $f \in C_b ((\mathbb{R}^d)^\mathbb{N} \times \mathcal{M}_1 ((\mathbb{R}^d)^\mathbb{N}))$.

It follows that
\begin{align*}
\kappa^h := \int_{\mathbb{S}^{d-1}} d \Omega \ \rho_\mathbb{P} (\Omega)  \delta_{\nu_\Omega^h} ,
\end{align*}
where $\rho_\mathbb{P}$ is given by
\begin{align*}
\rho_\mathbb{P} (\Omega) := \left( \int_{\mathbb{S}^{d-1}} \frac{d \Omega}{\left( \left< \Omega, \Sigma^{-1} \Omega \right> \right)^{\frac{d}{2}}}\right)^{-1} \frac{1}{\left( \left< \Omega, \Sigma^{-1} \Omega \right> \right)^{\frac{d}{2}}}
\end{align*}
\end{theorem}
\begin{proof}
Since
\begin{align*}
\lim_{n \to \infty} \left( h, \frac{S_n}{\sqrt{n}}\right) = (h,B_1)
\end{align*}
in law, where $B_1$ is a possibly non-correlated Gaussian random variable independent of $h$, it follows by Skorohod's representation theorem, see \cite[Chapter 17]{Klenke2020} that there exists a probability space and associated random variables such that this convergence in law can be elevated to almost sure convergence in the new space, and the new random variables agree in distribution with the old ones. By an abuse of notation, we will use the old variable notations for the new ones. Since $B_1 \not = 0$ almost surely, it follows that $|| S_n || \to \infty$ and $\widehat{S}_n \to \widehat{B_1}$, so long as $B_1 \not = 0$. In this new probability space, by \cref{thm:unbounded_convergence}, we have
\begin{align*}
\lim_{n \to \infty} d_{\operatorname{BL}_1} (\mu_n^h, \nu^h_{\widehat{B}_1}) = 0 .
\end{align*}
almost surely, and thus also in distribution in the actual probability space of interest. Since $B_1$ is independent of $h$, it follows that
\begin{align*}
\mathbb{E} f (h, \nu_{\widehat{B}_1}) = \mathbb{E} \int_{\mathbb{S}^{d-1}} P (d \Omega) f (h, \nu^h_\Omega) , 
\end{align*}
where $P (d \Omega)$ is distributed according to  the random variable $\widehat{B}_1$. The result follows by shifting to hyperspherical coordinates and integrating the radial factors away.
\end{proof}
\subsection{Newman-Stein metastate}
\noindent
For the construction of the Newman-Stein metastate, we will need the following conditioning result. This result is only required in dimension $d = 2$ due to the recurrence, or rather lack of transience, of the random walk.
\begin{lemma} \label{thm:zeroconditioning} Suppose that $h$ satisfies (A).

In addition, suppose that $\mathbb{E} |h_j(i)|^3 < \infty$ for all $(i,j) \in \mathbb{N} \times [d]$.

It follows that
\begin{align*}
\frac{1}{N} \sum_{n=1}^N \mathbbm{1}(|| S_n || \leq n^{\frac{1}{2} - \frac{1}{2d}}) = o (1)
\end{align*}
almost surely.
\end{lemma}
\begin{proof}
Denote the sequence inside the limit by $C_N$. Note that for every $N$, there exists $K$ such that $2^K \leq N \leq 2^{K+1}$, and for such a $K$, we have
\begin{align*}
\frac{C_{2^K}}{2} \leq C_N \leq 2 C_{2^{K+1}} .
\end{align*}
It follows that it is enough to prove that $C_{2^K} \to 0$ almost surely. Using Chebyshev's inequality, for every $\varepsilon > 0$, it follows that
\begin{align*}
\mathbb{P}(C_N > \varepsilon) \leq \frac{1}{\varepsilon N} \sum_{n=1}^N \mathbb{P}(|| S_n || \leq n^{\frac{1}{2} - \frac{1}{2d}}) .
\end{align*}
By the multivariate Berry-Esseen bounds, see \cite{Gotze1991}, we have
\begin{align*}
\mathbb{P}(|| S_n || \leq n^{\frac{1}{2} - \frac{1}{2d}}) \leq \mathbb{P}(|| G || \leq n^{- \frac{1}{2d}}) + O (n^{- \frac{1}{2}})
\end{align*} 
We have 
\begin{align*}
\mathbb{P}(|| G || \leq  n^{- \frac{1}{2d}}) = O (n^{- \frac{1}{2}}) ,
\end{align*}
so that
\begin{align*}
\mathbb{P}(|| S_n || \leq n^{\frac{1}{2}- \frac{1}{2d}}) = O (n^{- \frac{1}{2}}) .
\end{align*}
From this point onwards, one can proceed exactly as in the same proof for the $1$-dimensional model, see  \cite[Lemma 3.8.1]{Koskinen2023} of and the result follows.
\end{proof}
\noindent
The conditioning result \cref{thm:zeroconditioning} yields the following intermediate result concerning the almost sure asymptotic behaviour of the Newman-Stein metastates.
\begin{lemma} \label{thm:NS_asymptotic} Suppose that $h$ satisfies (A).

For dimensions $d \geq 3$, or for dimension $d = 2$ with the additional assumption that $\mathbb{E} |h_j(i)|^3 < \infty$ for all $(i,j) \in \mathbb{N} \times [d]$, it follows that
\begin{align*}
\lim_{n \to \infty} d_{\operatorname{BL}_1} \left( \overline{\kappa}_N^h, \frac{1}{N} \sum_{n=1}^N \delta_{\nu^h_{\hat{S}_n}}\right)  = 0 
\end{align*}
almost surely.
\end{lemma}
\begin{proof}
We have
\begin{align*}
d_{\operatorname{BL}_1} \left( \overline{\kappa}_N^h, \frac{1}{N} \sum_{n=1}^N \delta_{\nu^h_{\hat{S}_n}}\right) \leq \frac{1}{N} \sum_{n=1}^N d_{\operatorname{BL}_1} (\mu_n^h, \nu^h_{\widehat{S}_n}) .    
\end{align*}
In dimensions $d \geq 3$, by transience of the random walk, by \cref{thm:unbounded_convergence}, it follows that
\begin{align*}
\lim_{n \to \infty} d_{\operatorname{BL}_1} (\mu_n^h, \nu^h_{\widehat{S}_n}) = 0 
\end{align*}
almost surely, from which the result follows. For dimension $d = 2$ with the additional assumption, denote $A_n := \mathbbm{1}(||S_n|| > n^{\frac{1}{2} - \frac{1}{2d}})$. By the same argument as for dimension $d \geq 3$, it follows that
\begin{align*}
\lim_{n \to \infty} A_n d_{\operatorname{BL}_1} (\mu_n^h, \nu^h_{\widehat{S}_n}) = 0 
\end{align*}
almost surely, and by \cref{thm:zeroconditioning}, it follows that
\begin{align*}
\lim_{N \to \infty} \frac{1}{N}   \sum_{n=1}^N (1 - A_n) = 0 
\end{align*}
almost surely. We have
\begin{align*}
\left| \frac{1}{N} \sum_{n=1}^N d_{\operatorname{BL}_1} (\mu_n^h, \nu^h_{\widehat{S}_n}) \right| \leq \frac{1}{N} \sum_{n=1}^N A_n d_{\operatorname{BL}_1} (\mu_n^h, \nu^h_{\widehat{S}_n}) + 2  \frac{1}{N}   \sum_{n=1}^N (1 - A_n) ,   
\end{align*}
from which the result follows.
\end{proof}
\noindent
By \cref{thm:NS_asymptotic}, the behaviour of the Newman-Stein metastates is governed, in a sense, by the behaviour of the random empirical probability measure
\begin{align*}
\frac{1}{N} \sum_{n=1}^N \delta_{\widehat{S}_n} ,    
\end{align*}
which can be investigated by using functional central limit theorems. We will utilize some results presented in \cite{HeiLo2018}. We present the following result concerning the convergence in law of the Newman-Stein metastates.
\begin{theorem} \label{thm:NSdistconv}
Suppose that $h$ satisfies (A).

For dimensions $d \geq 3$, or for dimension $d = 2$ with the additional assumption that $\mathbb{E} |h_j(i)|^3 < \infty$ for all $(i,j) \in \mathbb{N} \times [d]$, it follows that
\begin{align*}
\lim_{N \to \infty} \overline{\kappa}^h_N = \int_0^1 dt \ \delta_{\nu^h_{\widehat{B}_t}} 
\end{align*}
in law, where $B_t$ is a possibly correlated Brownian motion independent of $h$.
\end{theorem}
\begin{proof} 
First, by \cref{thm:NS_asymptotic}, it follows that
\begin{align*}
\lim_{N \to \infty} d_{\operatorname{BL}_1} \left(\overline{\kappa}_N, \frac{1}{N} \sum_{n=1}^N \delta_{\nu_{\widehat{S}_n}} \right) = 0   
\end{align*}
almost surely, so we can continue with the later empirical measure. Using the collection of sets $\mathcal{A} (\delta)$ and the associated proofs from \cref{sec:disjointPartitions}, we have
\begin{align*}
\left| \frac{1}{N} \sum_{n=1}^N f (\nu_{\widehat{S}_n}) - \sum_{A \in \mathcal{A}(\delta)} \pi_N (A) f(\nu_{a(A)}) \right|  &= \left| \sum_{A \in \mathcal{A}(\delta)} \pi_N (A) \frac{\pi_N [\mathbbm{1} (\cdot \in A) f(\nu_{\cdot})]}{\pi_N (A)} - \sum_{A \in \mathcal{A}(\delta)} \pi_N (A) f(\nu_{a(A)}) \right|  \\
&\leq \sum_{A \in \mathcal{A}(\delta)} \pi_N (A) \frac{\pi_N [\mathbbm{1} (\cdot \in A) |f(\nu_{\cdot}) - f(\nu_{a(A)})|]}{\pi_N (A)} \\
&\leq \delta .
\end{align*}
almost surely, where we recall that $a$ is a some element in the non-empty interior of $A \in A(\delta)$, and 
\begin{align*}
\pi_N (A) := \frac{1}{N} \sum_{n=1}^N \mathbbm{1} (\widehat{S}_n \in A) .
\end{align*}
It follows that
\begin{align*}
d_{\operatorname{BL}_1} \left( \frac{1}{N} \sum_{n=1}^N \delta_{\nu_{\widehat{S}_n}},  \sum_{A \in \mathcal{A}(\delta)} \pi_N (A) \delta_{\nu_{a(A)}} \right) \leq \delta    
\end{align*}
almost surely, and thus 
\begin{align*}
\lim_{\delta \to 0^+}  d_{\operatorname{BL}_1} \left( \frac{1}{N} \sum_{n=1}^N \delta_{\nu_{\widehat{S}_n}},  \sum_{A \in \mathcal{A}(\delta)} \pi_N (A) \delta_{\nu_{a(A)}} \right) = 0  
\end{align*}
almost surely. By \cite[Theorem 4.13]{HeiLo2018}, we know that
\begin{align*}
\lim_{N \to \infty} (\pi_N(A_1 (\delta),..., \pi_N (A_I (\delta)))) = \left( \int_0^1 dt \ \mathbbm{1}(\widehat{B}_t \in A_1 (\delta)),..., \int_0^1 dt \ \mathbbm{1}(\widehat{B}_t \in A_I (\delta))\right)    
\end{align*}
in law, where $B_t$ is a possibly correlated Brownian motion independent of $h$, and we have enumerated the elements of $\mathcal{A}(\delta)$ by $i = 1,..., I$. It is important to note that the sets in $\mathcal{A}(\delta)$ have been precisely constructed to satisfy the requirements of \cite[Theorem 4.13]{HeiLo2018}.  It follows that
\begin{align*}
\lim_{N \to \infty}  \sum_{A \in \mathcal{A}(\delta)} \pi_N (A) \delta_{\nu_{a(A)}} = \sum_{A \in \mathcal{A}(\delta)} \int_0^1 dt \ \mathbbm{1}(\widehat{B}_t \in A) \delta_{\nu_{a(A)}} 
\end{align*}
in law, and by reversing the inequalities, it follows that
\begin{align*}
d_{\operatorname{BL}_1} \left( \sum_{A \in \mathcal{A}(\delta)} \int_0^1 dt \ \mathbbm{1}(\widehat{B}_t \in A) \delta_{\nu_{a(A)}}, \ \int_0^1 dt \ \nu_{\widehat{B}_t} \right) \leq  \delta  
\end{align*}
almost surely in the probability space of $B_t$, and thus 
\begin{align*}
  \lim_{\delta \to 0^+} d_{\operatorname{BL}_1} \left( \sum_{A \in \mathcal{A}(\delta)} \int_0^1 dt \ \mathbbm{1}(\widehat{B}_t \in A) \delta_{\nu_{a(A)}}, \ \int_0^1 dt \ \nu_{\widehat{B}_t} \right) = 0
\end{align*}
almost surely in the probability space of $B_t$. We can now chain together all the inequalities as follows. Let $\delta > 0$ be small but fixed. For any $f \in \operatorname{BL}_1 (\mathcal{M}_1(\mathbb{R}^d)^\mathbb{N})$ and $g \in \operatorname{BL}_1 (\mathbb{R})$, we have
\begin{align*}
\left| \mathbb{E} g \left( \overline{\kappa}_N [f]\right) - \mathbb{E} g \left(  \sum_{A \in \mathcal{A}(\delta)} \pi_N (A) f(\nu_{a(A)})\right)\right| 
&\leq \mathbb{E} d_{\operatorname{BL}_1} \left(\overline{\kappa}_N, \frac{1}{N} \sum_{n=1}^N \delta_{\nu_{\widehat{S}_n}} \right) \\
&+ \mathbb{E} d_{\operatorname{BL}_1} \left( \frac{1}{N} \sum_{n=1}^N \delta_{\nu_{\widehat{S}_n}},  \sum_{A \in \mathcal{A}(\delta)} \pi_N (A) \delta_{\nu_{a(A)}} \right) ,
\end{align*}
and, for the Brownian motion, we have
\begin{align*}
&\left| \mathbb{E} g \left( \int_0^1 dt \ f (\widehat{B}_t)\right) - \mathbb{E} g \left(  \sum_{A \in \mathcal{A}(\delta)} \int_0^1 dt \ \mathbbm{1}(\widehat{B}_t \in A) f(\nu_{a(A)})\right)\right| \\
&\leq   \mathbb{E} d_{\operatorname{BL}_1} \left( \sum_{A \in \mathcal{A}(\delta)} \int_0^1 dt \ \mathbbm{1}(\widehat{B}_t \in A) \delta_{\nu_{a(A)}}, \ \int_0^1 dt \ \nu_{\widehat{B}_t} \right) .
\end{align*}
Combining these two together, we find that
\begin{align*}
&\left| \mathbb{E} g \left( \overline{\kappa}_N [f]\right) - \mathbb{E} g \left( \int_0^1 dt \ f (\widehat{B}_t)\right) \right| \\ &\leq 2 \delta + \left| \mathbb{E} g \left(  \sum_{A \in \mathcal{A}(\delta)} \int_0^1 dt \ \mathbbm{1}(\widehat{B}_t \in A) f(\nu_{a(A)})\right) - \mathbb{E} g \left( \sum_{A \in \mathcal{A}(\delta)} \pi_N (A) f(\nu_{a(A)}) \right) \right| .  
\end{align*}
Using the convergence in law of the second term on the right, it follows that
\begin{align*}
\limsup_{N \to \infty}   \left| \mathbb{E} g \left( \overline{\kappa}_N [f]\right) - \mathbb{E} g \left( \int_0^1 dt \ f (\widehat{B}_t)\right) \right| \leq 2 \delta . 
\end{align*}
Letting $\delta \to 0^+$, the result follows.
\end{proof}
\noindent
Using similar techniques, we have the accompanying chaotic size dependence result.
\begin{theorem} \label{thm:NSclusterpoints}
Suppose that $h$ satisfies (A). 

For dimensions $d \geq 3$, or for dimension $d = 2$ with the additional assumption that $\mathbb{E} |h_j(i)|^3 < \infty$ for all $(i,j) \in \mathbb{N} \times [d]$, it follows that
\begin{align*}
\operatorname{clust} (\overline{\kappa}^h_N) = \left\{ \int_{\mathbb{S}^{d-1}} \eta(d \Omega) \ \delta_{\nu^h_{\Omega}} : \eta \in \mathcal{M}_1 (\mathbb{S}^{d-1}) \right\} 
\end{align*}
almost surely.
\end{theorem}
\begin{proof} First, by using \cref{thm:NS_asymptotic}, we can immediately approximate the Newman-Stein metastates by the probability measures
\begin{align*}
\overline{\kappa}'_N := \frac{1}{N} \sum_{n=1}^N \delta_{\nu_{\widehat{S}_n}}     
\end{align*}
from which we have that
\begin{align*}
 \operatorname{clust} (\overline{\kappa}^h_N) \subset \left\{ \int_{\mathbb{S}^{d-1}} \eta(d \Omega) \ \delta_{\nu^h_{\Omega}} : \eta \in \mathcal{M}_1 (\mathbb{S}^{d-1}) \right\} .
\end{align*}
For the other direction, by separability, there exists a countable dense subset $(\eta_i)$ of $\mathcal{M}_1 (\mathbb{S}^{d-1})$, and it is sufficient to construct convergent subsequences of $(\overline{\kappa}'_N)$ that converge to any $\eta_i$. For any $k \in \mathbb{N}$, it follows that
\begin{align*}
 \lim_{k \to \infty}  d_{\operatorname{BL}_1} \left( \overline{\kappa}'_N,  \sum_{A \in \mathcal{A}(\frac{1}{k})} \pi_N (A) \delta_{\nu_{a(A)}} \right) = 0  ,
\end{align*}
where
\begin{align*}
\pi_N (A) := \frac{1}{N} \sum_{n=1}^N \mathbbm{1}(\widehat{S}_n \in A)   .
\end{align*} 
We also naturally have
\begin{align*}
\lim_{k \to \infty}  d_{\operatorname{BL}_1} \left( \int_{\mathbb{S}^{d-1}} \eta_i (d \Omega) \ \delta_{\nu_{\Omega}},  \sum_{A \in \mathcal{A}(\frac{1}{k})} \eta_i(A) \delta_{\nu_{a(A)}} \right) = 0 .   
\end{align*}
Combining together all approximations, we have
\begin{align*}
\operatorname{d}_{\operatorname{BL}_1} (\overline{\kappa}_N, \eta [\delta_{\nu_\cdot}])   
&\leq \operatorname{d}_{\operatorname{BL}_1} (\overline{\kappa}_N, \overline{\kappa}_N') +  \operatorname{d}_{\operatorname{BL}_1} ( \overline{\kappa}_N', \eta_i [\delta_{\nu_\cdot}]) + \operatorname{d}_{\operatorname{BL}_1} ( \eta[\delta_{\nu_\cdot}], \eta_i [\delta_{\nu_\cdot}]) ,
\end{align*}
followed by
\begin{align*}
\operatorname{d}_{\operatorname{BL}_1} ( \overline{\kappa}_N', \eta_i [\delta_{\nu_\cdot}]) 
&\leq \operatorname{d}_{\operatorname{BL}_1} \left( \overline{\kappa}_N', \sum_{A \in \mathcal{A} (\frac{1}{k})} \pi_N (A) \delta_{\nu_{a(A)}}\right) + \operatorname{d}_{\operatorname{BL}_1} \left( \sum_{A \in \mathcal{A} (\frac{1}{k})} \pi_N (A) \delta_{\nu_{a(A)}}, \sum_{A \in \mathcal{A} (\frac{1}{k})} \eta_i (A) \delta_{\nu_{a(A)}} \right)  \\ &+  \operatorname{d}_{\operatorname{BL}_1} \left( \eta_i [\delta_{\nu_\cdot}], \sum_{A \in \mathcal{A} (\frac{1}{k})} \eta_i (A) \delta_{\nu_{a(A)}}\right) ,
\end{align*}
followed by the final inequality
\begin{align*}
 \operatorname{d}_{\operatorname{BL}_1} \left( \sum_{A \in \mathcal{A} (\frac{1}{k})} \pi_N (A) \delta_{\nu_{a(A)}}, \sum_{A \in \mathcal{A} (\frac{1}{k})} \eta_i (A) \delta_{\nu_{a(A)}} \right) \leq \sum_{A \in \mathcal{A} (\frac{1}{k})} |\pi_N (A) - \eta_i (A)| .
\end{align*}
Giving the appropriate bounds on the inequalities, it follows that
\begin{align*}
\operatorname{d}_{\operatorname{BL}_1} (\overline{\kappa}_N, \eta [\delta_{\nu_\cdot}]) 
&\leq \operatorname{d}_{\operatorname{BL}_1} (\overline{\kappa}_N, \overline{\kappa}_N') + \operatorname{d}_{\operatorname{BL}_1} ( \eta[\delta_{\nu_\cdot}], \eta_i [\delta_{\nu_\cdot}]) + \frac{2}{k} + \sum_{A \in \mathcal{A} (\frac{1}{k})} |\pi_N (A) - \eta_i (A)| .
\end{align*}
For a large but fixed $i$ and $k$, if there exists a random subsequence $(N_j)$ such that 
\begin{align*}
\lim_{j \to \infty}  \sum_{A \in \mathcal{A} (\frac{1}{k})} |\pi_{N_j} (A) - \eta_i (A)| = 0   
\end{align*}
almost surely, then it follows that
\begin{align*}
\limsup_{j \to \infty} \operatorname{d}_{\operatorname{BL}_1} (\overline{\kappa}_{N_j}, \eta [\delta_{\nu_\cdot}]) \leq \operatorname{d}_{\operatorname{BL}_1} ( \eta[\delta_{\nu_\cdot}], \eta_i [\delta_{\nu_\cdot}]) + \frac{2}{k}  ,
\end{align*}
and since the left hand side does not depend on $i$ or $k$, we can take their limits to obtain
\begin{align*}
\lim_{j \to \infty} \operatorname{d}_{\operatorname{BL}_1} (\overline{\kappa}_{N_j}, \eta [\delta_{\nu_\cdot}])   = 0 . 
\end{align*}
From this observation, we see that it remains to prove that for a large but fixed $i$ and $k$, there exists a random subsequence $(N_j)$ such that 
\begin{align*}
\lim_{j \to \infty}  \sum_{A \in \mathcal{A} (\frac{1}{k})} |\pi_{N_j} (A) - \eta_i (A)| = 0     
\end{align*}
almost surely. To this end, We wish to prove that
\begin{align*}
\mathbb{P} \left( (\pi_N (A) ) \in B ((\eta_i (A)), \varepsilon_1)  \text{ infinitely often}\right) = 1     
\end{align*}
for arbitrarily small $\varepsilon_1 > 0$. By the reverse Fatou lemma, it follows that
\begin{align*}
\mathbb{P} \left( (\pi_N (A) ) \in B ((\eta_i (A)), \varepsilon_1)  \text{ infinitely often}\right) &\geq \limsup_{N \to \infty}  \mathbb{P} \left( (\pi_N (A) ) \in B ((\eta_i (A)), \varepsilon_1)  \right) \\ &= \mathbb{P} \left( \left( \int_0^1 dt \ \mathbbm{1}(\widehat{B}_t \in A)\right) \in B ((\eta_i (A)), \varepsilon_1)\right) , 
\end{align*}
where we are using the notation $(\pi_N (A))$ to mean the sequence indexed by $A \in \mathcal{A}(\frac{1}{k})$. Now, since the event 
\begin{align*}
(\pi_N (A) ) \in B ((\eta_i (A)), \varepsilon_1)  \text{ infinitely often}  
\end{align*}
belongs to the exchangeable sigma algebra, it follows that if it has a positive probability, it must necessarily have probability $1$. It is then enough to show that 
\begin{align*}
\mathbb{P} \left( \left( \int_0^1 dt \ \mathbbm{1}(\widehat{B}_t \in A)\right) \in B ((\eta_i (A)), \varepsilon_1)\right)  > 0    
\end{align*}
for any $\varepsilon_1 > 0$ small but fixed. To do this, we will use the so-called forgery theorem or support theorem for Brownian motion which states that
\begin{align*}
\mathbb{P}(\sup_{t \in [0,1]} || B_t - g (t)|| < \varepsilon_2) > 0    
\end{align*}
for any such $g \in C_b ([0,1])$ such that $g(0) = 0$, and $\varepsilon_2 > 0$ is arbitrary. First, for ease of notation, we enumerate the sets of $\mathcal{A}(\frac{1}{k})$ by $A_l$ for $l \in \{1,2,..., L\}$, and we choose small but fixed $\varepsilon_3 > 0$ which can be chosen arbitrarily small as the construction proceeds. We start the construction of the function $g$ as follows, we set $g(0) := 0$. Next, we set 
\begin{align*}
g (t) = a(A_1) \in A_l, \ t \in [\varepsilon_3, \eta_i (A_1) - \varepsilon_3] ,    
\end{align*}
followed by
\begin{align*}
g(t) = a(A_l) \in A_l, \ t \in \left[ \sum_{l' = 1}^{l - 1}\eta_i (A_{l'}) + \varepsilon_3, \sum_{l' = 1}^{l}\eta_i (A_{l'}) - \varepsilon_3 \right] ,  
\end{align*}
where we must choose $\varepsilon_3 > 0$ small enough so that each of the intervals of definition above are disjoint. To complete the construction, we use the Tietze extension theorem to obtain a function $g : [0,1] \to \mathbb{R}$ which is bounded, continuous, satisfies $g(0) = 0$, and $g$ is piecewise constant on the given intervals of the construction. The extension is valid since the pre-constructed $g$ is continuous on each of the disjoint sets given, and the disjoint sets are all closed. Now, we let $\varepsilon_2 > 0$ be arbitrary but small, and, by the forgery theorem, we consider $B_t$ such that
\begin{align*}
\sup_{t \in [0,1]} || B_t - g(t)|| < \varepsilon_2 
\end{align*}
for the constructed $g$. First, note that
\begin{align*}
&\sup_{t \in \left[ \sum_{l' = 1}^{l - 1}\eta_i (A_{l'}) + \varepsilon_3, \sum_{l' = 1}^{l}\eta_i (A_{l'}) - \varepsilon_3 \right]} || B_t - g(t)|| < \varepsilon_2 \\
&\implies \sup_{t \in \left[ \sum_{l' = 1}^{l - 1} \eta_i (A_{l'}) + \varepsilon_3, \sum_{l' = 1}^{l}\eta_i (A_{l'}) - \varepsilon_3 \right]} || \widehat{B}_t - a(A_l)|| < \frac{2 \varepsilon_2}{1 - \varepsilon_2} .
\end{align*}
Recall that the point $a(A_l)$ belongs to the non-empty interior of $A_l$. For small enough $\varepsilon_2$, it follows that $\widehat{B}_t$ belongs to to $A_l$ on the given interval above. For such a $B_t$, we see that
\begin{align*}
\int_0^1 dt \ \mathbbm{1}(\widehat{B}_t \in A_l) \geq \int_{\sum_{l' = 1}^{l - 1}\eta_i (A_{l'}) + \varepsilon_3}^{\sum_{l' = 1}^{l}\eta_i (A_{l'}) - \varepsilon_3} dt \ \mathbbm{1}(\widehat{B}_t \in A_l) = \eta_i (A_l) - 2 \varepsilon_3 .
\end{align*}
For the upper bound, since for small enough $\varepsilon_2$ we know that $\widehat{B}_t$ belongs to $A_{l''}$ on the interval 
\begin{align*}
\left[ \sum_{l' = 1}^{l'' - 1}\eta_i (A_{l'}) + \varepsilon_3, \sum_{l' = 1}^{l''}\eta_i (A_{l'}) - \varepsilon_3 \right] ,
\end{align*}
it follows that
\begin{align*}
\int_0^1 dt \ \mathbbm{1}(\widehat{B}_t \in A_l) &\leq \sum_{l'' = 1}^{L} \int_{\sum_{l' = 1}^{l'' - 1}\eta_i (A_{l'}) + \varepsilon_3}^{\sum_{l' = 1}^{l''}\eta_i (A_{l'}) - \varepsilon_3} dt \ \mathbbm{1}(\widehat{B}_t \in A_l) + 2 \varepsilon_3 L \\
&= \int_{\sum_{l' = 1}^{l - 1}\eta_i (A_{l'}) + \varepsilon_3}^{\sum_{l' = 1}^{l}\eta_i (A_{l'}) - \varepsilon_3} dt \ \mathbbm{1}(\widehat{B}_t \in A_l) + 2 \varepsilon_3 L \\
&= \eta_i (A_l) + 2 \varepsilon_3 (L - 1) .
\end{align*}
It follows that
\begin{align*}
\left|\int_0^1 dt \ \mathbbm{1}(\widehat{B}_t \in A_l) - \eta_i (A_l) \right| < 2 \varepsilon_3 (L - 1) .    
\end{align*}
This holds for any $l \in \{ 1,2,...,L\}$. Now, to complete this step of the proof, we first select $\varepsilon_2 > 0$ small enough so that the various steps of the above inequalities hold, and then we select $\varepsilon_3 > 0$, depending also on $\varepsilon_1$, small enough to prove the following 
\begin{align*}
\left\{ \sup_{t \in [0,1]} || B_t - g (t) || < \varepsilon_2 \right\} \subset \left\{ \left( \int_0^1 dt \ \mathbbm{1}(\widehat{B}_t \in A_l)\right) \in B ((\eta_i(A_l)), \varepsilon_1) \right\}     
\end{align*}
which can be done by the given inequalities. By the forgery theorem, this implies that
\begin{align*}
0 < \mathbb{P} \left( \sup_{t \in [0,1]} || B_t - g (t) || < \varepsilon_2 \right) \leq \mathbb{P} \left( \left( \int_0^1 dt \ \mathbbm{1}(\widehat{B}_t \in A_l)\right) \in B ((\eta_i(A_l)), \varepsilon_1) \right) ,
\end{align*}
which by earlier remarks shows that
\begin{align*}
\mathbb{P} \left( (\pi_N (A) ) \in B ((\eta_i (A)), \varepsilon_1) \text{ infinitely often}\right) = 1 .    
\end{align*}
In particular, we can select $\varepsilon_1 = \frac{1}{p}$ for $p \in \mathbb{N}$ large enough. 

To complete the proof, we now re-index everything in a more transparent manner. First, we fix a diameter $k \in \mathbb{N}$, for each $k$, there exists a finite index set $L_k$ such that the sets $(A_{l_k})_{l_k \in L_k} \in \mathcal{A}(\frac{1}{k})$. By the given proofs, it follows that
\begin{align*}
\mathbb{P} \left( (\pi_N (A_{l_k}) )_{l_k \in L_k} \in B \left((\eta_i (A_{l_k}))_{l_k \in L_k}, \frac{1}{p} \right) \text{ infinitely often}\right) = 1   
\end{align*}
where $k,p \in \mathbb{N}^2$. It follows that
\begin{align*}
\mathbb{P} \left( \bigcap_{(k,p,i) \in \mathbb{N}^3} \left( (\pi_N (A_{l_k}) )_{l_k \in L_k} \in B \left((\eta_i (A_{l_k}))_{l_k \in L_k}, \frac{1}{p} \right) \text{ infinitely often}\right) \right)  = 1 .  
\end{align*}
Now, for the construction, fix $i$ and $k$. Choose $N_p \in \mathbb{N}$ such that $N_{p-1} < N_p$, and 
\begin{align*}
\left| \left| (\pi_{N_p} (A_{l_k}))_k - (\eta_i(A_{l_k}))_k\right| \right| \leq \frac{1}{p} .    
\end{align*}
It follows that
\begin{align*}
\lim_{p \to \infty} (\pi_{N_p} (A_{l_k}))_k  = (\eta_i(A_{l_k}))_k 
\end{align*}
almost surely, and the result follows.
\end{proof}
\section{Scaled random fields, overlaps, and metastates} \label{sec:scaledRandomOverlaps}
\noindent
In this section, we will use the same assumptions for the external random field as in the previous section given in \cref{assump:A}. We modify the Hamiltonian $H_n^h$ by introducing a weaker random field and redefining it as follows
\begin{align*}
H_n^{\frac{h}{\sqrt{n}}} (\phi) := - \frac{1}{2n} \sum_{i,j}^n \left< \phi (i), \phi(j) \right> - \frac{1}{\sqrt{n}} \sum_{i=1}^n \left< h (i), \phi(i) \right> .
\end{align*}
We are primarily interested in the overlap $R^{a,b}_n$ denoted by 
\begin{align*}
R_n^{a,b} := \frac{1}{n} \sum_{i=1}^n \left< \phi^a(i), \phi^b(i)\right> ,    
\end{align*}
and the random probability measure corresponding to the pushforward measure 
\begin{align*}
{R_n^{a,b}}_* (\mu_n^{\frac{h}{\sqrt{n}}} \otimes  \mu_n^{\frac{h}{\sqrt{n}}}) ,
\end{align*}
where $a$ and $b$ are labels to distinguish between which component of the tensor product acts on which variable. In this section, it will be enough to consider $a,b \in \{1,2,3\}$.

The representation of the finite-volume Gibbs states given by
\begin{align*}
\mu_n^{\frac{h}{\sqrt{n}}} [f] = \frac{1}{Z_n (\beta, \frac{h}{\sqrt{n}})} \int_{B_{2d} (0,1)} \frac{dx dy}{(1 - ||x ||^2 - || y ||^2)^{d+1}} e^{n \psi_n^{\beta, {\frac{h}{\sqrt{n}}}} (x,y)} \nu_n^{x,y, h} [f] ,    
\end{align*}
where we now have
\begin{align*}
\psi_n^{\beta, \frac{h}{\sqrt{n}}} (x,y) = \frac{\beta}{2} || x ||^2 + \frac{\beta}{\sqrt{n}} \left< m_n,x \right>  + \frac{\beta}{\sqrt{n}} \left< s_n,y \right> + \frac{d}{2} \ln (1 - ||x||^2 - || y ||^2) , 
\end{align*}
by redefinition. Note that the definition of $\nu_n^{x,y,h}$ is unchanged. In addition, we can also introduce the corresponding mixing probability measure $\alpha_n^{\frac{h}{\sqrt{n}}}$, now given by
\begin{align*}
\alpha_n^{\frac{h}{\sqrt{n}}} (dx, dy) := \frac{1}{Z_n (\beta, \frac{h}{\sqrt{n}})}\frac{dx dy}{(1 - ||x ||^2 - || y ||^2)^{d+1}} e^{n \psi_n^{\beta, {\frac{h}{\sqrt{n}}}} (x,y)}   
\end{align*}
so that
\begin{align*}
\mu_n^{\frac{h}{\sqrt{n}}} = \alpha_n^{\frac{h}{\sqrt{n}}} [\nu_n^{\cdot, \cdot, h}] .    
\end{align*}

To continue, we present the following result concerning the action of $\nu_n^{x,y,h}$ on the overlap.
\begin{lemma} \label{thm:overlapMC} Suppose that $h$ satisfies (A).

It follows that
\begin{align*}
\lim_{n \to \infty} \sup_{(x^a,y^a), (x^b, y^b) \in B_{2d} (0,1)} \sup_{f \in \operatorname{BL}_1 ([-1,1])} \left| (\nu_n^{x^a,y^a,h} \otimes \nu_n^{x^b,y^b,h}) [f (R_n^{a,b})] - f \left(\left< x^a, x^b \right> + \left< y^a, y^b \right> \right) \right| = 0 
\end{align*}
almost surely.
\end{lemma}
\begin{proof} We have
\begin{align*}
R_n^{a,b} = \frac{\left< \phi^a, \phi^b \right>}{n}  
= \frac{1}{n} \sum_{(i,j) \in [n] \times [d]} \left< \phi^a, e_{i,j,n} \right> \left< \phi^b, e_{i,j,n} \right>     .
\end{align*}
Using \cref{def:FVMCM} and the probabilistic representation, it follows that
\begin{align*}
 (\nu_n^{x^a,y^a} \otimes \nu_n^{x^b,y^b}) [f (R_n^{a,b})]  &= \mathbb{E} f \big( \sum_{j=1}^d \left( x_j^a x_j^b + y_j^a y_j^b + \sqrt{1 - || x^a ||^2 - ||y^a ||^2} \sqrt{1 - || x^b ||^2 - ||y^b ||^2}\right) \\ &\times \sum_{i = 3}^n \frac{G_j^a (i) G_j^b (i)}{|| \pi_{([n] \setminus \{1,2\} )\times [d]} (G^a) || \, || \pi_{([n] \setminus \{1,2\} )\times [d]} (G^b) ||} \big)   \\
 &= \mathbb{E} f \big( \left< x^a, x^b \right> + \left< y^a, y^b \right> + \sqrt{1 - || x^a ||^2 - ||y^a ||^2} \sqrt{1 - || x^b ||^2 - ||y^b ||^2} \\ &\times \sum_{j=1}^d \frac{|| \pi_{[n] \times \{ j \}} (G^a)||}{|| \pi_{([n] \setminus \{1,2\} )\times [d]} (G^a) || \, || \pi_{([n] \setminus \{1,2\} )\times [d]} (G^b) ||} G_j^c \big) ,
\end{align*}
where $\{ G^a, G^b \}$ are independent identically distributed standard Gaussian random variables on $\left( \mathbb{R}^d \right)^\mathbb{N}$, and $G^c$ is a standard Gaussian variable on $\mathbb{R}^d$. It follows that
\begin{align*}
&\left| (\nu_n^{x^a,y^a} \otimes \nu_n^{x^b,y^b}) [f (R_n^{a,b})] - f \left(\left< x^a, x^b \right> + \left< y^a, y^b \right> \right) \right| \\ &\leq \mathbb{E} \left| \sum_{j=1}^d \frac{||\pi_{[n] \times \{ j \}} (G^a) ||}{|| \pi_{([n] \setminus \{1,2\} )\times [d]} (G^a) || \, || \pi_{([n] \setminus \{1,2\} )\times [d]} (G^b) ||} G_j^c  \right| \\
&\leq \sqrt{ \left( \mathbb{E} \frac{1}{|| \pi_{([n] \setminus \{1,2\} )\times [d]} (G^a)||^2} \sum_{j=1} || \pi_{[n] \times \{ j \}} (G^a) ||^2  \right)  \left( \mathbb{E} \frac{1}{|| \pi_{([n] \setminus \{1,2\} )\times [d]} (G^b)||^2} \right) \left( \mathbb{E} \sum_{j=1}^d {(G_j^c)}^2 \right)} \\ &= \sqrt{d} \sqrt{\mathbb{E} \frac{1}{|| \pi_{([n] \setminus \{1,2\} )\times [d]} (G^b)||^2}} .
\end{align*}
Using the Laplace method, one can check that
\begin{align*}
\lim_{n \to \infty} n \mathbb{E} \frac{1}{|| \pi_{([n] \setminus \{1,2\} )\times [d]} (G^b) ||^2} < \infty
\end{align*}
exists and is finite, and the result follows. Note that there is trivially no $h$-dependence in the convergence.
\end{proof} 
\subsection{Triviality of the overlap distribution for non-scaled random fields}
\noindent
In this subsection, we will apply the previous result \cref{thm:overlapMC} accompanied by the results in \cref{sec:random}. We have the following triviality result.
\begin{theorem} \label{thm:overlapCSD}
Suppose that $h$ satisfies (A).

For dimension $d = 2$, it follows that
\begin{align*}
\operatorname{clust} ({R_n^{a,b}}_* (\mu_n^h \otimes \mu_n^h)) = \overline{\left\{ \int_{\mathbb{S}^{1}} \gamma^{z} (d \Omega^a) \int_{\mathbb{S}^{1}} \gamma^{z} (d \Omega^b) \ \delta_{({r^*)^2 \left< \Omega^a, \Omega^b \right>} + || y^* ||^2} : z \in P \right\}} 
\end{align*}
almost surely, where $P \subset \mathbb{R}^2$ is the set of recurrent values of the random walk $(S_n)$, $\gamma^z$ is a probability measure on $\mathbb{S}^{d-1}$ given by
\begin{align*}
\gamma^z (d \Omega) := \frac{d \Omega \ e^{\beta r^* \left< \Omega, z \right>}}{\int_{\mathbb{S}^{d-1}} d \Omega \  e^{\beta r^* \left< \Omega, z \right>}} ,    
\end{align*}
and
\begin{align*}
r^* = \sqrt{1 - \frac{d}{ \beta} - \mathbb{E} || h(i)||^2 }, \ (y^*)_j  =    \sqrt{\mathbb{E} h_j(i)^2} .
\end{align*}
For dimension $d \geq 3$, it follows that
\begin{align*}
\operatorname{clust} ({R_n^{a,b}}_* (\mu_n^h \otimes \mu_n^h)) = \{\delta_{({r^*)^2 + || y^* ||^2}}\} = \{ \delta_{1 - \frac{d}{\beta}}\} 
\end{align*}
almost surely.
\end{theorem}
\begin{proof}
By the considerations given in \cref{rmk:bounded_convergence_density}, for dimension $d = 2$, it follows that 
\begin{align*}
\operatorname{clust} (\alpha_n^h) = \overline{\left\{ \left( \int_{\mathbb{S}^1} d \Omega \ e^{ \beta r^* \left< z, \Omega \right>} \right)^{-1} \int_{\mathbb{S}^1} d \Omega \ e^{ \beta r^* \left< z, \Omega \right>} \delta_{r^* \Omega, y^*} : z \in P \} \cup \{ \delta_{r^* \Omega, y^*} : \Omega \in \mathbb{S}^{d-1} \right\}}
\end{align*}
almost surely, where $P \subset \mathbb{R}^d$ is the set of recurrent values of $(S_n)$, and, for dimensions $d \geq 3$, it follows that
\begin{align*}
\operatorname{clust} (\alpha_n^h) = \{ \delta_{r^* \Omega, y^*} : \Omega \in \mathbb{S}^{d-1}\}
\end{align*}
almost surely. If $(\alpha_{n_k}^h)$ is a convergent subsequence with a limit $\alpha^h$, then by the calculation given in \cref{sec:overlap_conergence}, along with the intermediate convergence result \cref{thm:overlapMC}, it follows that
\begin{align*}
\limsup_{k \to \infty} d_{\operatorname{BL}_1} \left( {R_{n_k}^{a,b}}_* (\mu_{n_k}^h \otimes \mu_{n_k}^h),  (\alpha^h \otimes \alpha^h) [\delta_{\left<x^a, x^b \right> + \left<y^a, y^b \right>}] \right) \leq 2 \limsup_{n \to \infty} d_{\operatorname{BL}_1} (\alpha_{n_k}^h, \alpha^h) = 0 .
\end{align*}
The result now follows by adapting the constructions given in the proof of \cref{thm:CSD} for random subsequences.
\end{proof}
\subsection{Overlap distribution for scaled random fields}
\noindent
For the scaled random fields, observe that the contents of \cref{thm:uniconvFVETF}, \cref{thm:MSregime}, and \cref{thm:free_energy} hold also for the scaled random fields by setting limit of the sample standard deviation vector to be vanishing. It follows that
\begin{align*}
\lim_{n \to \infty} \frac{1}{n} Z_n \left(\beta, \frac{h}{\sqrt{n}} \right) = \sup_{(x,y) \in B_{2d} (0,1)} \psi^{\beta,\frac{h}{\sqrt{\cdot}}} (x,y), \ \psi^{\beta,\frac{h}{\sqrt{\cdot}}} (x,y) := \frac{\beta}{2} || x ||^2 + \frac{d}{2} \ln (1 - || x ||^2 - || y ||^2)    
\end{align*}
almost surely, and the set of global maximizing points $M^* \left({\beta,\frac{h}{\sqrt{\cdot}}} \right)$ of the redefined $\psi^{\beta,\frac{h}{\sqrt{\cdot}}}$ is given by
\begin{align*}
M^* \left({\beta,\frac{h}{\sqrt{\cdot}}} \right) = \sqrt{1 - \frac{d}{\beta}} \mathbb{S}^{d-1} \times \{0 \} ,   
\end{align*}
whenever $1 - \frac{d}{\beta} > 0$. Now, when we say that $h$ satisfies (A), we mean that \cref{assump:A} hold as if the limit of the non-scaled random field sample standard deviation vector is vanishing. This is an important distinction because the external field components are still chosen to be non-degenerate so that the actual sample standard deviation vector does not vanish in the limit, but for the purposes of the computing the free energy, finding the global maximizers, and the asymptotics, we treat the model for convenience as if it were vanishing.

We proceed as we would with the non-scaled random field in the special case where we would allow the limit of the standard sample deviation vector be vanishing. We obtain the following representation.
\begin{lemma} \label{thm:overlapApproximation} Suppose that $h$ satisfies (A). 

It follows that 
\begin{align*}
\lim_{n \to \infty}\operatorname{d}_{\operatorname{BL}_1} ({R_n^{a,b}}_* (\mu_n^{\frac{h}{\sqrt{n}}} \otimes \mu_n^{\frac{h}{\sqrt{n}}}), {R_1^{a,b}}_* (r^* \gamma^{\frac{S_n}{\sqrt{n}}} \otimes r^* \gamma^{\frac{S_n}{\sqrt{n}}})) ) = 0 
\end{align*}
almost surely, where $\gamma^z$ is a probability measure on $\mathbb{S}^{d-1}$ given by
\begin{align*}
\gamma^z (d \Omega) := \frac{d \Omega \ e^{\beta r^* \left< \Omega, z \right>}}{\int_{\mathbb{S}^{d-1}} d \Omega \  e^{\beta r^* \left< \Omega, z \right>}}     
\end{align*}
for $z \in \mathbb{R}^d$, the constant $r^*$ is given by
\begin{align*}
r^* = \sqrt{1 - \frac{d}{ \beta}} ,
\end{align*}
and $r^* \gamma^z$ is the probability measure on $r^* \mathbb{S}^{d-1}$ given by its action
\begin{align*}
r^* \gamma^z [g] := \left( \int_{\mathbb{S}^{d-1}} d \Omega \  e^{\beta r^* \left< \Omega, z \right>} \right)^{-1} \int_{\mathbb{S}^{d-1}} d \Omega \  e^{\beta r^* \left< \Omega, z \right>} g (r^* \Omega ) 
\end{align*}
on $g \in C_b (\mathbb{R}^d)$.
\end{lemma}
\begin{proof} Observe that the results proved in \cref{thm:uniconvFVETF}, \cref{thm:MSregime}, \cref{thm:free_energy},  and \cref{thm:mixturerep2} are precisely the same as for the scaled random field if we set the limit of the sample standard deviation vector to vanish, and change the integral with respect to the measure $\nu_n^{x,y}$ with the corresponding integral with respect to the overlap. For the first step, we will first simplify or tide up the form of the mixing probability measures for this application, and then use a concentration result. For this particular case, we have 
\begin{align*}
r^* = \sqrt{1 - \frac{d}{\beta}}, \ y^* = 0 ,    
\end{align*}
and by using this in combination with the overlap convergence calculation in \cref{sec:overlap_conergence}, it follows that
\begin{align*}
\lim_{n \to \infty} d_{\operatorname{BL}_1} \left( {R_n^{a,b}}_* (\mu_n^{\frac{h}{\sqrt{n}}} \otimes \mu_n^{\frac{h}{\sqrt{n}}}), {R_1^{a,b}}_* (r^* \rho_n \otimes r^* \rho_n )\right) = 0 ,
\end{align*}
where
\begin{align*}
 \rho_n (d \Omega) := \frac{d \Omega \ \int_{B_{d+1} (0,1)} dr dy \ \mathbbm{1}(r > 0) e^{\left(n - \frac{2(d+1)}{d} \right) \psi_n^{\beta,\frac{h}{\sqrt{n}}} (r \Omega, y)}}{\int_{\mathbb{S}^{d-1}} d \Omega \ \int_{B_{d+1} ((r^*,0),\delta)} dr dy \ e^{\left(n - \frac{2(d+1)}{d} \right) \psi_n^{\beta,\frac{h}{\sqrt{n}}} (r \Omega, y)}} .
\end{align*}
For the concentration, note that
\begin{align*}
\lim_{n \to \infty} \frac{1}{n} \ln \frac{\int_{\mathbb{S}^{d-1} } d \Omega \int_{B_{d+1} ((r^*,0),\delta)} dr dy \  e^{\left(n - \frac{2(d+1)}{d} \right) \psi_n^{\beta,\frac{h}{\sqrt{n}}} (r \Omega, y)}}{\int_{\mathbb{S}^{d-1} } d \Omega \int_{B_{d+1} (0,1) \setminus B_{d+1} ((r^*,0),\delta)} dr dy \ \mathbbm{1}(r > 0) e^{\left(n - \frac{2(d+1)}{d} \right) \psi_n^{\beta, \frac{h}{\sqrt{n}}} (r \Omega, y)}}  > 0 ,  
\end{align*}
for any $\delta > 0$. Now by applying the concentration inequality given in \cref{eq:concentrationineq}, it follows that
\begin{align*}
\lim_{n \to \infty} d_{\operatorname{BL}_1} \left( {R_1^{a,b}}_* (r^* \rho_n \otimes r^* \rho_n), {R_1^{a,b}}_* (r^* \rho_n^\delta \otimes r^* \rho_n^\delta)\right) = 0 ,    
\end{align*}
where
\begin{align*}
\rho_n^\delta (\Omega) = \frac{\int_{B_{d+1} ((r^*,0),\delta)} dr dy \ e^{n(d)  \psi_n^{\beta, \frac{h}{\sqrt{n}}} (r \Omega, y)}}{Q_n^\delta} ,      
\end{align*}
and
\begin{align*}
n(d) = n - \frac{2(d+1)}{d}, \ Q_n^\delta =   \int_{\mathbb{S}^{d-1}} d \Omega \int_{B_{d+1} ((r^*,0),\delta)} dr dy \ e^{\left(n - \frac{2(d+1)}{d} \right) \psi_n^{\beta, \frac{h}{\sqrt{n}}} (r \Omega, y)}  . 
\end{align*}
In the next step, we will now prove a Laplace-type approximation for the density $\rho_n^\delta$. For the purposes of this proof, denote the uniform limit of $\psi_n$ to be the function $\psi$ on $d+1$ variables given by
\begin{align*}
\psi(r, y) = \frac{\beta}{2} r^2 + \frac{d}{2} \ln (1 - r^2 - ||y|| ^2) .    
\end{align*}
Here we have omitted the dependence of the input variables on the angle of the limit. We rewrite the density $\rho_n$ as follows 
\begin{align*}
\rho_n^\delta (\Omega) = \frac{\int_{B_{d+1} ((r^*,0),\delta)} dr dy \ \mathbbm{1}(r > 0)e^{n(d)  (\psi (r, y) - \psi(r^*,0))} e^{n(d) \beta r \left< \frac{m_n}{\sqrt{n}}, \Omega \right> + n(d) \beta \left< \frac{s_n}{\sqrt{n}}, y \right>} }{\int_{\mathbb{S}^{d-1}}d \Omega \int_{B_{d+1} ((r^*,0),\delta)} dr dy \ \mathbbm{1}(r > 0)e^{n(d)  (\psi (r, y) - \psi(r^*,0))} e^{n(d) \beta r \left< \frac{m_n}{\sqrt{n}}, \Omega \right> + n(d) \beta \left< \frac{s_n}{\sqrt{n}}, y \right>} } .    
\end{align*}
We apply the change of variables $(r,y) \mapsto \left(r^* + \frac{r}{\sqrt{n(d)}}, \frac{y}{\sqrt{n(d)}} \right)$ and cancel like terms in the numerator and denominator to obtain
\begin{align*}
\rho_n^\delta (\Omega) = \frac{\int_{B_{d+1} (0,\delta \sqrt{n(d)})} dr dy \ e^{n(d)  (\psi (r^* + \frac{r}{\sqrt{n(d)}}, y) - \psi(r^*,0))} e^{n(d) \beta \left( r^* + \frac{r}{ \sqrt{n(d)}} \right) \left< \frac{m_n}{\sqrt{n}}, \Omega \right> + n(d) \beta \left< \frac{s_n}{\sqrt{n}}, \frac{y}{\sqrt{n(d)}} \right>} }{\int_{\mathbb{S}^{d-1}}d \Omega \int_{B_{d+1} (0,\delta \sqrt{n(d)})} dr dy \ e^{n(d)  (\psi (r^* + \frac{r}{\sqrt{n(d)}}, y) - \psi(r^*,0))} e^{n(d) \beta \left( r^* + \frac{r}{ \sqrt{n(d)}} \right) \left< \frac{m_n}{\sqrt{n}}, \Omega \right> + n(d) \beta \left< \frac{s_n}{\sqrt{n}}, \frac{y}{\sqrt{n(d)}} \right>}} .   
\end{align*}
To save space, we denote $g_n^\delta (r,y;\Omega)$ to be the mapping given by
\begin{align*}
 g_n^\delta (r,y;\Omega) := \mathbbm{1}((r,y) \in B(0, \delta \sqrt{n(d)}))e^{n(d)  (\psi (r^* + \frac{r}{\sqrt{n(d)}}, y) - \psi(r^*,0))} e^{n(d) \beta \frac{r}{ \sqrt{n(d)}} \left< \frac{m_n}{\sqrt{n}}, \Omega \right> + n(d) \beta \left< \frac{s_n}{\sqrt{n}}, \frac{y}{\sqrt{n(d)}} \right>} .   
\end{align*}
Since the mapping $\psi$ has a negative definite Hessian at $(r^*,0)$, it follows that
\begin{align*}
\lim_{n \to \infty} g_n^\delta (r,y;\Omega) = e^{\left< (r,y), \frac{H[\psi] (r^*,0)}{2} (r,y)\right>} e^{\beta \left< s, y \right>} .    
\end{align*}
In addition, by considering third order derivatives of $\psi$, for large enough $n$, it follows that there exists constants $a,b,c > 0$ such that
\begin{align*}
g_n^\delta (r,y;\Omega)\leq e^{\left< (r,y), (\frac{H[\psi] (r^*,0)}{2} + a I) (r,y)\right>  + b |r| + c ||y||},      
\end{align*}
where $\frac{H[\psi] (r^*,0)}{2} + a I$ is still negative definite. We can apply dominated convergence, and it follows that
\begin{align*}
\lim_{n \to \infty} \int_{\mathbb{R}^{d + 1}} dr dy \ g_n^\delta (r,y;\Omega) = \int_{\mathbb{R}^{d+1}} dr dy \ e^{\left< (r,y), \frac{H[\psi] (r^*,0)}{2} (r,y)\right>} e^{\beta \left< s, y \right>} .
\end{align*}
Next, treating $\Omega$ like a standard vector on $\mathbb{R}^d$ for the purposes of differentiation, we have
\begin{align*}
\left| \left| \frac{d}{d \Omega} \int_{\mathbb{R}^{d + 1}} dr dy \ g_n^\delta (r,y;\Omega) \right| \right| \leq \frac{n(d) || m_n ||}{\sqrt{n} \sqrt{n(d)}} \int_{\mathbb{R}^{d + 1}} dr dy \ |r| g_n^\delta (r,y;\Omega) .
\end{align*}
Also note that
\begin{align*}
g_n^\delta (r,y;\Omega) \leq \mathbbm{1}((r,y) \in B(0, \delta, \sqrt{n(d)}))e^{n(d)  (\psi (r^* + \frac{r}{\sqrt{n(d)}}, y) - \psi(r^*,0))} e^{n(d) \beta \frac{r || m_n||}{ \sqrt{n(d)} \sqrt{n}} + n(d) \beta \left< \frac{s_n}{\sqrt{n}}, \frac{y}{\sqrt{n(d)}} \right>} .
\end{align*}
Since this upper bound is independent of $\Omega$, and we can still use dominated convergence and the bounds given by the third derivatives of $\psi$ in precisely the same way as before, it follows that
\begin{align*}
\lim_{n \to \infty} \sup_{\Omega \in \mathbb{S}^{d-1}} \left| \left| \frac{d}{d \Omega} \int_{\mathbb{R}^{d + 1}} dr dy \ g_n^\delta (r,y;\Omega) \right| \right| = 0 .    
\end{align*}
Denoting $G_n^\delta (\Omega)$ to be
\begin{align*}
G_n^\delta (\Omega) := \int_{\mathbb{R}^{d + 1}} dr dy \ g_n^\delta (r,y;\Omega) .    
\end{align*}
it follows that the sequence of functions $G_n^\delta (\Omega)$ are uniformly bounded on $\mathbb{S}^{d-1}$. One can verify by the same $\Omega$-independent upper-bound of $g_n^\delta(r,y;\Omega)$ that the sequence of functions $G_n (\Omega)$ is also uniformly bounded in $\mathbb{S}^{d-1}$, and thus, by the Arzela-Ascoli theorem, it follows that the pointwise convergence of the integrals can be elevated to uniform convergence
\begin{align*}
\lim_{n \to \infty} \sup_{\Omega \in \mathbb{S}^{d-1}} \left| G_n^\delta (\Omega) - \int_{\mathbb{R}^{d+1}} dr dy \ e^{\left< (r,y), \frac{H[\psi] (r^*,0)}{2} (r,y)\right>} e^{\beta \left< s, y \right>} \right| = 0 .    
\end{align*}
Denote the uniform $\Omega$-independent limit by $C > 0$. Returning to the initial overlap, we have 
\begin{align*}
&{R_1^{a,b}}_* (r^* \rho_n^\delta \otimes r^* \rho_n^\delta) [f] = \int_{\mathbb{S}^{d-1}} \rho_n^{\delta} (d \Omega^a)  \int_{\mathbb{S}^{d-1}} \rho_n^{\delta} (d \Omega^b) \ f \left({(r^*)}^2 \left< \Omega^a, \Omega^b \right> \right) \\
&= \frac{1}{Q_n^\delta} \int_{\mathbb{S}^{d-1}} \gamma^{\frac{S_n}{\sqrt{n}}} (d \Omega^a) \ G_n^\delta (\Omega^a)  \frac{1}{Q_n^\delta}\int_{\mathbb{S}^{d-1}} \gamma^{\frac{S_n}{\sqrt{n}}} (d \Omega^b) \ G_n^\delta (\Omega^b) f \left({(r^*)}^2 \left< \Omega^a, \Omega^b \right> \right) 
\end{align*}
for arbitrary $f \in \operatorname{BL}_1 (\mathbb{R})$, where we relabel
\begin{align*}
Q_n^\delta := \int_{\mathbb{S}^{d-1}} \gamma^{\frac{S_n}{\sqrt{n}}} (d \Omega) \ G^\delta_n (\Omega) .
\end{align*}
Since $G_n^\delta$ converges uniformly to the constant $C$, and the convergence does not depend on the particular $f$, the result follows.
\end{proof}
\noindent
By applying this approximation, and the relevant limit theorems for random walks, we have the following collection of results.
\begin{theorem} \label{thm:overlapCSDScaled}
Suppose that $h$ satisfies (A).

We have
\begin{align*}
\operatorname{clust} ({R_n^{a,b}}_* (\mu_n^{\frac{h}{\sqrt{n}}} \otimes \mu_n^{\frac{h}{\sqrt{n}}})) = \overline{\{ {R_1^{a,b}}_* (r^*\gamma^z \otimes r^*\gamma^z) : z \in \mathbb{R}^d \}}  
\end{align*}
almost surely, where
\begin{align*}
r^* = \sqrt{1 - \frac{d}{\beta}} ,     
\end{align*}
and we denote $r^* \gamma^z$ the probability measure on $r^* \mathbb{S}^{d-1}$ given by its action
\begin{align*}
r^* \gamma^z [g] := \left( \int_{\mathbb{S}^{d-1}} d \Omega \ e^{\beta r^* \left< z, \Omega \right>}\right)^{-1}    \int_{\mathbb{S}^{d-1}} d \Omega \ e^{\beta r^* \left< z, \Omega \right>} f (r^* \Omega) 
\end{align*}
on $g \in C_b (\mathbb{R}^d)$.

We have
\begin{align*}
\lim_{n \to \infty} {R_n^{a,b}}_* (\mu_n^{\frac{h}{\sqrt{n}}} \otimes \mu_n^{\frac{h}{\sqrt{n}}}) =  {R_1^{a,b}}_* (r^* \gamma^{B_1} \otimes r^* \gamma^{B_1})     
\end{align*}
in law, where $B_1$ is a possibly correlated $d$-dimensional Gaussian random variable independent of $h$.
\end{theorem}
\begin{proof}
Using \cref{thm:overlapApproximation}, it follows that
\begin{align*}
\lim_{n \to \infty} d_{\operatorname{BL}_1} ({R_n^{a,b}}_* (\mu_n^{\frac{h}{\sqrt{n}}} \otimes \mu_n^{\frac{h}{\sqrt{n}}} ), {R_1^{a,b}}_* (r^* \gamma^{\frac{S_n}{\sqrt{n}}} \otimes r^* \gamma^{\frac{S_n}{\sqrt{n}}} ))  = 0 .  
\end{align*}
almost surely. Since the set of limit points of $(\frac{S_n}{\sqrt{n}})$ is given by the entirety of $\mathbb{R}^d$, we can obtain any $z \in \mathbb{R}^d$ as a convergent limit of some subsequence $(\frac{S_{n_k}}{\sqrt{n_k}})$, and along this subsequence we have
\begin{align*}
\lim_{k \to \infty} d_{\operatorname{BL}_1} ({R_n^{a,b}}_* (\mu_{n_k}^{\frac{h}{\sqrt{n_k}}} \otimes \mu_{n_k}^{\frac{h}{\sqrt{n_k}}} ), {R_1^{a,b}}_* (r^* \gamma^{z} \otimes r^* \gamma^{z} )) = 0 .
\end{align*}
almost surely. In the other direction, if the subsequence $(\frac{S_{n_k}}{\sqrt{n_k}})$ is bounded, then it contains a convergent subsubsequence and we reapply the result we just used, and if instead there is an unbounded  subsubsequence then, by compactness, the subsubsequence $(\widehat{S}_{n_{k_j}})$ contains a convergent subsubsubsequence with a limit $\Omega \in \mathbb{S}^{d-1}$. Along this subsubsubsequence, using the Laplace method to prove concentration of the probability measure $\gamma^{\frac{S_n}{\sqrt{n}}}$ to the measure $\delta_{\Omega}$, it follows that
\begin{align*}
\lim_{l \to \infty} d_{\operatorname{BL}_1} \left( {R_{n_{k_{j_l}}}^{a,b}}_* (\mu_{n_{k_{j_l}}}^{\frac{h}{\sqrt{{n_{k_{j_l}}}}}} \otimes \mu_{n_{k_{j_l}}}^{\frac{h}{\sqrt{{n_{k_{j_l}}}}}}), \delta_{{r^*}^2} \right)) = 0 
\end{align*}
almost surely. The limit point result follows.

For the convergence in law, we replicate the proof of \cref{thm:AWMS} by using \cref{thm:overlapApproximation}. As before, we elevate convergence in law to almost sure convergence in a new probability space to obtain
\begin{align*}
\lim_{n \to \infty} \left( h, \frac{S_n}{\sqrt{n}} \right) = (h,B_1)    
\end{align*}
almost surely for a possibly uncorrelated $d$-dimensional Gaussian random variable, where we are abusing the notation by referring to the new random variables with the same labels as the old ones. In this new probability space, for $B_1 \not = 0$, it follows that
\begin{align*}
 \lim_{n \to \infty} d_{\operatorname{BL}_1} ({R_n^{a,b}}_* (\mu_n^{\frac{h}{\sqrt{n}}} \otimes \mu_n^{\frac{h}{\sqrt{n}}}), {R_1^{a,b}}_* (r^* \gamma^{B_1} \otimes r^* \gamma^{B_1})) = 0 .  
\end{align*}
almost surely. Converting back to convergence in law the result follows.
\end{proof}
\subsection{Metastates for scaled random fields}
\noindent
In addition, we can directly prove the following results concerning the metastates of the scaled random field model.
\begin{theorem} \label{thm:metastatesScaled}
Suppose that $h$ satisfies (A).

We have
\begin{align*}
\operatorname{clust} (\mu_n^{\frac{h}{\sqrt{n}}}) = \overline{ \left\{ \int_{\mathbb{S}^{d-1}} \gamma^{z} (d \Omega) \nu^0_\Omega : z \in \mathbb{R}^d \right\}}    
\end{align*}
almost surely. 

We have 
\begin{align*}
\lim_{n \to \infty}\left(h,  \mu_n^{\frac{h}{\sqrt{n}}} \right) = \left( h, \int_{\mathbb{S}^{d-1}} \gamma^{B_1} (d \Omega) \nu^0_\Omega \right)
\end{align*}
in law, where $B_1$ is a possibly correlated $d$-dimensional Gaussian random variable independent of $h$.

We have
\begin{align*}
\lim_{N \to \infty} \frac{1}{N} \sum_{n=1}^n \delta_{\mu_n^{\frac{h}{\sqrt{n}}}} = \int_0^1 dt \ \delta_{\int_{\mathbb{S}^{d-1}} \gamma^{\frac{B_t}{\sqrt{t}}} (d \Omega) \nu^0_\Omega}    
\end{align*}
in law, where $B_t$ is a possibly correlated $d$-dimensional Brownian motion independent of $h$.
\end{theorem}
\begin{proof}
The asymptotics derived for the overlap distribution in \cref{thm:overlapApproximation} hold also for the finite-volume Gibbs states in the following form
\begin{align*}
\lim_{n \to \infty} d_{\operatorname{BL}_1} (\mu_n^{\frac{h}{\sqrt{n}}}, \overline{\nu}^0_\frac{S_n}{\sqrt{n}}) =  0 
\end{align*}
almost surely The first result, concerning chaotic size dependence, follows by using this approximation along with the limiting point result $\operatorname{clust} (\frac{S_n}{\sqrt{n}}) = \mathbb{R}^d$ in the exact same way as for the proof of \cref{thm:overlapCSD}.

For the second result, we elevate the convergence in law
\begin{align*}
\lim_{n \to \infty} \left( h, \frac{S_n}{\sqrt{n}}\right) = (h, B_1) 
\end{align*}
for $B_1$ a possibly correlated $d$-dimensional Gaussian independent of $h$, to almost sure convergence in a new probability space using Skorohod's representation theorem. In this new probability space, by an abuse of notation, it follows that
\begin{align*}
\lim_{n \to \infty} d_{\operatorname{BL}_1} (\mu_n^{\frac{h}{\sqrt{n}}}, \overline{\nu}^0_{B_1}) =  0 
\end{align*}
almost surely, and by converting back to the old probability space in distribution, the result follows.

For the final result, we have
\begin{align*}
d_{\operatorname{BL}_1}  \left( \kappa_N^{\frac{h}{\sqrt{\cdot}}} , \frac{1}{N} \sum_{n=1}^N \delta_{\overline{\nu}^0_{\frac{S_n}{\sqrt{n}}}} \right) \leq \frac{1}{N} \sum_{n=1}^N d_{\operatorname{BL}_1} (\mu_n^{\frac{h}{\sqrt{n}}},  \overline{\nu}^0_{\frac{S_n}{\sqrt{n}}})    
\end{align*}
It follows that
\begin{align*}
\lim_{N \to \infty}  d_{\operatorname{BL}_1}  \left( \kappa_N^{\frac{h}{\sqrt{\cdot}}} , \frac{1}{N} \sum_{n=1}^N \delta_{\overline{\nu}^0_{\frac{S_n}{\sqrt{n}}}} \right) = 0   
\end{align*}
almost surely, and by the functional central limit theorem, it follows that
\begin{align*}
\lim_{N \to \infty} \frac{1}{N} \sum_{n=1}^N \delta_{\overline{\nu}^0_{\frac{S_n}{\sqrt{n}}}}  = \int_0^1 dt \  \delta_{\overline{\nu}^0_{\frac{B_t}{\sqrt{t}}}} 
\end{align*}
in law, from which the result follows.
\end{proof}

\appendix

\section{Delta function computations} \label{sec:deltaComputations}
We decompose the Hamiltonian and particle number functions as follows
\begin{align*}
H_n^h (\phi) = \sum_{j=1}^d \left(- \frac{1}{2 n} \left( \sum_{i=1}^n \phi_j(i) \right)^2 - \sum_{i=1}^n h_j(i) \phi_j (i) \right), \ N_n(\phi) = \sum_{j=1}^d \sum_{i=1}^n \phi_j (i)^2 .
\end{align*} 
Observe that
\begin{align*}
\sum_{i=1}^n h_j (i)\phi_j (i) =  \left( \frac{1}{n} \sum_{i'=1}^n h_j (i')  \right) \sum_{i=1}^n \phi_j (i) + \sum_{i=1}^n \left( h_j(i) - \frac{1}{n} \sum_{i'=1}^n h_j (i') \right) \phi_j (i) 
\end{align*}
for any $j$. Now, recall the vectors in $(\mathbb{R}^d)^n$ given by
\begin{align*}
(e^h_{1,j',n})_j (i) := \frac{(\delta_{j'})_j (i)}{\sqrt{n}} , \ (e^h_{2,j',n})_j (i)  := \frac{(\delta_{j'})_j (i)(h_j (i) - (m_n)_j)}{\sqrt{n} (s_n)_j} . 
\end{align*}
Using these vectors, it follows that
\begin{align*}
H_n^h (\phi) = - \frac{1}{2} \sum_{j=1}^d \left< \phi, e^h_{1,j,n} \right>^2 - \sum_{j=1}^d (m_n)_j \sqrt{n} \left< \phi, e^h_{1,j,n} \right> - \sum_{j=1}^d (s_n)_j \sqrt{n} \left< \phi, e^h_{2,j,n}\right> ,
\end{align*}
and
\begin{align*}
N_n (\phi) = \sum_{j=1}^d \left< \phi, e^h_{1,j,n} \right>^2 + \sum_{j=1}^d \left< \phi, e^h_{2,j,n}\right>^2 + \sum_{3 \leq i \leq n, j \in [d]} \left< \phi, e^h_{i,j,n}\right>^2 .
\end{align*}
One can show that the following identity holds
\begin{align*}
\frac{n^{\frac{nd-2}{2}}}{2} \int_{\mathbb{S}^{nd - 1}} d \Omega \  e^{- \beta H_n^h (\sqrt{n} \Omega)} f (\sqrt{n} \Omega) = \lim_{\varepsilon \to 0} \frac{1}{\sqrt{2 \pi \varepsilon^2}} \int_{ \left( \mathbb{R}^d \right)^n} d \phi \ e^{- \frac{1}{2 \varepsilon^2} \left( N_n (\phi) - n \right)^2} e^{- \beta H_n^h (\phi)} f (\phi) 
\end{align*}
Using this identity, and the change of variables corresponding to changing coordinates to the basis represented by $\{e^h_{i,j,n}\}_{(i,j) \in [n] \times [d]}$, it follows that 
\begin{align*}
&\frac{1}{\sqrt{2 \pi \varepsilon^2}} \int_{ \left( \mathbb{R}^d \right)^n} d \phi \ e^{- \frac{1}{2 \varepsilon^2} \left( N_n (\phi) - n \right)^2} e^{- \beta H_n^h(\phi)} f (\phi) \\ &= \frac{1}{\sqrt{2 \pi \varepsilon^2}} \int_{ \mathbb{R}^d \times \mathbb{R}^d \times \left( \mathbb{R}^d \right)^{n-2}} dx dy d \phi' \ e^{- \frac{1}{2 \varepsilon^2} \left( || x ||^2 + || y ||^2 + || \phi' ||^2 - n \right)^2} e^{\frac{\beta}{2} || x||^2 + \beta \sqrt{n} \left< m_n, x \right> + \beta \sqrt{n} \left< s_n, y \right>} \\
&\times f \left( \sum_{j=1}^d (x_j e^h_{1,j,n} + y_j e^h_{2,j,n}) + \sum_{3 \leq i \leq n, j \in [d]} \phi'_j (i) e^h_{i,j,n} \right) .
\end{align*}
Changing the order of integration, we have
\begin{align*}
&\frac{1}{\sqrt{2 \pi \varepsilon^2}} \int_{ \left( \mathbb{R}^d \right)^n} d \phi \ e^{- \frac{1}{2 \varepsilon^2} \left( N_n (\phi) - n \right)^2} e^{- \beta H_n^h (\phi)} f (\phi) \\ &= \int_{ \mathbb{R}^d \times \mathbb{R}^d} dx dy \ e^{\frac{\beta}{2} || x||^2 + \beta \sqrt{n} \left< m_n, x \right> + \beta \sqrt{n} \left< s_n, y \right>} \  \frac{1}{\sqrt{2 \pi \varepsilon^2}} \int_{ \left( \mathbb{R}^d \right)^{n-2}} d \phi' \ e^{- \frac{1}{2 \varepsilon^2} \left(|| \phi' ||^2 - (n - || x||^2 - || y ||^2) \right)^2}  \\
&\times f \left( \sum_{j=1}^d (x_j e^h_{1,j,n} + y_j e^h_{2,j,n}) + \sum_{3 \leq i \leq n, j \in [d]} \phi'_j (i) e^h_{i,j,n} \right)  .
\end{align*}
Changing to hyperspherical coordinates, and taking the limit on both sides as $\varepsilon \to 0$, it follows that
\begin{align*}
&\frac{n^{\frac{nd-2}{2}}}{2} \int_{\mathbb{S}^{nd - 1}} d \Omega \  e^{- \beta H_n^h (\sqrt{n} \Omega)} f (\sqrt{n} \Omega) \\ &= \int_{\mathbb{R}^d \times \mathbb{R}^d} dx dy \ \mathbbm{1}(|| x||^2 + || y ||^2 < n) e^{\frac{\beta}{2} || x||^2 + \beta \sqrt{n} \left< m_n, x \right> + \beta \sqrt{n} \left< s_n, y \right>} \\ &\times \frac{(n - || x ||^2 - || y ||^2)^{\frac{(n-2)d - 2}{2}}}{2} \\ &\int_{\mathbb{S}^{(n-2)d - 1}} d \Omega \ f \left( \sum_{j=1}^d (x_j e^h_{1,j,n} + y_j e^h_{2,j,n}) + \sqrt{n - || x ||^2 - || y ||^2}\sum_{3 \leq i \leq n, j \in [d]} \Omega_j (i) e^h_{i,j,n} \right) .
\end{align*}
We rescale the variables $(x,y) \mapsto \sqrt{n} (x,y)$ to obtain
\begin{align*}
&\frac{n^{\frac{nd-2}{2}}}{2} \int_{\mathbb{S}^{nd - 1}} d \Omega \  e^{- \beta H_n^{h} [\sqrt{n} \Omega]} f (\sqrt{n} \Omega) \\ &= \frac{n^{\frac{nd - 2}{2}}}{2}\int_{\mathbb{R}^d \times \mathbb{R}^d} dx dy \ \mathbbm{1}(|| x||^2 + || y ||^2 < 1) e^{n \left( \frac{\beta}{2} || x||^2 + \beta \left< m_n, x \right> + \beta  \left< s_n, y \right> \right)} \\ &\times (1 - || x ||^2 - || y ||^2)^{\frac{(n-2)d - 2}{2}} \\ &\int_{\mathbb{S}^{(n-2)d - 1}} d \Omega \ f \left( \sum_{j=1}^d (x_j \sqrt{n} e^h_{1,j,n} + y_j \sqrt{n} e^h_{2,j,n}) + \sqrt{n} \sqrt{1 - || x ||^2 - || y ||^2}\sum_{3 \leq i \leq n, j \in [d]} \Omega_j (i) e^h_{i,j,n} \right) .
\end{align*}
Ultimately, we have 
\begin{align*}
&\int_{ \left( \mathbb{R}^d \right)^n} d \phi \ \delta (N_n (\phi) - n) e^{- \beta H_n^h (\phi)}  f (\phi) \\ &= \frac{n^{\frac{nd-2}{2}}}{2} \int_{\mathbb{S}^{nd - 1}} d \Omega \  e^{- \beta H_n^h (\sqrt{n} \Omega)} f (\sqrt{n} \Omega) \\ &= \frac{n^{\frac{nd-2}{2}} |\mathbb{S}^{(n-2)d - 1}|}{2} \int_{B_{2d} (0,1)} \frac{dx dy}{(1 - || x ||^2 - || y ||^2)^{d+1}}\  e^{n \psi_n^{\beta,h} (x,y)} \nu_n^{x,y,h} [f] .
\end{align*}

\section{Coordinate transformations for tilting functions} \label{sec:coordinateTransforms}
We construct a number of orthogonal transformations which will simplify and clarify the form of the finite-volume tilting functions. In the following, we will construct orthogonal transformations $O_n^h$ and $U_n^h$ which utilize the unit vectors $\frac{m_n}{|| m_n||}$ and $\frac{s_n}{|| s_n ||}$ respectively. The same exact construction will also hold if we replace these unit vectors by their limits $\frac{m}{|| m||}$ and $\frac{s}{|| s ||}$, and in this case we will refer to those orthogonal transformations as $O^h$ and $U^h$ respectively. If $m = 0$, then one should only use hyperspherical coordinates for the construction of $O^h$ as follows.

We begin by changing coordinates on the sphere appropriately. Let $\left\{ \frac{m_n}{|| m_n||}, e_2,..., e_d \right\} \subset \mathbb{R}^d$ be an orthonormal basis, with the convention that $e_1 := \frac{m_n}{|| m_n||}$. Denote $O_n^h : \mathbb{R}^d \to \mathbb{R}^d$ to be the orthogonal change of coordinates given by
\begin{align*}
(O_n^h (x))_j = \left< x, e_j \right> .
\end{align*}
For the spherical element $\Omega$, let us fix the notation and coordinates of the sphere by setting
\begin{align*}
\Omega_1 (\theta, \varphi_2,..., \varphi_{d-1}) &= \cos (\theta) \\
\Omega_j (\theta, \varphi_2,..., \varphi_{d-1}) &= \sin (\theta) \cos (\varphi_2) \cos(\varphi_3)... \cos (\varphi_{j-1}) ,
\end{align*}
where $\theta, \varphi_{2},..., \varphi_{d-2} \in [0, \pi]$, and $\varphi_{d-1} \in [0, 2 \pi)$.
The tilting function can be written as follows
\begin{align*}
\psi_n^{\beta,h} (x,y) = \frac{\beta}{ 2} || O_n^h (x)||^2 + \beta || m_n || (O_n^h(x))_1 + \beta \left< s_n, y \right> + \frac{d}{2} \ln (1 - || x ||^2 - || y ||^2) .
\end{align*}
In primed coordinates $x' := x' (x) = O_n (x)$, it follows that
\begin{align*}
\psi_n^{\beta} ({(O_n^h)}^{-1} x', y) = \frac{\beta}{2} || x' ||^2 + \beta || m_n || x'_1 + \beta \left< s_n, y \right> + \frac{d}{2} \ln (1 - || x' ||^2 - || y ||^2) .
\end{align*}
In hyperspherical coordinates $x' = r' \Omega'$, it follows that
\begin{align*}
\psi_n^{\beta,h} (r' {(O_n^h)}^{-1}  \Omega' (\theta', \varphi_2',..., \varphi_{d-1}'), y) = \frac{\beta}{2} {r'}^2 + \beta || m_n || r' \cos (\theta') + \beta \left< s_n, y \right> + \frac{d}{2} \ln (1 - {r'}^2 - || y ||^2) .
\end{align*}
For the variable $y$, let $ \left\{ \frac{s_n}{|| s_n ||}, e_2',...,e_d' \right\} \subset \mathbb{R}^d$ be an orthonormal basis with the convention that $e_1' := \frac{s_n}{|| s_n ||}$. Denote $U_n^h : \mathbb{R}^d \to \mathbb{R}^d$ to be the orthogonal change of coordinates given by
\begin{align*}
(U_n^h(y))_j := \left< y, e_j' \right> .
\end{align*}
If we denote $y' := y'(y) = U_n^h (y)$, then it follows that
\begin{align*}
&\psi_n^{\beta,h} (r'{(O_n^h)}^{-1}  \Omega' (\theta', \varphi_2',..., \varphi_{d-1}'),{(U_n^h)}^{-1}  (y')) \\ &= \frac{\beta}{2} {r'}^2 + \beta || m_n || r' \cos (\theta') + \beta || s_n || y'_1 + \frac{d}{2} \ln \left(1 - {r'}^2 - {y_1'}^2 - \sum_{j=2}^d {y_j'}^2 \right) .
\end{align*}
Let us define another tilting function $\Psi_n^{\beta,h} $ in accordance with this change of variables to be
\begin{align*}
\Psi_n^{\beta,h} (r', \theta', y'_1) := \frac{\beta}{2} {r'}^2 + \beta || m_n || r' \cos (\theta') + \beta || s_n || y'_1 + \frac{d}{2} \ln (1 - {r'}^2 - {y'_1}^2) ,
\end{align*}
and one should note that
\begin{align*}
\psi_n^{\beta,h} (r' {(O_n^h)}^{-1}  \Omega'( \theta', \varphi_2',..., \varphi_{d-1}'), {(U_n^h)}^{-1}  (y_1',0,...,0) ) = \Psi_n^{\beta,h} (r', \theta', y_1') .
\end{align*}
One should also note that
\begin{align*}
{(O_n^h)}^{-1}  \Omega' &= \sum_{j=1}^d \left< {(O_n^h)}^{-1}  \Omega', e_j \right> e_j  \\ &= \sum_{j=1}^d (\Omega')_j e_j \\
 &= \cos (\theta') \frac{m_n}{|| m_n ||} + \sum_{j=2}^{d} \sin (\theta') \cos (\varphi_2') ... \cos (\varphi_{j-1}')  e_j .
\end{align*}
In particular, we see that
\begin{align*}
(O_n^h)^{-1} \Omega' (0, \varphi_2',..., \varphi_{d-1}') = \frac{m_n}{|| m_n ||} .
\end{align*}
From here on out, we will omit the apostrophe from the notation for the angles, and simply refer to $\theta, \varphi_2,..., \varphi_{d-1}$, in addition, we will refer to the $y_1'$ as $y$ simply, and it should be understood from context that is  now a real-valued variable. 
\section{Local-to-global inequality} \label{sec:localToGlobalInequality}
We present a useful local-to-global inequality. First, we have the following useful property for any  $f \in \operatorname{BL}_1 ((\mathbb{R}^d)^\mathbb{N})$
\begin{align*}
|f(\phi) - f (\pi_{[k] \times [d]} (\phi), \pi_{(\mathbb{N} \setminus [k]) \times [d]} (0))| \leq \frac{1}{C(d)} \sum_{(i,j) \in (\mathbb{N} \setminus [k]) \times [d]} \frac{1}{2^{i + j}} .
\end{align*}
Given any two probability measures $\mu,\nu$ on $(\mathbb{R}^d)^\mathbb{N}$, it follows that
\begin{align*}
\left| \mu[f] - \nu [f] \right| &\leq |\mu [ f (\pi_{[k] \times [d]}, \pi_{(\mathbb{N} \setminus [k]) \times [d]} (0))] - \nu [f (\pi_{[k] \times [d]}, \pi_{(\mathbb{N} \setminus [k]) \times [d]} (0))]|  + \frac{2}{C(d)} \sum_{(i,j) \in (\mathbb{N} \setminus [k]) \times [d]} \frac{1}{2^{i + j}} \\
&\leq \frac{1}{C(d)} \sum_{(i,j) \in [k] \times [d]} \frac{\Gamma_k [|\phi_j (i) - \phi'_j (i)|]}{2^{i + j}}  + \frac{2}{C(d)} \sum_{(i,j) \in (\mathbb{N} \setminus [k]) \times [d]} \frac{1}{2^{i + j}} ,
\end{align*}
where $ \Gamma_k$ is a coupling of $\mu|_{[k] \times [d]}$ and $\nu |_{[k] \times [d]}$, 
and we have used the $1$-Lipschitz property of $f$ to obtain the first term 
in the following way. To be explicit, a coupling $\Gamma_k$ is a probability measure on $(\mathbb{R}^d)^k \times (\mathbb{R}^d)^k$ such that the marginal distribution of the first component is given by $\mu|_{[k] \times [d]}$ and the marginal distribution of the second component is given by $\nu|_{[k] \times [d]}$. Since the above construction does not explicitly depend on the chosen function, it follows that
\begin{align} \label{thm:couplingBound}
\sup_{f \in \operatorname{BL}_1 ((\mathbb{R}^d)^\mathbb{N})} \left| \mu[f] - \nu [f] \right| \leq  \frac{1}{C(d)} \sum_{(i,j) \in [k] \times [d]} \frac{\Gamma_k [|\phi_j (i) - \phi'_j (i)|]}{2^{i + j}}  + \frac{2}{C(d)} \sum_{(i,j) \in (\mathbb{N} \setminus [k]) \times [d]} \frac{1}{2^{i + j}}  ,   
\end{align}
where the index $k$ and the coupling can be freely chosen.
\section{Concentration inequality} \label{sec:analyticConcentration}
We make a brief remark on a generic form of a concentration inequality. Observe that
\begin{align} \label{eq:concentrationineq}
\left| \mathbb{E} f (X) - \frac{1}{\mathbb{P}(X \in A)}\mathbb{E} \mathbbm{1}(X \in A) f (X) \right| \leq \frac{2 || f ||_\infty}{1 + \frac{\mathbb{P}(X \in A)}{\mathbb{P}(X \not \in A)}} ,
\end{align}
where $f$ is a bounded measurable function, in an appropriate sense. To obtain a further upper bound on this quantity, we require a lower bound for $\mathbb{P}(X \in A)$ and an upper bound for $\mathbb{P}(X \not \in A)$. Whenever we decompose a probability measure in this way, we are assuming that conditioning on the set $A$ yields some benefit such as improved control over $f$ in whatever context the integration appears, and the remark on the bounding essentially states that to show smallness of the right hand side of the inequality, it is sufficient to give bounds of the type stated. 
\section{Tightness and convergence of random probability measures} \label{sec:tightnessConvergenceRandomMeasures}
In this appendix, we briefly describe the conditions under which tightness of random probability measures holds and in what contexts. For a thorough treatment of random measure theory, see \cite{Kallenberg2017}, and for a more detailed account with results and proofs on the particulars of what are discussed here, see \cite[Appendix]{Koskinen2023}. 

For this appendix $X$ will be a generic Polish space $\mathcal{M} := \mathcal{M}_1 (X)$ is the Polish space of probability measures on $X$ metrized by the (Levy)-Prokhorov metric. We say that a sequence of random probability measures $(\mu_n)$ converges almost surely to another random probability measure $\mu$ if 
\begin{align*}
\lim_{n \to \infty} d_{\operatorname{LP}}(\mu_n, \mu) = 0    
\end{align*}
almost surely, or, equivalently, if 
\begin{align*}
\lim_{n \to \infty} \mu_n[f] = \mu[f]    
\end{align*}
for any $f \in C_b (X)$ almost surely. Note that $f \in C_b (X)$ can equivalently be replaced by a number of smaller classes of functions than $C_b(X)$, and in this article we will typically use $f \in \operatorname{BL}_1 (X)$.

We say that a sequence of random probability measures $(\mu_n)$ converges in law to another random probability measure $\mu$ if 
\begin{align*}
\lim_{n \to \infty} \mathbb{E} f (\mu_n) = \mathbb{E} f (\mu)    
\end{align*}
for any $f \in C_b (\mathcal{M})$, or $f \in \operatorname{BL}_1 (\mathcal{M})$. We should also mention the following specific property of convergence in law of random probability measures, namely, if the sequence of laws $(\mathcal{L} \mu_n)$ of $(\mu_n)$ is tight, then convergence in law is equivalent to the convergence in law of the sequence of real-valued random variables $(\mu_n[f])$ to $\mu[f]$ for any $f \in \operatorname{BL}_1 (X)$.

For the empirical measures $(\kappa_N)$ given by
\begin{align*}
\kappa_N := \frac{1}{N} \sum_{n=1}^N \delta_{\mu_n} ,    
\end{align*}
by virtue of being random probability measures themselves, almost sure convergence to another random probability measure $\kappa$ is equivalent to stating that
\begin{align*}
\lim_{N \to \infty} \kappa_N [f] := \lim_{N \to \infty}  \frac{1}{N} \sum_{n=1}^N f (\mu_n) = \kappa[f]   
\end{align*}
for any $f \in \operatorname{BL}_1 (\mathcal{M})$ almost surely. By the previous characterization, given the tightness of $(\mathcal{L} \kappa_N)$, convergence in law is equivalent to convergence in law of $(\kappa_N[f])$ for any $f \in \operatorname{BL}_1 (X)$.

For the tightness property, we introduce the intensity measures $\mathbb{E} \mu_n$ which are defined by their functionals $f \mapsto \mathbb{E} \mu_n [f]$ for $f \in C_b (X)$. If the sequence of intensity measures $(\mathbb{E} \mu_n)$ is tight, it follows that the sequence of laws $(\mathcal{L} \mu_n)$ is tight. We differentiate between almost sure tightness of $(\mu_n)$ and tightness of $(\mathcal{L} \mu_n)$ by always stating that the second one is with respect to the laws of the random probability measures.

For the empirical measures, if $(\mu_n)$ is tight almost surely, then it follows that $(\kappa_N)$ is tight almost surely. If $(\mathbb{E} \mu_n)$ is tight then it follows that $(\mathcal{L} \kappa_N)$ is tight.

From these statements, we see that if $(\mu_n)$ is tight almost surely, then so is $(\kappa_N)$, and if $(\mathbb{E} \mu_n)$ is tight, then so are $(\mathcal{L} \mu_n)$ and $(\mathcal{L} \kappa_N)$.
\section{Disjoint partitions of the unit sphere} \label{sec:disjointPartitions}
We have the following construction of a particular type of partition of the unit sphere.
\begin{lemma} \label{thm:disjointPartitionsUnitSphere}
Let $\delta > 0$ be small but fixed.

There exists a measurable finite disjoint partition $\mathcal{A}(\delta)$ of $\mathbb{S}^{d-1}$ such that for any $A \in \mathcal{A}(\delta)$ the following properties hold:
\begin{enumerate}
\item  We have
\begin{align*}
\operatorname{int} (A) \not = \emptyset  ,  
\end{align*}
where $\operatorname{int} (\cdot)$ is the interior of the set with respect to $\mathbb{S}^{d-1}$.
\item We have
\begin{align*}
\operatorname{diam} (A) \leq \delta    , 
\end{align*}
where $\operatorname{diam}$ is the diameter of the set $A$ considered either in the standard Euclidean metric on $\mathbb{R}^d$ or $\mathbb{S}^{d-1}$.
\item We have
\begin{align*}
\sigma_{d-1}( \partial A) = 0 ,    
\end{align*}
where $\sigma_{d-1}$ is the surface measure on $\mathbb{S}^{d-1}$, and $\partial (\cdot)$ is the boundary with respect to $\mathbb{S}^{d-1}$. 
\end{enumerate}
\end{lemma}
\begin{proof}
We consider the so called recursive zonal equal area partitions $\text{EQ}(d - 1, N)$ of $\mathbb{S}^{d-1}$ into $N$ equal area sets, and the set of recursive zonal equal area partitions $\text{EQ}(d-1) := \{\text{EQ}(d-1,N) : N \in \mathbb{N} \}$ presented in \cite[Definition 1.3, Definition 1.4]{Leopardi2006}. We list out the key properties of these partitions. First, each partition $P := \text{EQ}(d - 1, N)$ consists of sets closed sets $R$ such that
\begin{align*}
\bigcup_{R \in P} R = \mathbb{S}^{d-1} ,    
\end{align*}
the number of sets in $P$ is given by $|P| = N$,  the surface measure of each set satisfies
\begin{align*}
\sigma_{d-1} (R) = \frac{\sigma_{d-1} (\mathbb{S}^{d-1})}{N} ,    
\end{align*}
and the surface measure of the boundary of each set satisfies
\begin{align*}
\sigma_{d-1} (\partial R) = 0 .    
\end{align*}
As remarked in \cite{Leopardi2006}, the last property implies that the intersection $R_1 \cap R_2$ is either empty or non-empty but contained in the set $\partial R_1 \cap \partial R_2$ which is of $0$ measure. We note also that
\begin{align*}
R = \operatorname{int}(R) \cup \partial R,    
\end{align*}
and since the sets on the right hand side are disjoint, it follows that
\begin{align*}
\sigma_{d-1} (\operatorname{int} (R)) = \sigma_{d-1} (R) > 0 ,    
\end{align*}
which implies that they also have non-empty interiors. The key theorem is given by \cite[Theorem 1.6]{Leopardi2006}, which states that there exists $K > 0$, for all partitions in $\text{EQ} (d - 1)$, such that 
\begin{align*}
\operatorname{diam} (R) \leq \frac{K}{|P|^{\frac{1}{d - 1}}}    
\end{align*}
for any $R \in P$.

For our construction, since $K > 0$ is fixed, we choose $N \in \mathbb{N}$, and hence $P := \text{EQ}(d - 1,N)$, large enough so that
\begin{align*}
\frac{K}{N^{\frac{1}{d - 1}}} \leq \delta .    
\end{align*}
For the given $P$, let us enumerate the sets $R \in P$, by $(R_i)_{i=1}^N$. We set 
\begin{align*}
A_1 := R_1, \ A_i = R_i \setminus (\cup_{j=1}^{i-1} R_j) .    
\end{align*}
By construction, the sets $(A_i)$ form a measurable finite disjoint partition of $\mathbb{S}^{d-1}$. To confirm the properties of the sets, since $A_i \supset R_i \setminus \partial R_i$, it follows that
\begin{align*}
 \operatorname{int}(A_i) &= \operatorname{int}(R_i) \cap \operatorname{int}((\partial R_i)^c) \\ &= \operatorname{int}(R_i) \cap \operatorname{int}(\operatorname{int}(R_i) \cup R_i^c) \\&\supset \operatorname{int}(R_i) \cap \left(  \operatorname{int}(R_i) \cup R_i^c \right) \\&= \operatorname{int}(R_i) \not = \emptyset .
\end{align*}
Next, we have
\begin{align*}
\operatorname{diam}(A_i) \leq \operatorname{diam}(R_i) \leq \frac{K}{N^{\frac{1}{d-1}}} \leq \delta,    
\end{align*}
and, finally, we have
\begin{align*}
\sigma_{d-1} (\partial A_i) \leq \sigma_{d-1} (\partial R_i) = 0 .    
\end{align*}
The sets $(A_i)$ constructed in this way satisfy the conditions given in the result.
\end{proof}
\noindent
We have the accompanying approximation result for probability measures on the unit sphere.
\begin{lemma} \label{thm:densityConstruction}There exists a countable collection $(\eta_i)$ of probability measures on $\mathbb{S}^{d-1}$
\begin{align*}
\mathcal{M}_1 (\mathbb{S}^{d-1}) := \overline{\left\{ \sum_{A \in \mathcal{A}(\frac{1}{j})} \eta_i (A) \delta_{a(A)} : i \in \mathbb{N}, \ k \in \mathbb{N} \right\}    } ,
\end{align*}
where the element $a(A)$ in the sum is a(the) fixed element of $A \in \mathcal{A}(\frac{1}{k})$.
\end{lemma}
\begin{proof}
Let $\eta$ be an arbitrary probability measure, and let $(\eta_i)$ be a countable dense subset of the probability measures on the unit sphere. That the right hand side of the equality in the result is a subset of the left is immediate.

For the other direction, for any $\varepsilon > 0$ there exists $i$ large enough such that
\begin{align*}
d_{\operatorname{BL}_1}(\eta, \eta_i) < \frac{1}{2} \varepsilon .   
\end{align*}
Select $k \in \mathbb{N}$ such that $\frac{1}{k} < \frac{1}{2} \varepsilon$. For any $f \in \operatorname{BL}_1 (\mathbb{S}^{d-1})$, it follows that
\begin{align*}
\left| \eta_i[f] - \sum_{A \in \mathcal{A} (\frac{1}{k})} \eta_i (A) f(a(A)) \right| \leq \sum_{A \in \mathcal{A} (\delta)} \eta_i (A) \left| \frac{\eta_i [\mathbbm{1}(\cdot \in A) (f - f (a(A)))]}{\eta_i (A)}\right| \leq \frac{1}{k} < \frac{1}{2} \varepsilon .
\end{align*}
Combining the two together, it follows that for any $\varepsilon > 0$ there exists $(i,k) \in \mathbb{N} \times \mathbb{N}$ such that
\begin{align*}
d_{\operatorname{BL}_1} \left(\eta, \sum_{A \in \mathcal{A} (\frac{1}{k})} \eta_i (A) \delta_{a(A)}  \right) < \varepsilon ,
\end{align*}
and the result follows.
\end{proof}
\section{Overlap convergence} \label{sec:overlap_conergence}
Recall the following representation of the finite-volume Gibbs states given in \cref{def:mixture_representation_2}
\begin{align*}
\mu_n^h = \alpha_n^h [\nu_n^{\cdot, \cdot, h}] ,    
\end{align*}
where $\alpha_n^h$ is the mixing probability measure with the $\beta$ dependence omitted. In the tensor product, it follows that
\begin{align*}
{R_n^{a,b}}_* (\mu_n^h \otimes \mu_n^h) = (\alpha_n^h \otimes \alpha_n^h) [{R_n^{a,b}}_* (\nu_n^{x^a,y^a,h} \otimes \nu_n^{x^b,y^b, h})] ,    
\end{align*}
from which we see that
\begin{align*}
&d_{\operatorname{BL}_1} \left( {R_n^{a,b}}_* (\mu_n^h \otimes \mu_n^h), (\alpha_n^h \otimes \alpha_n^h) [\delta_{\left<x^a, x^b \right> + \left<y^a, y^b \right>}]\right)  \\ &\leq 2 \sup_{(x^a,y^a), (x^b, y^b) \in B_{2d}(0,1)} d_{\operatorname{BL}_1} ({R_n^{a,b}}_* (\nu_n^{x^a,y^a,h} \otimes \nu_n^{x^b,y^b, h}), \delta_{\left<x^a, x^b \right> + \left<y^a, y^b \right>}) .
\end{align*}
In addition, we observe that
\begin{align*}
d_{\operatorname{BL}_1} \left( (\alpha_n^h \otimes \alpha_n^h) [\delta_{\left<x^a, x^b \right> + \left<y^a, y^b \right>}], (\alpha^h \otimes \alpha^h) [\delta_{\left<x^a, x^b \right> + \left<y^a, y^b \right>}]\right) \leq 2 d_{\operatorname{BL}_1} (\alpha_n^h, \alpha^h)
\end{align*}
for any probability measure $\alpha^h$ on $B_{2d}(0,1)$, which could possibly also include an omitted $n$-dependence. In totality, it follows that
\begin{align*}
&d_{\operatorname{BL}_1} \left( {R_n^{a,b}}_* (\mu_n^h \otimes \mu_n^h),  (\alpha^h \otimes \alpha^h) [\delta_{\left<x^a, x^b \right> + \left<y^a, y^b \right>}] \right) \\
&\leq 2 d_{\operatorname{BL}_1} (\alpha_n^h, \alpha^h) + 2 \sup_{(x^a,y^a), (x^b, y^b) \in B_{2d}(0,1)} d_{\operatorname{BL}_1} ({R_n^{a,b}}_* (\nu_n^{x^a,y^a,h} \otimes \nu_n^{x^b,y^b, h}), \delta_{\left<x^a, x^b \right> + \left<y^a, y^b \right>}) .
\end{align*}
Since
\begin{align*}
\lim_{n \to \infty} \sup_{(x^a,y^a), (x^b, y^b) \in B_{2d}(0,1)} d_{\operatorname{BL}_1} ({R_n^{a,b}}_* (\nu_n^{x^a,y^a,h} \otimes \nu_n^{x^b,y^b, h}), \delta_{\left<x^a, x^b \right> + \left<y^a, y^b \right>}) = 0 
\end{align*}
by \cref{thm:overlapMC}, it follows that
\begin{align*}
\limsup_{n \to \infty} d_{\operatorname{BL}_1} \left( {R_n^{a,b}}_* (\mu_n^h \otimes \mu_n^h),  (\alpha^h \otimes \alpha^h) [\delta_{\left<x^a, x^b \right> + \left<y^a, y^b \right>}] \right) \leq 2 \limsup_{n \to \infty} d_{\operatorname{BL}_1} (\alpha_n^h, \alpha^h) . 
\end{align*}
\section*{Declarations}
\noindent There are no competing interests. Data sharing not applicable to this article as no datasets were generated or analysed during the current study.
\section*{Acknowledgements}
\noindent Kalle Koskinen gratefully acknowledges financial support of the European Research Council through the ERC StG MaTCh, grant agreement n. 101117299, the financial support of the Academy of Finland, via an Academy project (project No. 339228) and the Finnish {\em centre of excellence in Randomness and STructures\/} (project No. 346306), which also made possible the visit of Christof K\"ulske to the University of Helsinki to start the project. 
\printbibliography  

\end{document}